\documentclass[review]{elsarticle}

\biboptions{sort&compress}

\usepackage{ulem}

\usepackage{lineno,hyperref}
\usepackage{amsmath}

\usepackage[utf8]{inputenc}
\usepackage[T1]{fontenc}

\usepackage{float}
\usepackage{multirow}
\usepackage{longtable}
\usepackage{booktabs}
\usepackage{array}
\usepackage{pdflscape}
\usepackage{multirow}
\usepackage{wrapfig}
\usepackage{color}
\usepackage{subcaption}
\usepackage{adjustbox}
\usepackage{csquotes}
\usepackage{soul}

\newcommand{\Cc}[1]{{\mathcal C}_{#1}}

\newcommand{\C}[1]{{\mathcal C}_{#1}}

\definecolor{DRed}{rgb}{0.8,0,0.1}
\definecolor{DBlue}{rgb}{0,0,0.8}

\newcommand{\B}{{\mathcal B}}

\graphicspath{{Plots/}}

\usepackage{fancyhdr}
\fancypagestyle{firstpage}{%
	
	\lhead{}
	\rhead{\small BARI-TH/21-732}
}
\begin{document}

\begin{frontmatter}

\title{$b\to s\ell^+\ell^-$ Global Fits after $R_{K_S}$ and $R_{K^{*+}}$}

\author[1,2]{Marcel Algueró\corref{mycorrespondingauthor}}
\cortext[mycorrespondingauthor]{Corresponding author}
\ead{malguero@ifae.es}
\author[3]{Bernat Capdevila}
\author[4]{Sébastien Descotes-Genon}
\author[1,2]{Joaquim Matias}
\author[4,5]{Martín Novoa-Brunet}

\address[1]{Grup de F\'isica Te\`orica (Departament de F\'isica), Universitat Aut\`onoma de Barcelona, E-08193 Bellaterra (Barcelona), Catalunya.}
\address[2]{Institut de F\'isica d'Altes Energies (IFAE), The Barcelona Institute of Science and Technology, Campus UAB, E-08193 Bellaterra (Barcelona), Catalunya.}
\address[3]{Universit\`a di Torino and INFN Sezione di Torino, Via P. Giuria 1, Torino I-10125, Italy.}
\address[4]{Universit\'e Paris-Saclay, CNRS/IN2P3, IJCLab, 91405 Orsay, France.}
\address[5]{Istituto Nazionale di Fisica Nucleare, Sezione di Bari, Via Orabona 4,
I-70126 Bari, Italy}

\begin{abstract}
 We present an up-to-date complete
 model-independent global fit to $b\to s\ell\ell$ observables that confirms patterns of New Physics able to explain the data. We include the recent LHCb measurements of $R_K$,  $R_{K_S}$, $R_{K^{*+}}$, $B_s \to \phi\mu^+\mu^-$ and $B_s\to\mu^+\mu^-$ in our analysis, which now includes 254 observables. This updates our previous analyses and strengthens their two main outcomes. First, the presence of right-handed couplings encoded in the Wilson coefficients $\Cc{9'\mu}$ and $\Cc{10'\mu}$ 
remains a viable possibility. Second, a lepton flavour universality violating (LFUV) left-handed lepton coupling ($\Cc{9\mu}^{\rm V}=-\Cc{10\mu}^{\rm V}$), often preferred from the model building point of view, accommodates the data better if lepton-flavour universal New Physics is allowed, in particular in $\Cc{9}^{\rm U}$. We observe that the LFUV observable $Q_5$ offers a very interesting possibility to separate both types of scenarios.
\end{abstract}

% \begin{keyword}
% \tbd{SAME AS THE ONES OF ADDENDUM of EPJC}
% \end{keyword}

\end{frontmatter}

\thispagestyle{firstpage}
\section{Introduction}

The flavour anomalies in $b\to s\ell\ell$ processes are currently among the most promising signals of New Physics (NP)~\cite{Bifani:2018zmi,Albrecht:2021tul,London:2021lfn}. This has been reinforced by the recent LHCb updates of quantities assessing the violation of lepton-flavour universality (LFU). On the one hand,
we have the ratio $R_K$~\cite{Aaij:2021vac}:
\begin{equation}
R_{K}=\frac{\B(B^+\to K^+\mu^+\mu^-)}{\B(B^+\to K^+e^+e^-)}\qquad R^{[1.1,6]}_{K,{\rm{LHCb}}}=0.846_{-0.039 \, -0.012}^{+0.042 \, +0.013}
\end{equation}
with an extended statistics corresponding to 9 fb$^{-1}$, reaching the level of statistical evidence (above 3 standard deviations).
On the other hand, similar quantities have been recently measured for the experimentally challenging modes~\cite{LHCb:2021lvy}
\begin{equation}
R_{K_S}=\frac{\B(B^0\to K_S\mu^+\mu^-)}{\B(B^0\to K_Se^+e^-)}\qquad
R_{K^{*+}}=\frac{\B(B^+\to K^{*+}\mu^+\mu^-)}{\B(B^+\to K^{*+}e^+e^-)}
\end{equation}
with the results
\begin{equation}
R^{[1.1,6]}_{K_S,{\rm{LHCb}}}=0.66_{-0.14\,-0.04}^{+0.20\,+0.02} \qquad
R^{[0.045,6]}_{K^{*+},{\rm{LHCb}}}=0.70_{-0.13\,-0.04}^{+0.18\,+0.03}
\end{equation}
in agreement  each with the SM  below the 2$\sigma$ level
%
%at the 1.4 and 1.5 $\sigma$ levels respectively,
%
but consistent with the downward trend compared to the predictions of the Standard Model (SM).
Indeed, in the SM, these ratios are protected from hadronic contributions and are known to be 1 up to (tiny) electromagnetic corrections and (simple) kinematic mass effects.

The deviations observed in these modes can be efficiently and consistently analysed in a model-independent effective field theory (EFT) framework~(see, for instance,  Refs.~\cite{Descotes-Genon:2015uva,Capdevila:2017bsm,Alguero:2018nvb,Alguero:2019ptt,Geng:2021nhg,Altmannshofer:2021qrr,Hurth:2020ehu,Alok:2019ufo,Ciuchini:2020gvn, Datta:2019zca,Hurth:2020rzx}), where short-distance physics (SM and NP) is encoded in the Wilson coefficients of higher-dimension operators.\footnote{It is interesting to point out that the results in Ref.~\cite{Hurth:2020ehu} are very similar to the ones found in the analysis presented in this article. Although they use a similar set of observables (with the addition of baryon decays), the analyses differ through the treatment of hadronic uncertainties (form factors, charm-loop contributions). This similarity illustrates the robustness of the results with respect to different assumptions on hadronic uncertainties.}

This tool has proven particularly helpful in identifying NP scenarios (or patterns of NP) that could explain the data at the level of the EFT, providing guidelines for the construction of phenomenologically viable NP models.

In this context, we present here the latest theoretical and experimental update of our previous works in Refs.~\cite{Capdevila:2017bsm,Alguero:2018nvb,Alguero:2019ptt} to serve as an accurate guideline for model building, as well as an overview of observables relevant for the near future. We follow the same theoretical and statistical approach as in our previous works, updating and adding new experimental inputs and their corresponding SM predictions. It is important at this point to check if the inclusion of this new data alters some of our earlier conclusions,  in particular concerning
best-fit points and confidence intervals that are required for model building as well as the hierarchy of the various NP scenarios that are favoured by the current global fits. It turns out 
that our conclusions remain unchanged and are thus very robust. We will therefore discuss the outcome of our updated global fits 
 but we refer the interested reader to  Ref.~\cite{Alguero:2019ptt} for a more detailed  interpretation of our results as well as the differences with respect to other approaches~\cite{Geng:2021nhg,Altmannshofer:2021qrr,
 Hurth:2020ehu,
 Ciuchini:2020gvn}.
 
The structure of this article is the following. In Section \ref{section2} we list the additional and updated measurements included. Section \ref{section3} is devoted to the methodology of the global fit, with updated results presented in Section \ref{section4}. The link between neutral and charged anomalies using a scenario involving LFUV and LFU NP is discussed in Section \ref{section5}. An overview of the main results and conclusions is given in Section \ref{section6}, together with a proposal to disentangle the main two solutions of the global fit.  Finally, the list of 
experimental inputs and SM predictions for the observables included in our fits  is discussed in \ref{app:predictions}.

\section{Observables}\label{section2}

We consider the same observables and theoretical inputs as in Ref.~\cite{Alguero:2019ptt}, taking into account the following updated measurements (replacing the previous ones):
\begin{itemize}
\item The experimental values of $R_K$, $R_{K_S}$ and $R_{K^{*+}}$ from the LHCb collaboration already discussed in the introduction~\cite{Aaij:2021vac,LHCb:2021lvy}. We also take into account their update of $R_K$~\cite{LHCb:2021trn} as well as the branching ratios
for $B^{0,+}\to K^{0,+}\mu^+\mu^-$ updated by the Belle collaboration~\cite{Abdesselam:2019lab} (the Belle measurements of $R_{K^{(*)}}$ correspond to a combination of the charged and neutral channels $B^{0,+}\to K^{(*)0,+}\ell^+\ell^-$).

\item The experimental value of the branching ratio $\B(B_s\to\mu^+\mu^-)$ from the LHCb collaboration~\cite{LHCb:2021vsc}, which is combined with the results from CMS~\cite{Sirunyan:2019xdu} and ATLAS~\cite{Aaboud:2018mst}, leading to the average $\B(B_s\to\mu^+\mu^-)=2.85_{-0.31}^{+0.34} \times 10^{-9}$~\cite{Hurth:2021nsi}. This is to be compared with the most updated theoretical computation~\cite{MisiakOrsay}.

\item The angular distribution of $B^+\to K^{*+}\mu^+\mu^-$ \cite{LHCb:2020gog} using the optimised observables $P_i$~\cite{Descotes-Genon:2013vna} measured by LHCb, as well as the longitudinal polarisation and forward-backward asymmetry measured by the CMS collaboration~\cite{CMS:2020oqb}.
Compared to the neutral case, our computation for the charged case takes into account the different spectator quark not only by modifying 
the mass and lifetime, but also the annihilation and hard-spectator interactions following Ref.~\cite{Beneke:2004dp}.

\item The angular distribution of $B^+\to K^{+}\mu^+\mu^-$ from the CMS collaboration~\cite{Sirunyan:2018jll}.

\item The angular analysis of $B\to K^*e^+e^-$ at low $q^2$ from the LHCb collaboration~\cite{Aaij:2020umj}. The bins of this analysis are different from the previous ones~\cite{Aaij:2015dea}, but the measurements are correlated since the latter analysis includes the data of the former, leading us to discard Ref.~\cite{Aaij:2015dea}.

\item The new angular analysis and branching ratio of $B_s\to \phi\mu^+\mu^-$ from the LHCb collaboration~\cite{LHCb:2021xxq,LHCb:2021zwz} superseding the previous LHCb analysis~\cite{Aaij:2015esa}. We focus on CP-averaged quantities, as we will consider only CP-conserving New Physics.
\end{itemize}

We do not consider here the baryon mode $\Lambda_b\to \Lambda\mu^+\mu^-$~\cite{LHCb:2018jna}, as there is a known issue with the normalisation provided by the $\Lambda_b$ production fraction which may distort the results~\cite{Blake:2019guk,London:2021lfn}. We think that it is important that LHCb reanalyses this normalization without relying on combinations of LEP and Tevatron studies performed at different energies, so that corrected results of this mode could be included in future global analyses of $b\to s\ell\ell$ transitions in a completely safe way.

\section{Fit approach} \label{section3}

Our starting point is the weak effective Hamiltonian~\cite{Grinstein:1987vj,Buchalla:1995vs} in which heavy degrees of freedom (the top quark, the $W$ and $Z$ bosons, the Higgs boson and any potential heavy new particles) have been integrated out in short-distance Wilson coefficients $\C{i}$, leaving only a set of operators ${\mathcal O}_i$ describing the physics at long distances:
\begin{equation}
{\mathcal H}_{\rm eff}=-\frac{4G_F}{\sqrt{2}} V_{tb}V_{ts}^*\sum_i \C{i}  {\mathcal O}_i + h.c.,
\end{equation}
up to small corrections proportional to $V_{ub}V_{us}^*$ in the SM, included in our numerical analysis. 

In the SM, the Hamiltonian contains 10 main operators with specific chiralities due to the $V-A$ structure of the weak interactions. In presence of NP, additional operators may become of importance. For the processes considered here, we focus our attention on the operators
${\mathcal O}_{7^{(\prime)},9^{(\prime)}\ell,10^{(\prime)}\ell}$ and their associated Wilson coefficients  $\C{7^{(\prime)}},\C{9^{(\prime)}\ell},\C{10^{(\prime)}\ell}$ where $\ell=e$ or $\mu$. 
$\C{7^{(\prime)}}$ describe the interaction strength of bottom ($b$) and strange ($s$) quarks with the photon while $\C{9\ell,10\ell}$ and $\C{9'\ell,10'\ell}$ encode the interaction strength of $b$ and $s$ quarks with charged leptons.
$\C{9\ell,10\ell}$ and $\C{9'\ell,10'\ell}$ are equal for muons and electrons in the SM but NP can add different contributions to muon operators compared to the electron ones. For $\C7$ and $\C{9\ell,10\ell}$ we split SM and NP contributions like ${\mathcal C}_{i\ell}={\mathcal C}_{i\ell}^{\rm SM}+ {\mathcal C}_{i\ell}^{\rm  NP}$. The Wilson coefficients of the chirally-flipped operators are zero in the SM, apart from $\C{7'}$ which features a small SM contribution of $O(m_s/m_b)$.

%(the SM contributions to chirally-flipped operators are negligible). 

 Our evaluation of the various observables follows the same approach as in Ref.~\cite{Descotes-Genon:2015uva} with the updates of the theoretical inputs discussed in Refs.~\cite{Capdevila:2017bsm,Alguero:2019ptt}. Attention must naturally be paid to hadronic uncertainties~\cite{Jager:2012uw,Descotes-Genon:2014uoa,Jager:2014rwa,Ciuchini:2015qxb,Capdevila:2017ert,Ciuchini:2017mik,Arbey:2018ics}, which stems from two different sources in exclusive $b\to s\ell\ell$ decays such as $B \to K^{(*)}\ell^+\ell^-$ and $B_s\to\phi\ell^+\ell^-$. First, form factors must be determined through different methods at large recoil of the final hadron (light-cone sum rules involving either   light-meson~\cite{Ball:2004rg,Bharucha:2015bzk} or $B$-meson~\cite{Khodjamirian:2010vf,Khodjamirian:2012rm,Gubernari:2018wyi,Gubernari:2020eft} distribution amplitudes) or low recoil (lattice QCD~\cite{Bouchard:2013eph,Horgan:2013hoa}). Second, the non-local contribution from $c\bar{c}$ loops can be tackled similarly either at low recoil, through quark-hadron duality arguments for observables averaged over a large dilepton invariant mass~\cite{Grinstein:2002rm,Bobeth:2010wg,Beylich:2011aq,Bobeth:2012vn}, or large recoil, using various approaches (order-of-magnitude estimates, light-cone sum rule computations~\cite{Khodjamirian:2010vf,Khodjamirian:2012rm}, interpolation from the unphysical region below the photon pole up to the lowest charmonium resonances~\cite{Gubernari:2020eft,Bobeth:2017vxj}, \ldots). Obviously, the uncertainties of the theoretical predictions for these observables (within the SM or any NP scenario) are partly dependent on these assumptions. However, it is quite striking to notice that different analyses based on different underlying assumptions for these hadronic uncertainties may yield different numerical values for statistical quantities (significances, pulls, \ldots) but they have repeatedly led to very similar patterns of favoured scenarios, best-fit points and confidence regions for NP contributions to Wilson coefficients (see for instance Refs.~\cite{Geng:2021nhg,
Altmannshofer:2021qrr,Hurth:2020ehu,
Hurth:2021nsi, Ciuchini:2021smi}).

In practice, we perform fits to obtain information on the values of the parameters collectively denoted here as $\theta$, which represent the unknown NP contributions from the different scenarios that we estimate (e.g. $\Cc{9\mu}^{\rm NP}$, $\Cc{9\mu}^{\rm NP}=-\Cc{10\mu}^{\rm NP}$, etc). We work within a frequentist framework based on a gaussian approximation for the likelihood function $\mathcal{L}(\theta)$ where theoretical and experimental uncertainties are treated on the same footing:
\begin{equation}\label{eq:chi2function}
\begin{split}
-2\ln\mathcal{L}(\theta)&=\chi^2(\theta) \\
&=\sum_{i,j=1}^{N_{\rm obs}}\left(O^{th}(\theta)-O^{exp}\right)_i\left(V^{th}(\theta)+V^{exp}\right)_{ij}\left(O^{th}(\theta)-O^{exp}\right)_j,
\end{split}
\end{equation}
with $N_{\rm obs}$ the total number of observables in the fit, $O^{th}_i(\theta)$ the central value of the theory prediction for the $i$-th observable, $O^{exp}_i$ the experimental measurement (i.e. the central value quoted by experiments) of the same observable and $V^{th}_{ij}$ and $V^{exp}_{ij}$ the theoretical and experimental covariance matrices respectively.

On the one hand, the experimental covariance matrix contains all the available information on the errors and correlations among the measurements of the relevant observables released by the different experiments. Whenever the correlations are not available, we take those measurements as uncorrelated. In the case of asymmetric uncertainties (such as $R_K$), in order to be consistent with the gaussian approximation of the likelihood function, we symmetrise the errors by taking the largest uncertainty, with no change in the central value. On the other hand, the theoretical covariance matrix is estimated by performing a multivariate gaussian scan over all the nuisance parameters entering the calculation of theory predictions which we do not fit through the minimisation procedure.

The central values of the unknown parameters in our analyses are estimated by means of the \textit{method of maximum likelihood} (ML). By construction of the likelihood, the ML estimators $\hat{\theta}$ coincide with the best-fit points obtained by minimising the $\chi^2$ function: 
\begin{equation}
\left.\frac{\partial\chi^2}{\partial\theta_i}\right|_{\hat{\theta}}=0 \qquad {\rm such\,  that} \qquad \chi_{\rm min}=\chi^2(\hat{\theta}),
\end{equation}
for $i=1,...,n$, with $n$ being the number of parameters.  The minimisation is performed numerically using MIGRAD from the Python package \texttt{iMinuit}~\cite{iminuit}. 
For computational reasons, the theoretical covariance is assumed to depend mildly on the NP parameters, hence we take $V^{th}(\theta)$ in Eq.~(\ref{eq:chi2function}) at the SM point. We checked that our results remain unchanged if we repeat the fits with the $V^{th}(\theta)$ evaluated at different NP points, confirming the validity of our approximation.  This is in agreement with the results of Refs.~\cite{Descotes-Genon:2015uva,Bhom:2020lmk,Altmannshofer:2021qrr}, where the impact of accounting for the correlated theoretical uncertainties at each point in the Wilson coefficient parameter space was analysed in full detail.

In order to provide a complete description of the parameters, we also assess their errors and correlations. This information is encoded in the likelihood function and can be accessed through the Rao-Cramér-Fréchet formula for the inverse $V^{-1}$ of the covariance matrix $V_{ij}={\rm cov}(\hat{\theta}_i,\hat{\theta}_j)$ of the estimators
\begin{equation}
\left(V^{-1}\right)_{ij}=-\left.\frac{\partial^2\ln\mathcal{L}}{\partial\theta_i\partial\theta_j}\right|_{\hat{\theta}}=\frac{1}{2}\left.\frac{\partial^2\chi^2}{\partial\theta_i\partial\theta_j}\right|_{\hat{\theta}}.
\end{equation}
 In practice, the likelihood's Hessian matrix is numerically computed by MIGRAD as one of the outputs of the minimisation routine. Instead, for the computation of confidence intervals we use \texttt{iMinuit}’s MINOS algorithm~\cite{iminuit}. 

To quantify the level of agreement between a given hypothesis and the data, we compute the corresponding $p$-value of \textit{goodness-of-fit}:
\begin{equation}
p=\int_{\chi^2_{\rm min}}^\infty d\chi^2\,f(\chi^2;n_{\rm dof}),
\end{equation}
where $n_{\rm dof}=N_{\rm obs}-n$. Finally, to compare the descriptions offered by two different nested hypotheses $H_0$ and $H_1$ (with $n_{\rm{H_0}}$, $n_{\rm{H_1}}$ the respective number of degrees of freedom and $n_{\rm{H_0}}<n_{\rm{H_1}}$), we compute their relative Pull, measured in units of Gaussian standard deviations ($\sigma$):
\begin{equation}
{\rm Pull}_{H_0H_1}=\sqrt{2}\,{\rm Erf}^{-1}\left[F(\Delta\chi^2_{H_0H_1};n_{\rm{H_0 H_1}})\right] ,
\end{equation}
with $\Delta\chi^2_{H_0H_1}=\chi^2_{H_0,{\rm min}}-\chi^2_{H_1,{\rm min}}$, $n_{\rm{H_0H_1}}=n_{\rm{H_1}}-n_{\rm{H_0}}$, $F$ the $\chi^2$ cumulative distribution function and ${\rm Erf}^{-1}$ the inverse error function. Most of the time, we compare a given NP scenario with the SM case, denoting the result as ${\rm Pull}_{\rm SM}$ unless there is a risk of ambiguity.
Our statistical interpretation, based on Wilks' theorem~\cite{Wilks:1938dza}, assumes that the large number of observables leads to a statistical question where the linear/Gaussian approximation holds and that all observables have a similar sensitivity to all Wilson coefficients, so that the number of degrees of freedom can be computed as described above. This issue has been recently discussed in Refs.~\cite{Isidori:2021vtc,Isidori:2021tzd}  (see also earlier discussions on this topic in Refs.~\cite{Arbey:2018ics,Hurth:2018kcq}). These studies suggest that the effective number of degrees of freedom to be actually considered could be lower than  what a naive computation would indicate, due to a weak sensitivity of the $\chi^2$ function to some of the Wilson coefficients. In that case, our interpretation would be conservative, since it yields higher p-values and lower pulls than with the smaller effective number of degree of freedom advocated in these references.

\section{Fit Results}\label{section4}

\begin{figure}[!ht]
\begin{center}
\includegraphics[width=0.3\textwidth]{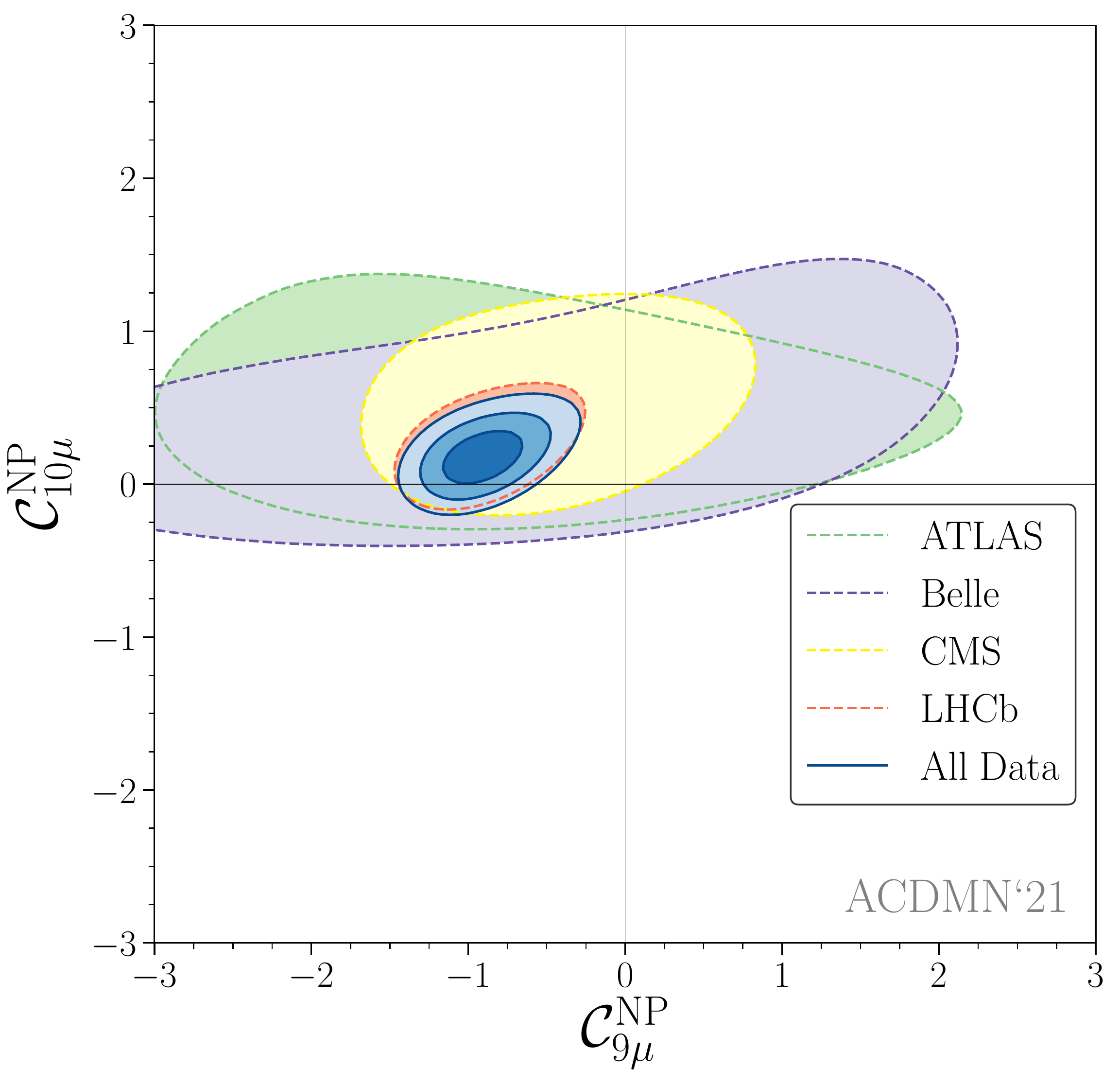}\hspace{2mm}
\includegraphics[width=0.3\textwidth]{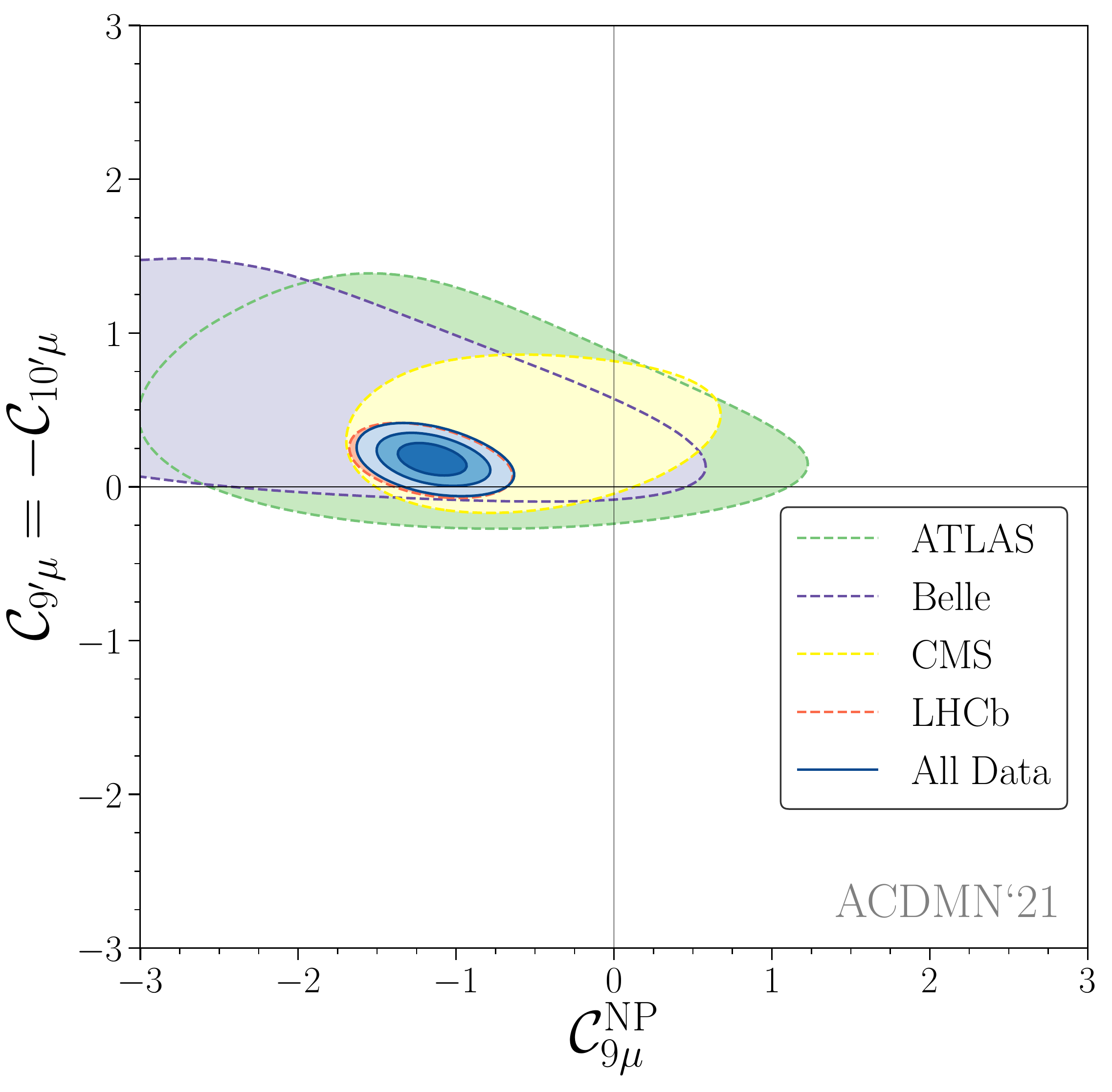}\hspace{2mm}
\includegraphics[width=0.3\textwidth]{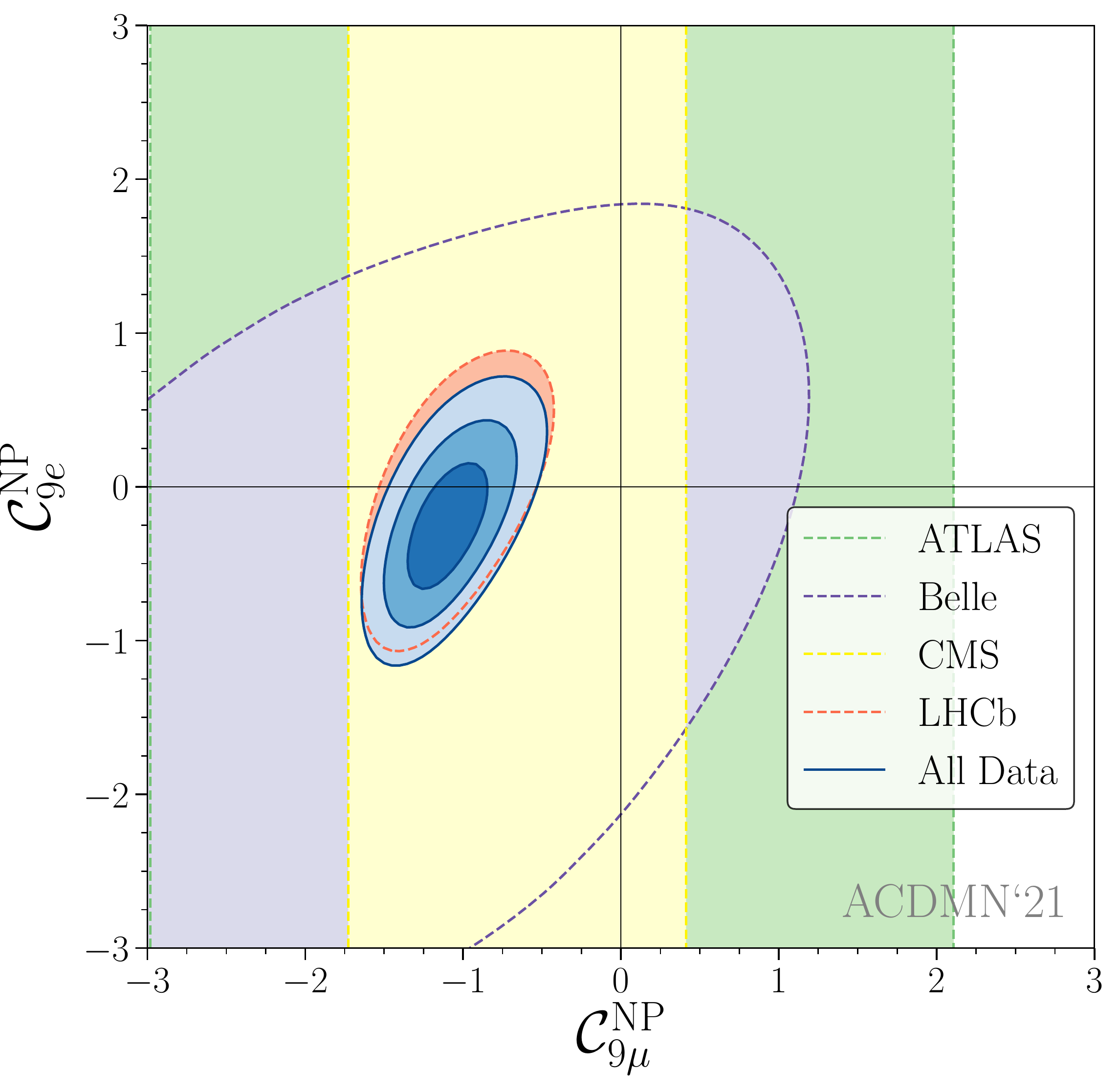}
\includegraphics[width=0.3\textwidth]{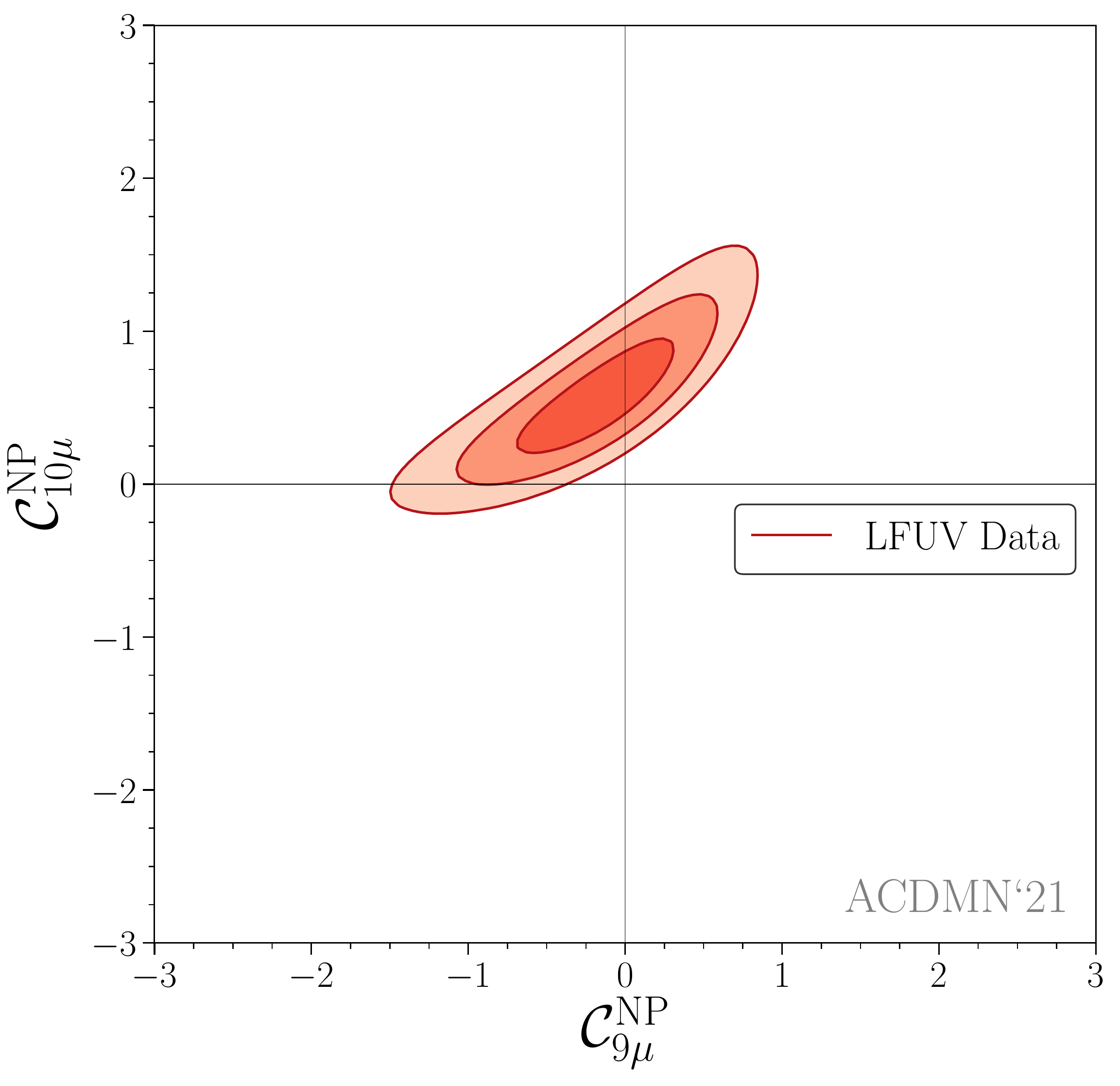}\hspace{2mm}
\includegraphics[width=0.3\textwidth]{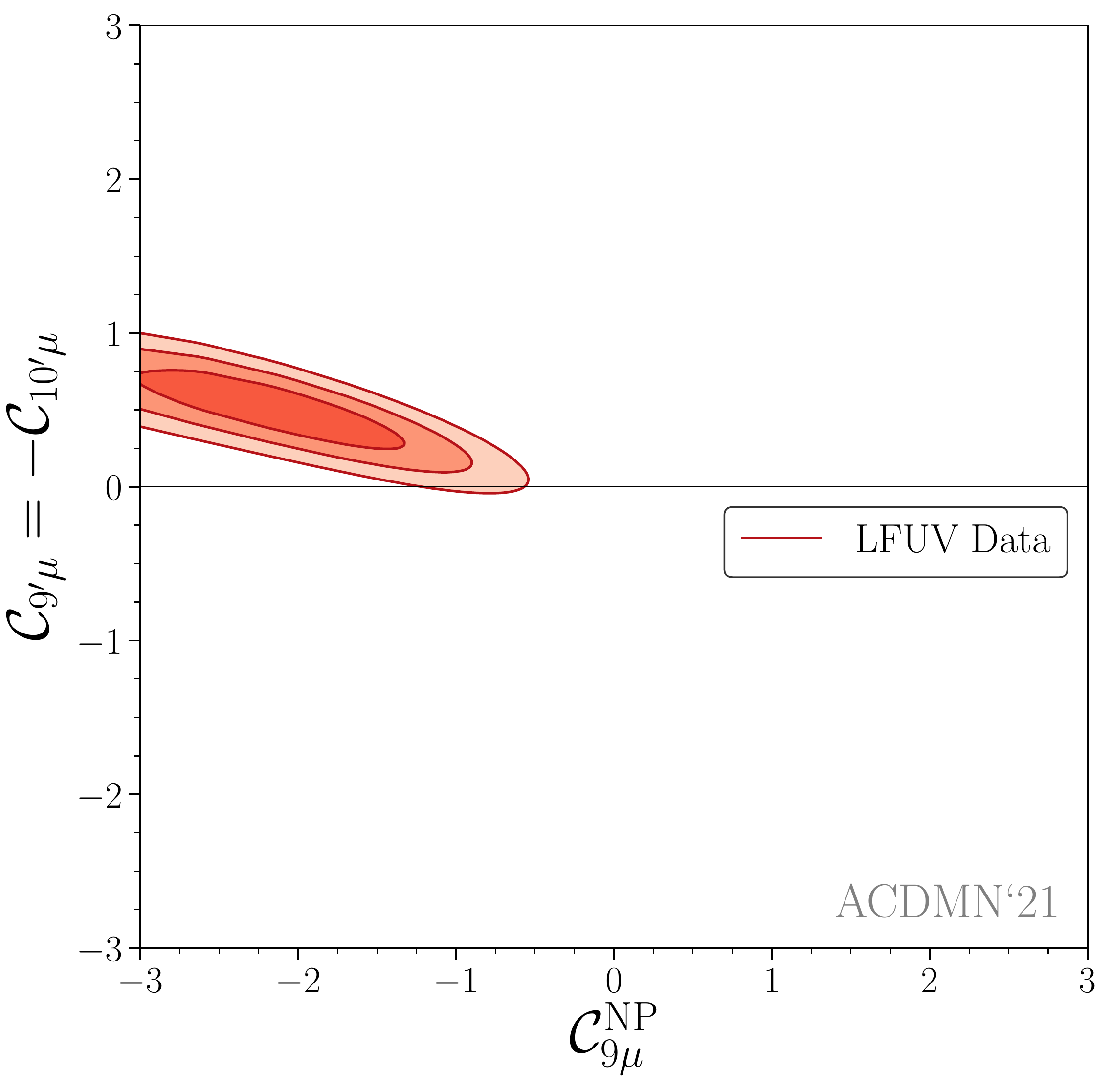}\hspace{2mm}
\includegraphics[width=0.3\textwidth]{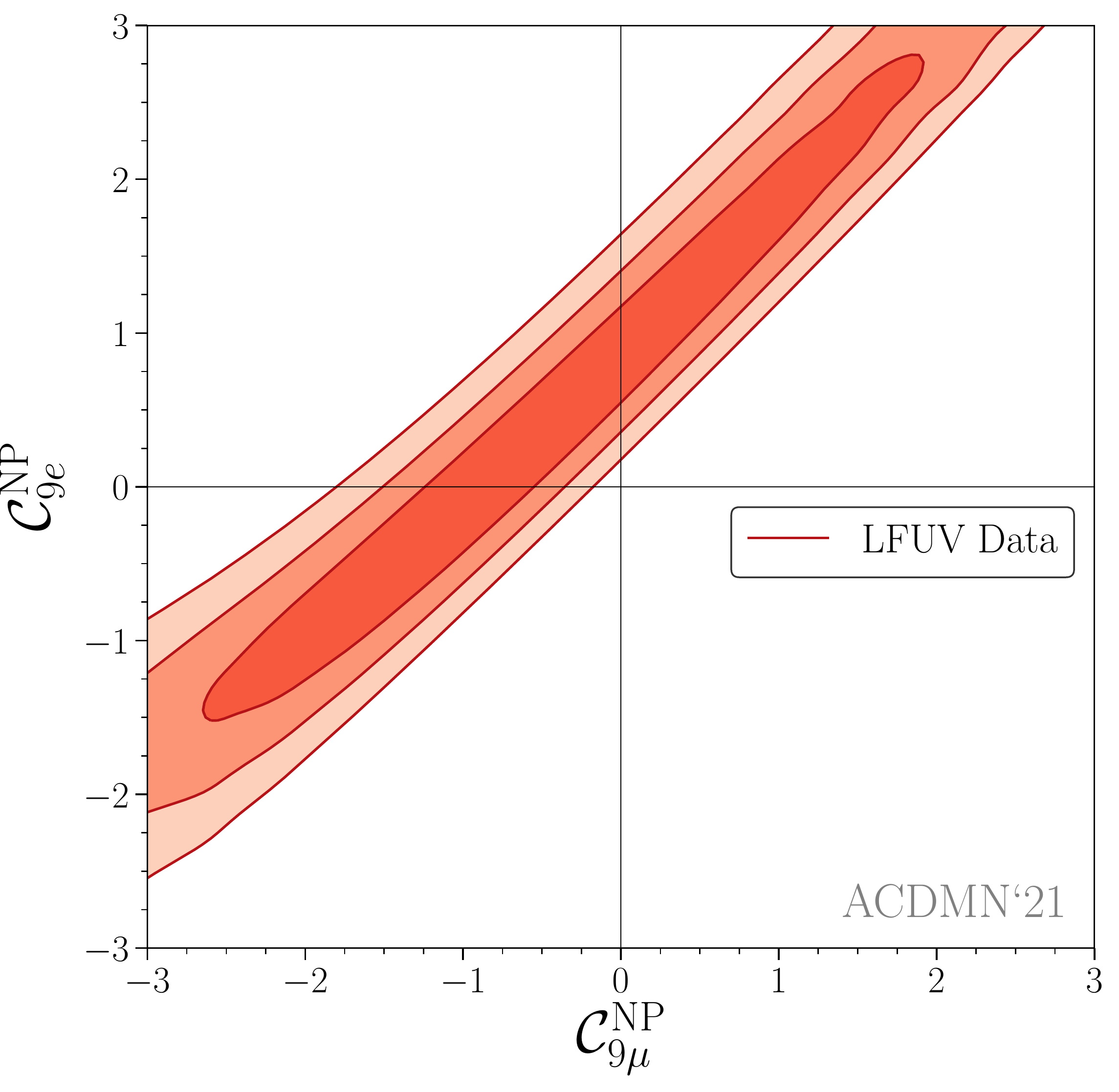}
\end{center}
\caption{From left to right: Allowed regions in the $(\Cc{9\mu}^{\rm NP},\Cc{10\mu}^{\rm NP})$, $(\Cc{9\mu}^{\rm NP},\Cc{9^\prime\mu}=-\Cc{{10}^\prime\mu})$ and $(\Cc{9\mu}^{\rm NP},\Cc{9e}^{\rm NP})$ planes for the corresponding 2D hypotheses, using all available data (fit ``All'') upper row or LFUV fit  lower row. Dashed lines represent  the 3~$\sigma$ regions while the solid lines represent 1, 2 and 3~$\sigma$ regions.}
\label{fig:FitResultAll}
\end{figure}

We start by considering the fits to NP scenarios which affect muon modes only.
Tabs.~\ref{tab:results1D}-\ref{tab:Fit6D} and Fig.~\ref{fig:FitResultAll} update the corresponding tables and figures of Ref.~\cite{Alguero:2019ptt} based on fits to the full set of data (``All'', 254 observables\footnote{We detail the full list of the observables present in our fits in the appendix, where we also provide their theoretical predictions within the SM, as well as the individual tension with respect to the experimental value. In the LFUV fits we include the observables $Q_4$ and $Q_5$ (measured by Belle) instead of $P'_{4e,\mu}$, $P'_{5e,\mu}$.}) or restricted to quantities assessing LFUV (``LFUV'', 24 observables). The results are similar to those in Ref.~\cite{Alguero:2019ptt}. 
%\section*{1D Fits}
\begin{table*}[!ht] %\scriptsize
    \centering
    \begin{adjustbox}{width=1.\textwidth,center=\textwidth}
\begin{tabular}{c||c|c|c|c||c|c|c|c} %\scriptsize
 & \multicolumn{4}{c||}{All} &  \multicolumn{4}{c}{LFUV}\\
\hline
1D Hyp.   & Best fit& 1 $\sigma$/2 $\sigma$   & Pull$_{\rm SM}$ & p-value & Best fit & 1 $\sigma$/ 2 $\sigma$  & Pull$_{\rm SM}$ & p-value\\
\hline\hline
\multirow{2}{*}{$\Cc{9\mu}^{\rm NP}$}    & \multirow{2}{*}{-1.01} &    $[-1.15,-0.87]$ &    \multirow{2}{*}{7.0}   & \multirow{2}{*}{24.0\,\%}
&   \multirow{2}{*}{-0.87}   &$[-1.11,-0.65]$&   \multirow{2}{*}{4.4}  & \multirow{2}{*}{40.7\,\%}  \\
 &  & $[-1.29,-0.72]$ &  & &  &  $[-1.37,-0.45]$ & \\
 \multirow{2}{*}{$\Cc{9\mu}^{\rm NP}=-\Cc{10\mu}^{\rm NP}$}    &   \multirow{2}{*}{-0.45} &    $[-0.52,-0.37]$ &   \multirow{2}{*}{6.5}  & \multirow{2}{*}{16.9\,\%}
 &  \multirow{2}{*}{-0.39}   &   $[-0.48,-0.31]$ & \multirow{2}{*}{5.0}   & \multirow{2}{*}{73.5\,\%}  \\
 &  & $[-0.59,-0.30]$ &  & & & $[-0.56,-0.23]$  &    \\
 \multirow{2}{*}{$\Cc{9\mu}^{\rm NP}=-\Cc{9'\mu}$}     & \multirow{2}{*}{-0.92} &    $[-1.07,-0.75]$   &  \multirow{2}{*}{5.7}  & \multirow{2}{*}{8.2\,\%}
 &  \multirow{2}{*}{-1.60}   &    $[-2.10,-0.98]$  & \multirow{2}{*}{3.2} & \multirow{2}{*}{8.4\,\%} \\
 &  & $[-1.22,-0.59]$ &  & & & $[-2.49,-0.46]$ &    \\
%\hline
% \multirow{2}{*}{$\Cc{9\mu}^{\rm NP}=-3 \Cc{9e}^{\rm NP}$} & \multirow{2}{*}{-0.86}  & $[-0.99,-0.73]$ & \multirow{2}{*}{6.7}  & \multirow{2}{*}{20.1\,\%}
%  &   \multirow{2}{*}{-0.65} &    $[-0.82,-0.49]$ & \multirow{2}{*}{4.4}  & \multirow{2}{*}{40.9\,\%} \\
% & & $[-1.13,-0.60]$ &  & & & $[-1.00,-0.34]$     & \\

\end{tabular} \end{adjustbox}
\caption{Most prominent 1D patterns of NP in $b\to s\mu\mu$. Pull$_{\rm SM}$ is quoted in units of standard deviation. The $p$-value of the SM hypothesis is $0.44\%$ for the fit ``All'' and $0.91\%$ for the fit LFUV.} 
\label{tab:results1D}
\end{table*}

%\section*{2D Fits}

\begin{table*}[!ht] 
    \centering
   \begin{adjustbox}{width=0.8\textwidth,center=\textwidth}
%\scriptsize
\begin{tabular}{c||c|c|c||c|c|c} %\scriptsize
 & \multicolumn{3}{c||}{All} &  \multicolumn{3}{c}{LFUV}\\
\hline
 2D Hyp.  & Best fit  & Pull$_{\rm SM}$ & p-value & Best fit & Pull$_{\rm SM}$ & p-value\\
\hline\hline
$(\Cc{9\mu}^{\rm NP},\Cc{10\mu}^{\rm NP})$ & $(-0.92,+0.17)$ & 6.8 & 25.6\,\% & $(-0.16,+0.55)$ & 4.7 & 71.2\,\% \\
$(\Cc{9\mu}^{\rm NP},\Cc{7^{\prime}})$  & $(-1.02,+0.01)$ & 6.7 & 22.8\,\% & $(-0.88,-0.04)$ & 4.1 & 37.5\,\% \\
$(\Cc{9\mu}^{\rm NP},\Cc{9^\prime\mu})$  & $(-1.12,+0.36)$ & 6.9 & 27.4\,\% & $(-1.82,+1.09)$ & 4.5 & 60.2\,\% \\
$(\Cc{9\mu}^{\rm NP},\Cc{10^\prime\mu})$  & $(-1.15,-0.26)$ & 7.1 & 31.8\,\% & $(-1.88,-0.59)$ & 5.0 & 88.1\,\% \\ 
\hline
$(\Cc{9\mu}^{\rm NP}, \Cc{9e}^{\rm NP})$ & $(-1.11,-0.26)$ & 6.7 & 23.8\,\% & $(-0.52,+0.34)$ & 4.0 & 35.3\,\%  \\
\hline
Hyp. 1 & $(-1.01,+0.31)$ & 6.7 & 24.0\,\% & $(-1.60,+0.32)$ & 4.5 & 62.5\,\% \\
Hyp. 2 & $(-0.89,+0.06)$ & 5.4 & 8.0\,\% & $(-1.95,+0.25)$ & 3.6 & 20.4\,\% \\
Hyp. 3 & $(-0.45,+0.04)$ & 6.2 & 15.9\,\% & $(-0.39,-0.14)$ & 4.7 & 70.2\,\% \\
Hyp. 4  & $(-0.47,+0.07)$ & 6.3 & 16.8\,\% & $(-0.48,+0.15)$ & 4.8 & 79.6\,\% \\
Hyp. 5  & $(-1.15,+0.17)$ & 7.1 & 31.1\,\% & $(-2.13,+0.50)$ & 5.0 & 89.4\,\% \\
\end{tabular}
\end{adjustbox}
\caption{Most prominent 2D patterns of NP in $b\to s\mu\mu$. The last five rows correspond to Hypothesis 1: $(\Cc{9\mu}^{\rm NP}=-\Cc{9^\prime\mu} , \Cc{10\mu}^{\rm NP}=\Cc{10^\prime\mu})$,  2: $(\Cc{9\mu}^{\rm NP}=-\Cc{9^\prime\mu} , \Cc{10\mu}^{\rm NP}=-\Cc{10^\prime\mu})$, 3: $(\Cc{9\mu}^{\rm NP}=-\Cc{10\mu}^{\rm NP} , \Cc{9^\prime\mu}=\Cc{10^\prime\mu}$), 4: $(\Cc{9\mu}^{\rm NP}=-\Cc{10\mu}^{\rm NP} , \Cc{9^\prime\mu}=-\Cc{10^\prime\mu})$ and 5: $(\Cc{9\mu}^{\rm NP} , \Cc{9^\prime\mu}=-\Cc{10^\prime\mu})$.}
\label{tab:results2D}
\end{table*}

%\newpage

%\section*{6D Fit}

\begin{table*}[!ht] %\scriptsize
    \centering
    \begin{adjustbox}{width=1.\textwidth,center=\textwidth}
\begin{tabular}{c||c|c|c|c|c|c}%\scriptsize
 & $\Cc7^{\rm NP}$ & $\Cc{9\mu}^{\rm NP}$ & $\Cc{10\mu}^{\rm NP}$ & $\Cc{7^\prime}$ & $\Cc{9^\prime \mu}$ & $\Cc{10^\prime \mu}$  \\
\hline\hline
Best fit & +0.00 & -1.08 & +0.15 & +0.00 & +0.16 & -0.18 \\ \hline
1 $\sigma$ & $[-0.02,+0.01]$ & $[-1.25,-0.90]$ & $[+0.02,+0.28]$ & $[-0.01,+0.02]$ & $[-0.20,+0.53]$ &$[-0.36,+0.02]$ 
\\
2 $\sigma$ & $[-0.04,+0.03]$ & $[-1.41,-0.72]$ & $[-0.10,+0.42]$ & $[-0.03,+0.03]$ & $[-0.56,+0.92]$ &$[-0.54,+0.22]$
\end{tabular}
\end{adjustbox}
\caption{ 1 and 2~$\sigma$ confidence intervals for the NP contributions to Wilson coefficients in
the 6D hypothesis allowing for NP in $b\to s\mu\mu$ operators dominant in the SM and their chirally-flipped counterparts, for the fit ``All''. The Pull$_{\rm SM}$ is $6.3\sigma$ and the \textit{p}-value is $27.8\%$.}
\label{tab:Fit6D}
\end{table*}

\begin{figure}[!ht]
\begin{center}
\includegraphics[width=0.315\textwidth]{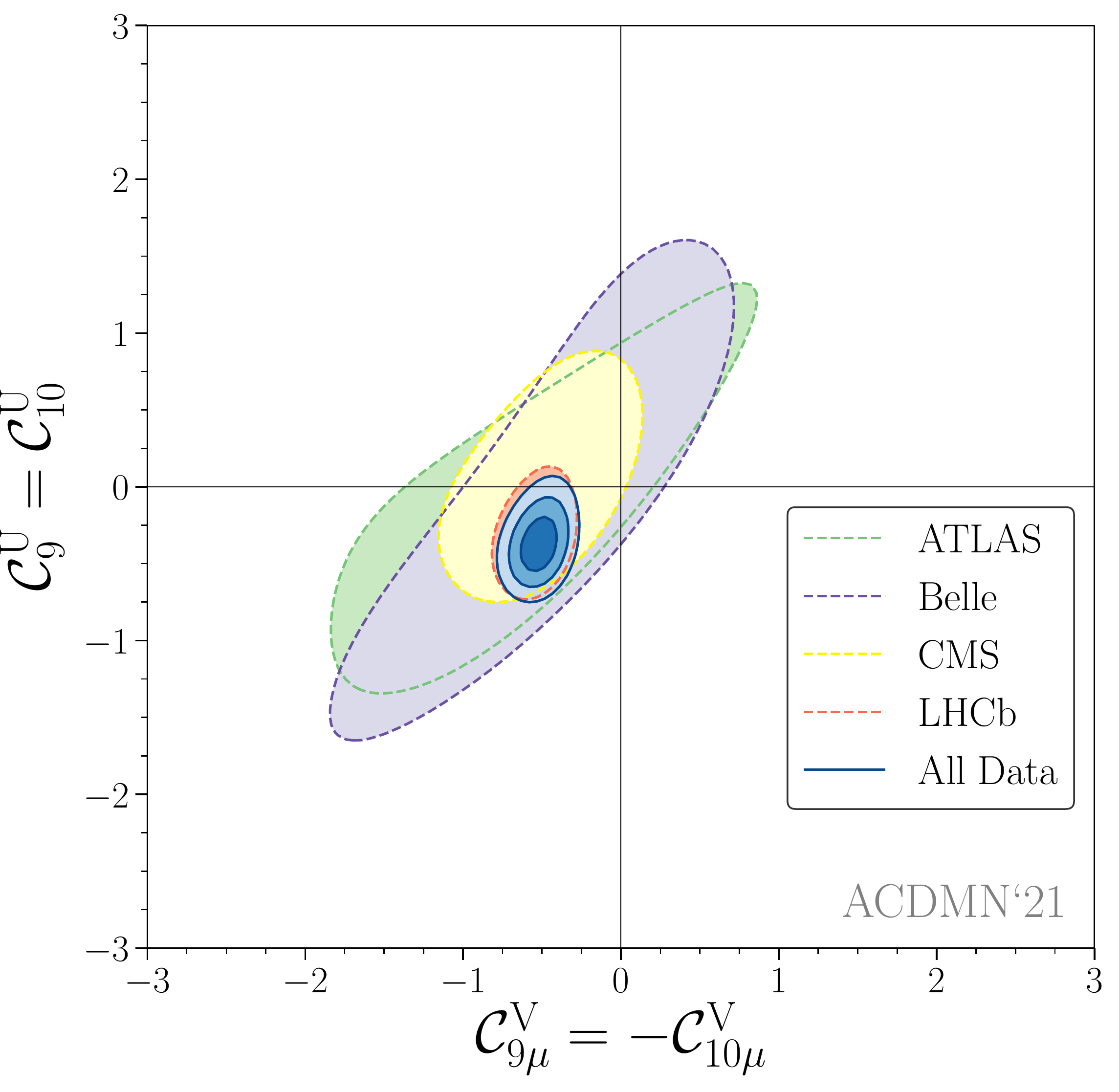}\hspace{2mm}
\includegraphics[width=0.315\textwidth]{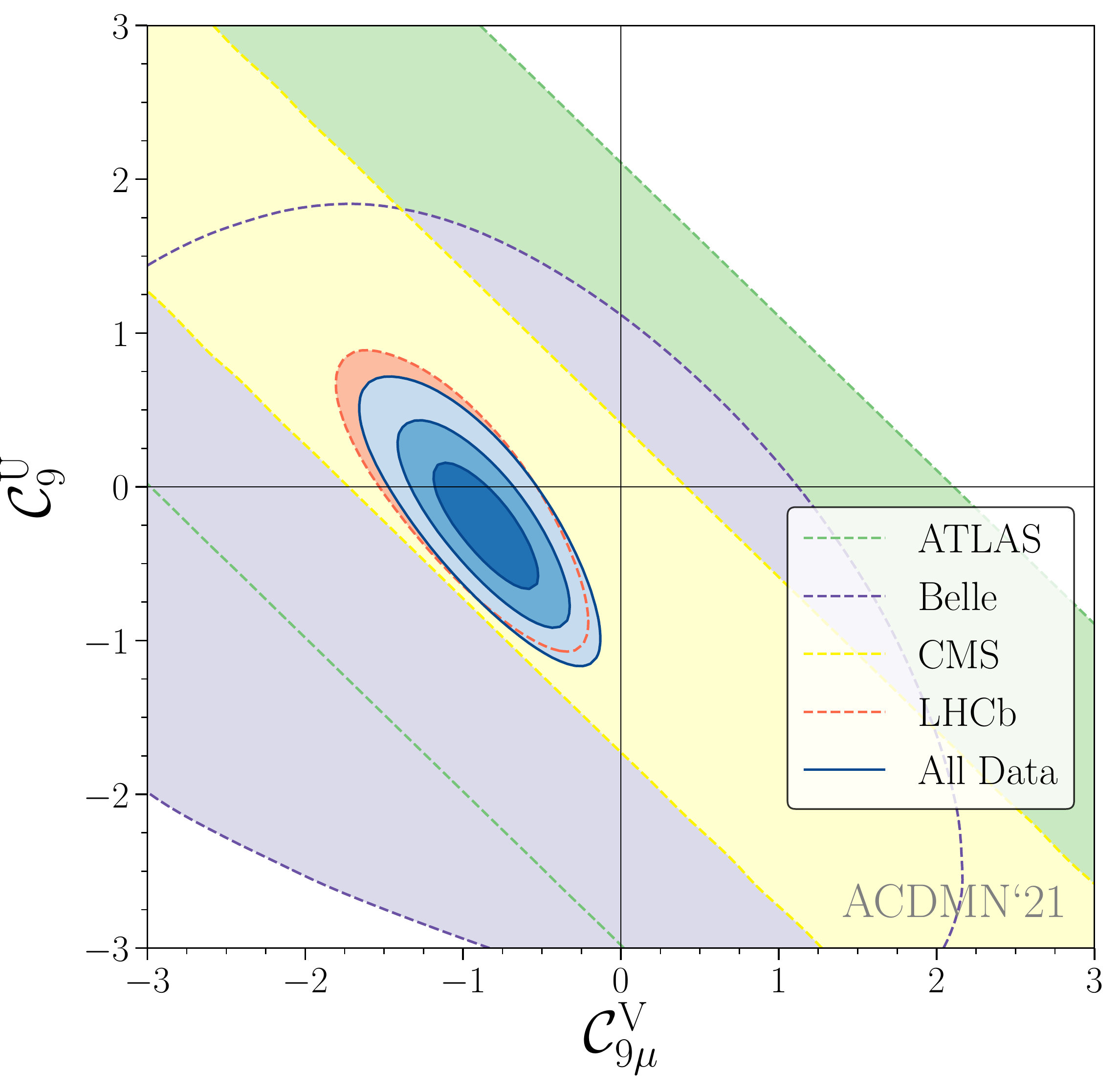}\\
\includegraphics[width=0.315\textwidth]{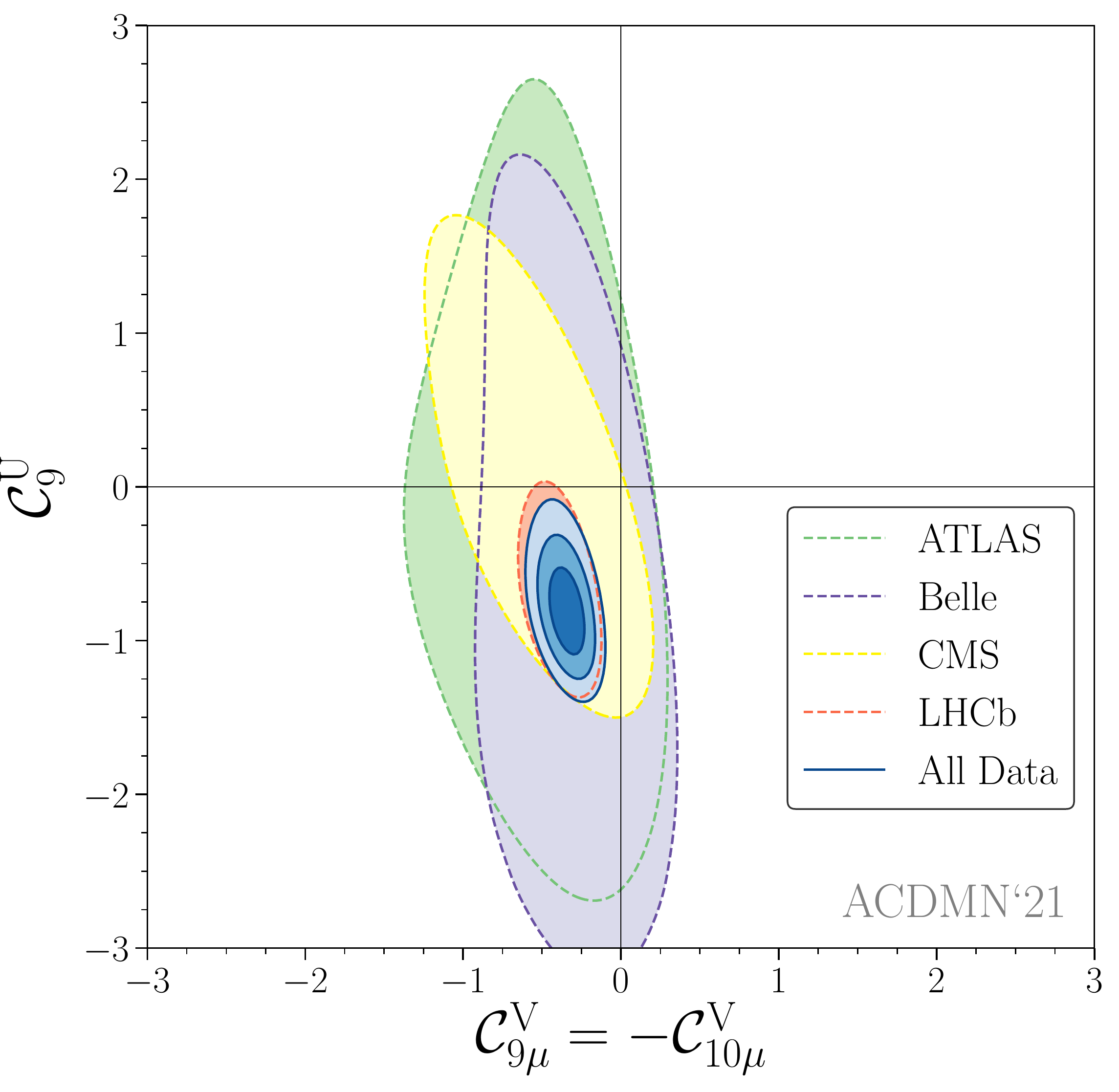}\hspace{2mm}
\includegraphics[width=0.315\textwidth]{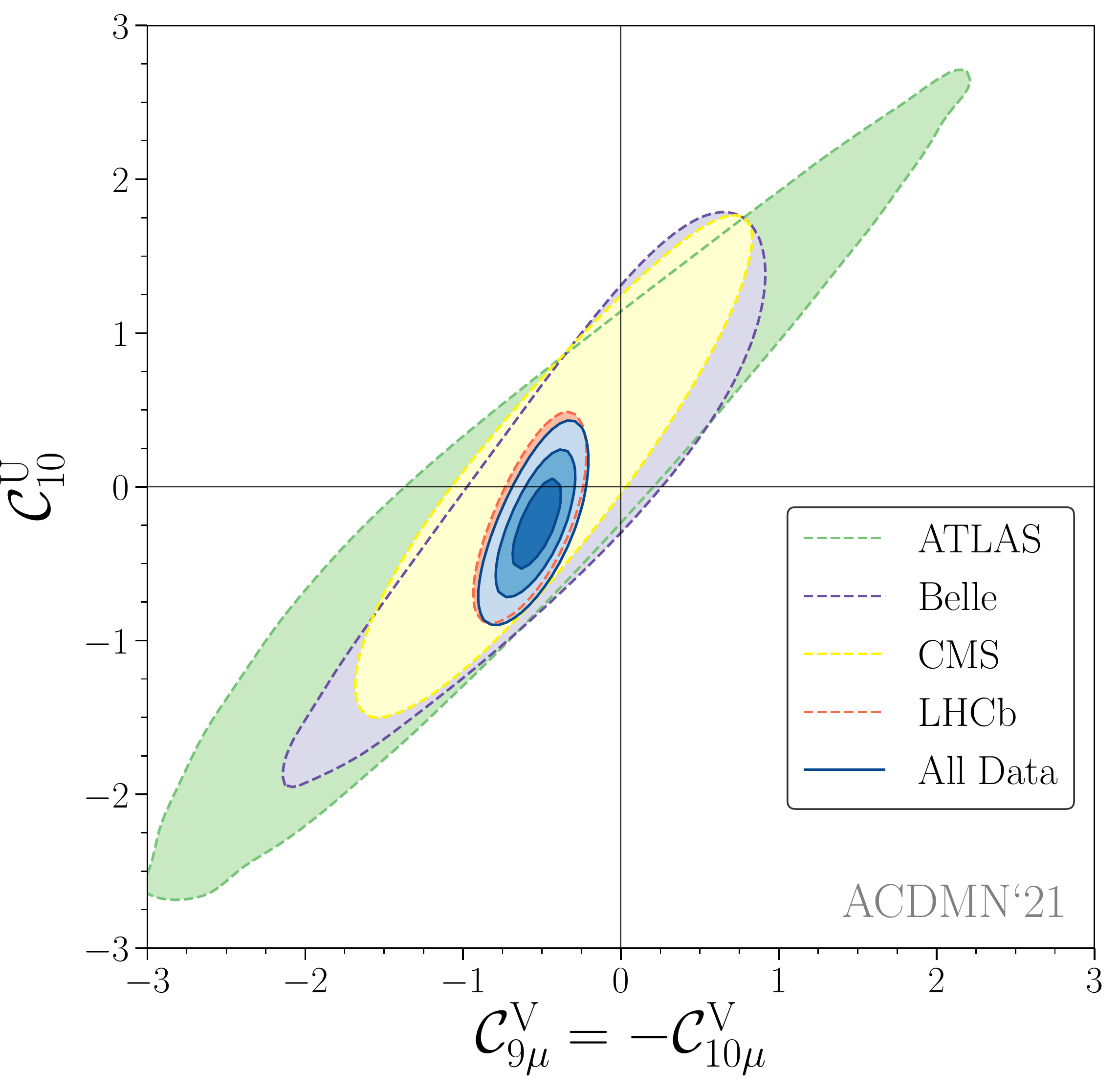}\\
\includegraphics[width=0.315\textwidth]{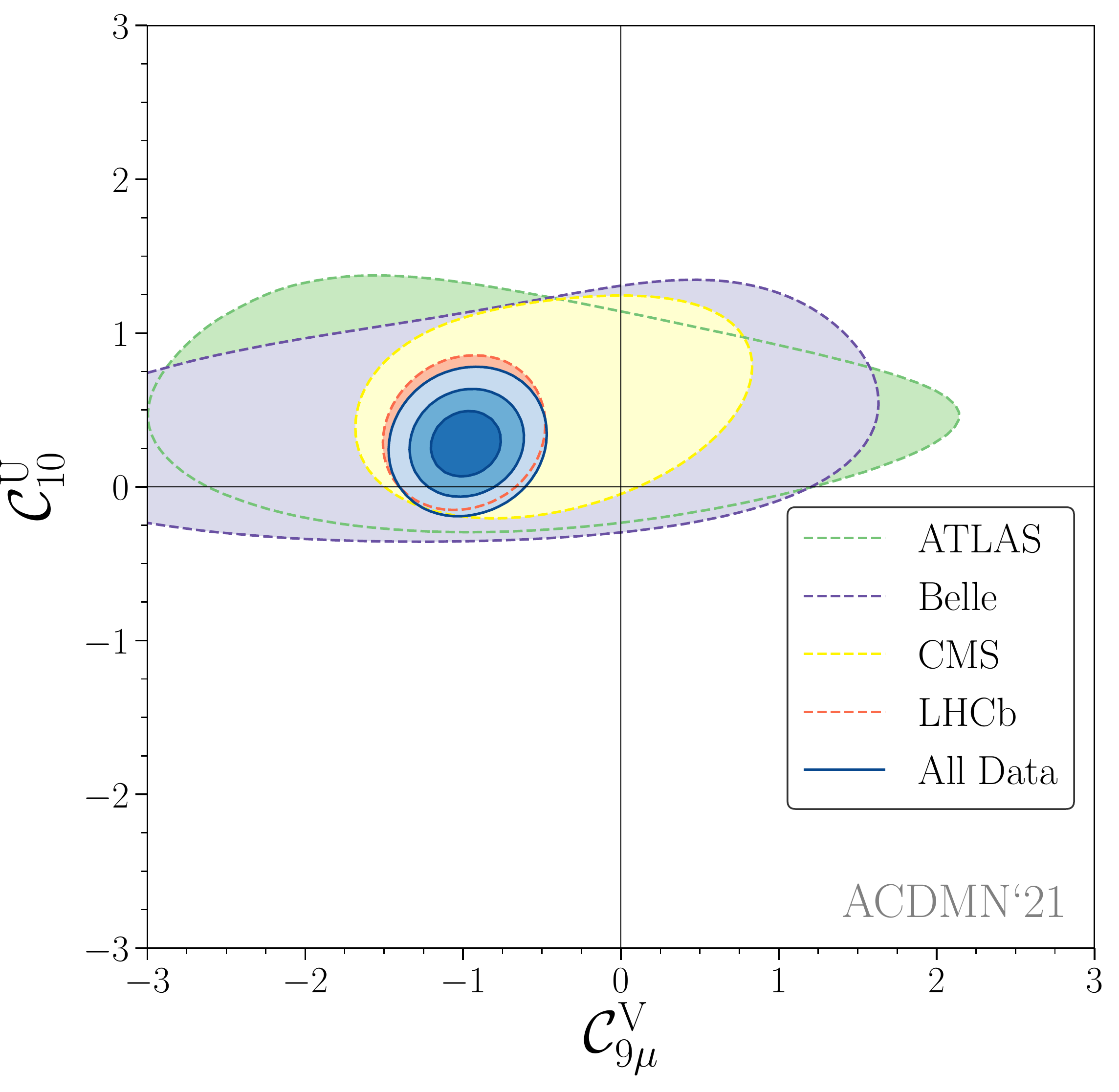}\hspace{2mm}
\includegraphics[width=0.315\textwidth]{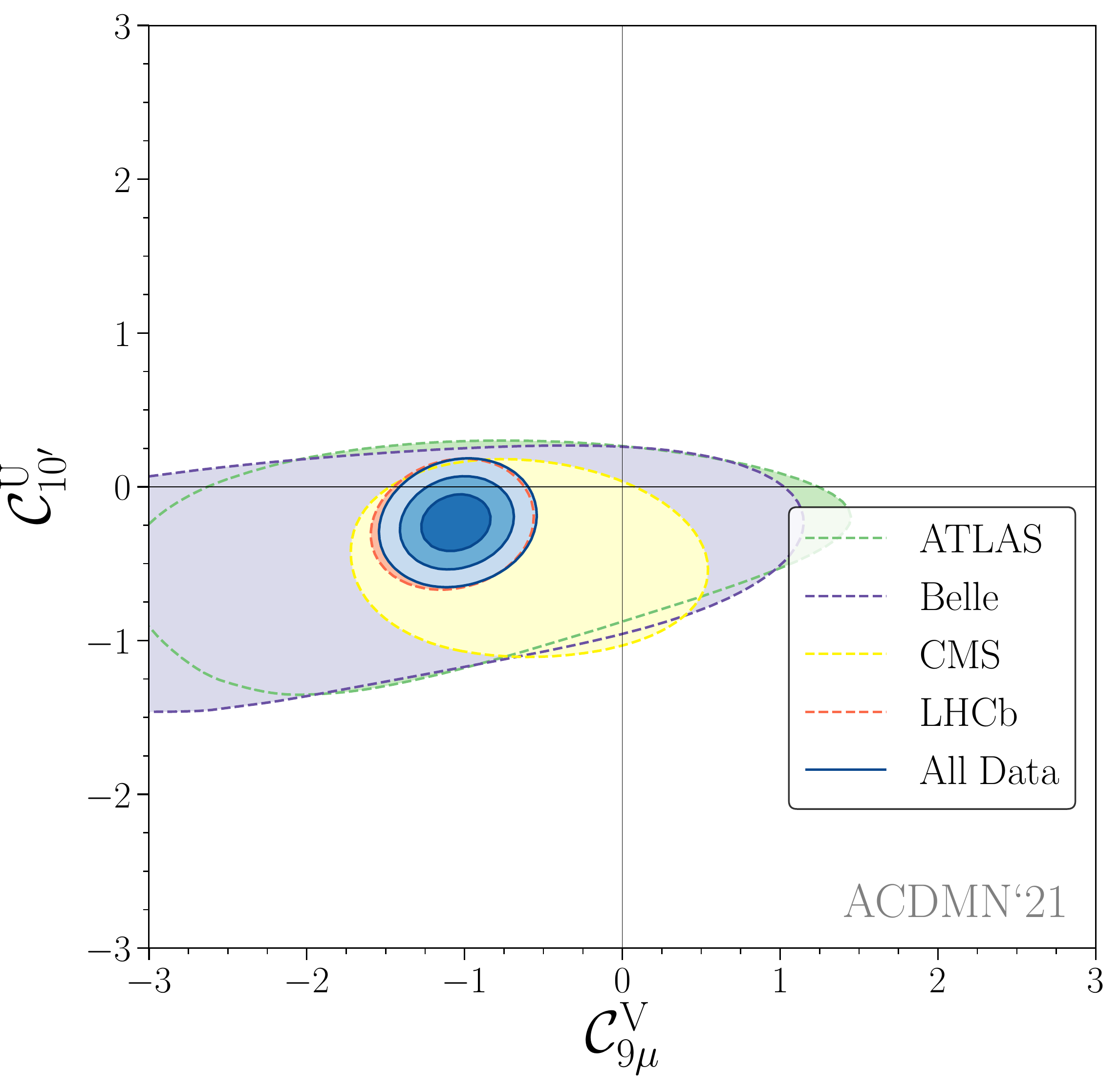}\hspace{2mm}
\includegraphics[width=0.315\textwidth]{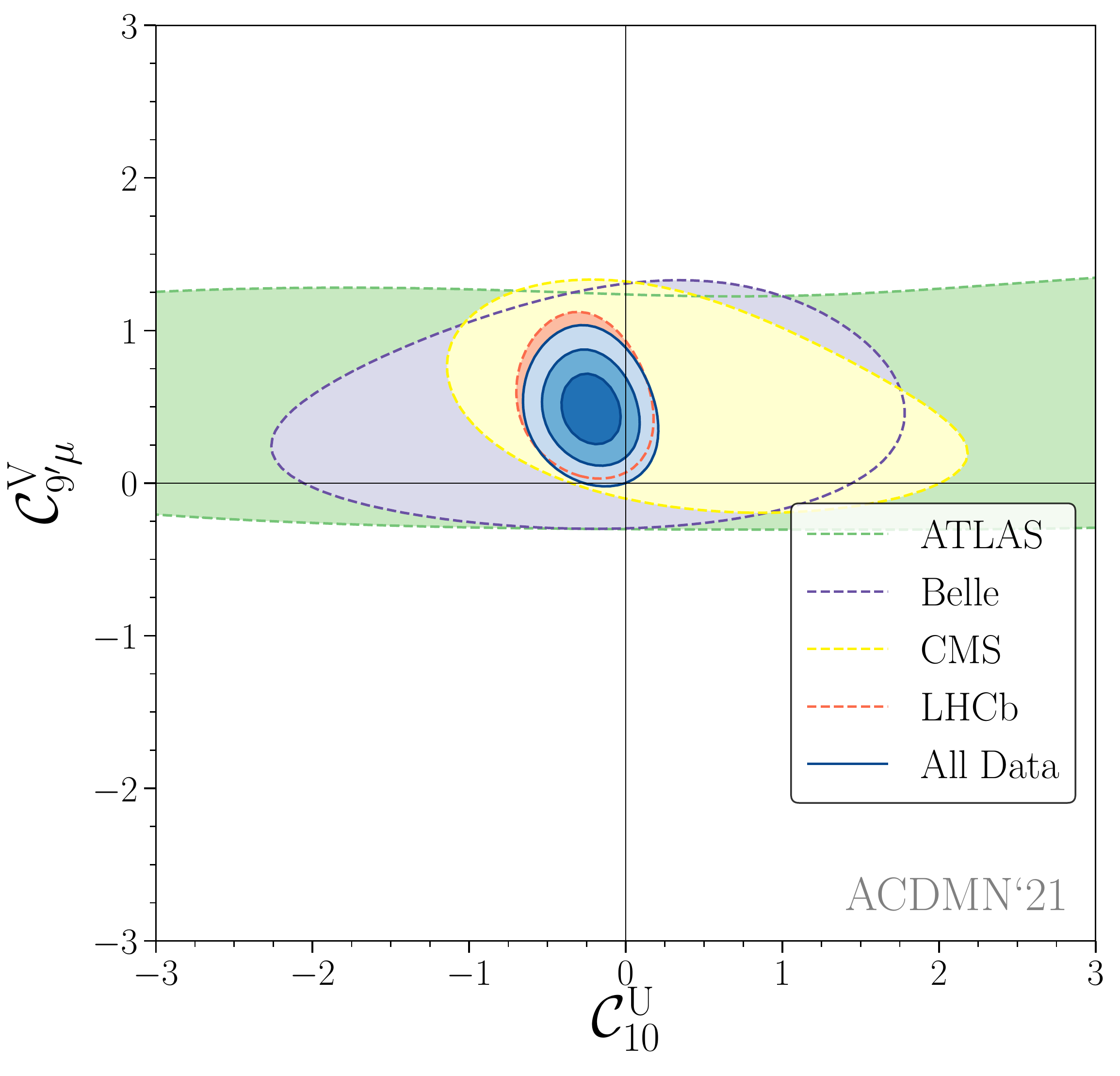}
\end{center}
\caption{From left to right : Allowed regions for the 2D scenarios presented in Tab.~\ref{Fit3Dbis}. Scenarios 6 and 7 on the upper row, 8 and 9 in the middle row and 10 to 12 in the bottom row using all available data (fit ``All''). Dashed lines represent  the 3~$\sigma$ regions while the solid lines represent 1, 2 and 3~$\sigma$ regions.}
\label{fig:FitResultLFU}
\end{figure}

We turn to scenarios that allow also for the presence of lepton flavour universal NP~\cite{Alguero:2018nvb,Alguero:2019pjc} in addition to LFUV contributions to muons only. We define the separation between the two types of NP by considering the following shifts to the value of the Wilson coefficients
\begin{equation}
\Cc{ie}=\Cc{i}^{\rm U}\,,\qquad \Cc{i\mu}=\Cc{i}^{\rm U}+\Cc{i\mu}^{\rm V}\,,
\end{equation}
(with $i=9^{(\prime)},10^{(\prime)}$) for $b\to see$ and $b\to s\mu\mu$ transitions respectively.
We update the scenarios considered in Ref.~\cite{Alguero:2019ptt} in Tab.~\ref{Fit3Dbis} and Fig.~\ref{fig:FitResultLFU}.
Interestingly, when we perform the 10-dimensional fit allowing for NP in both muon and electron coefficients (i.e. $\C7,\C{9\ell},\C{10\ell}$ and $\C{7'},\C{9'\ell},\C{10'\ell}$ for both $\ell = e$ and $\mu$), we obtain almost the same results as in Tab.~\ref{tab:Fit6D} for the muon coefficients, whereas the electron coefficients are only very loosely constrained, indicating the need for more data on electronic modes. We obtain a Pull$_{\rm SM}$ of $6.0\sigma$ ($p$-value of $28.3\%$) for this 10-dimensional fit.

\begin{table*}[!ht]
    \centering
    \begin{adjustbox}{width=0.8\textwidth,center=\textwidth}
\begin{tabular}{lc||c|c|c|c|c}
\multicolumn{2}{c||}{Scenario} & Best-fit point & 1 $\sigma$ & 2 $\sigma$ & Pull$_{\rm SM}$ & p-value \\
\hline\hline
\multirow{ 3}{*}{Scenario 5} &$\Cc{9\mu}^{\rm V}$ & $-0.55$ & $[-1.02,-0.11]$ & $[-1.56,+0.32]$ &
\multirow{ 3}{*}{6.6} & \multirow{ 3}{*}{25.2\,\%} \\
&$\Cc{10\mu}^{\rm V}$ & $+0.49$ & $[+0.08,+0.84]$ & $[-0.44,+1.15]$ & \\
&$\Cc{9}^{\rm U}=\Cc{10}^{\rm U}$ & $-0.35$ & $[-0.73,+0.07]$ & $[-1.06,+0.60]$ &\\
\hline
\multirow{ 2}{*}{Scenario 6}&$\Cc{9\mu}^{\rm V}=-\Cc{10\mu}^{\rm V}$ & $-0.52$ & $[-0.59,-0.44]$ & $[-0.67,-0.37]$ &
\multirow{ 2}{*}{6.9} & \multirow{ 2}{*}{26.6\,\%} \\
&$\Cc{9}^{\rm U}=\Cc{10}^{\rm U}$ & $-0.38$ & $[-0.50,-0.26]$ & $[-0.60,-0.13]$ &\\
\hline
\multirow{ 2}{*}{Scenario 7}&$\Cc{9\mu}^{\rm V}$ & $-0.85$ & $[-1.07,-0.63]$ & $[-1.30,-0.42]$ &
\multirow{ 2}{*}{6.7} & \multirow{ 2}{*}{23.8\,\% }  \\
&$\Cc{9}^{\rm U}$ & $-0.26$ & $[-0.52,+0.01]$ & $[-0.79,+0.30]$  &\\
\hline
\multirow{ 2}{*}{Scenario 8}&$\Cc{9\mu}^{\rm V}=-\Cc{10\mu}^{\rm V}$ & $-0.34$ & $[-0.41,-0.27]$ & $[-0.49,-0.20]$ &
\multirow{ 2}{*}{$7.2$} & \multirow{ 2}{*}{34.5\,\%} \\
&$\Cc{9}^{\rm U}$ & $-0.82$ & $[-0.99,-0.63]$ & $[-1.16,-0.42]$ &\\
\hline\hline
\multirow{ 2}{*}{Scenario 9}&$\Cc{9\mu}^{\rm V}=-\Cc{10\mu}^{\rm V}$ & $-0.53$ & $[-0.63,-0.43]$ & $[-0.74,-0.33]$ &
\multirow{ 2}{*}{6.3} & \multirow{ 2}{*}{17.5\,\%} \\
&$\Cc{10}^{\rm U}$ & $-0.24$ & $[-0.44,-0.05]$ & $[-0.63,+0.15]$ &\\
\hline
\multirow{ 2}{*}{Scenario 10}&$\Cc{9\mu}^{\rm V}$ & $-0.98$ & $[-1.13,-0.84]$ & $[-1.27,-0.69]$ &
\multirow{ 2}{*}{6.9} & \multirow{ 2}{*}{27.9\,\%} \\
&$\Cc{10}^{\rm U}$ & $+0.27$ & $[+0.13,+0.42]$ & $[-0.01,+0.56]$ &\\
\hline
\multirow{ 2}{*}{Scenario 11}&$\Cc{9\mu}^{\rm V}$ & $-1.06$ & $[-1.20,-0.91]$ & $[-1.34,-0.76]$ &
\multirow{ 2}{*}{6.9} & \multirow{ 2}{*}{27.4\,\%} \\
&$\Cc{10'}^{\rm U}$ & $-0.23$ & $[-0.35,-0.10]$ & $[-0.47,+0.02]$ &\\
\hline
\multirow{ 2}{*}{Scenario 12}&$\Cc{9'\mu}^{\rm V}$ & $+0.49$ & $[+0.34,+0.65]$ & $[+0.19,+0.81]$ &
\multirow{ 2}{*}{3.2} & \multirow{ 2}{*}{1.4\,\%} \\
&$\Cc{10}^{\rm U}$ & $-0.25$ & $[-0.38,-0.13]$ & $[-0.50,-0.00]$ &\\
\hline
\multirow{ 4}{*}{Scenario 13}&$\Cc{9\mu}^{\rm V}$ & $-1.11$ & $[-1.27,-0.96]$ & $[-1.41,-0.79]$ &
\multirow{ 4}{*}{6.7} & \multirow{ 4}{*}{29.6\,\%} \\
&$\Cc{9'\mu}^{\rm V}$ & $+0.37$ & $[+0.13,+0.60]$ & $[-0.11,+0.84]$ &\\
&$\Cc{10}^{\rm U}$ & $+0.28$ & $[+0.10,+0.47]$ & $[-0.08,+0.66]$ &\\
&$\Cc{10'}^{\rm U}$ & $+0.03$ & $[-0.15,+0.21]$ & $[-0.33,+0.40]$ &\\
\end{tabular}
\end{adjustbox}
\caption{Most prominent patterns for LFU and LFUV NP contributions from Fit ``All''.
Scenarios 5 to 8 were introduced in Ref.~\cite{Alguero:2018nvb}.  Scenarios 9 (motivated by 2HDMs~\cite{Crivellin:2019dun}) and 10 to 13  (motivated by $Z^\prime$ models with vector-like quarks~\cite{Bobeth:2016llm}) were introduced in Ref.~\cite{Alguero:2019ptt}.}\label{Fit3Dbis} 
\end{table*}

\section{Favoured scenarios and connection with other observables}\label{section5}

\begin{figure}[!ht]
    \centering
    \includegraphics[width=0.9\textwidth]{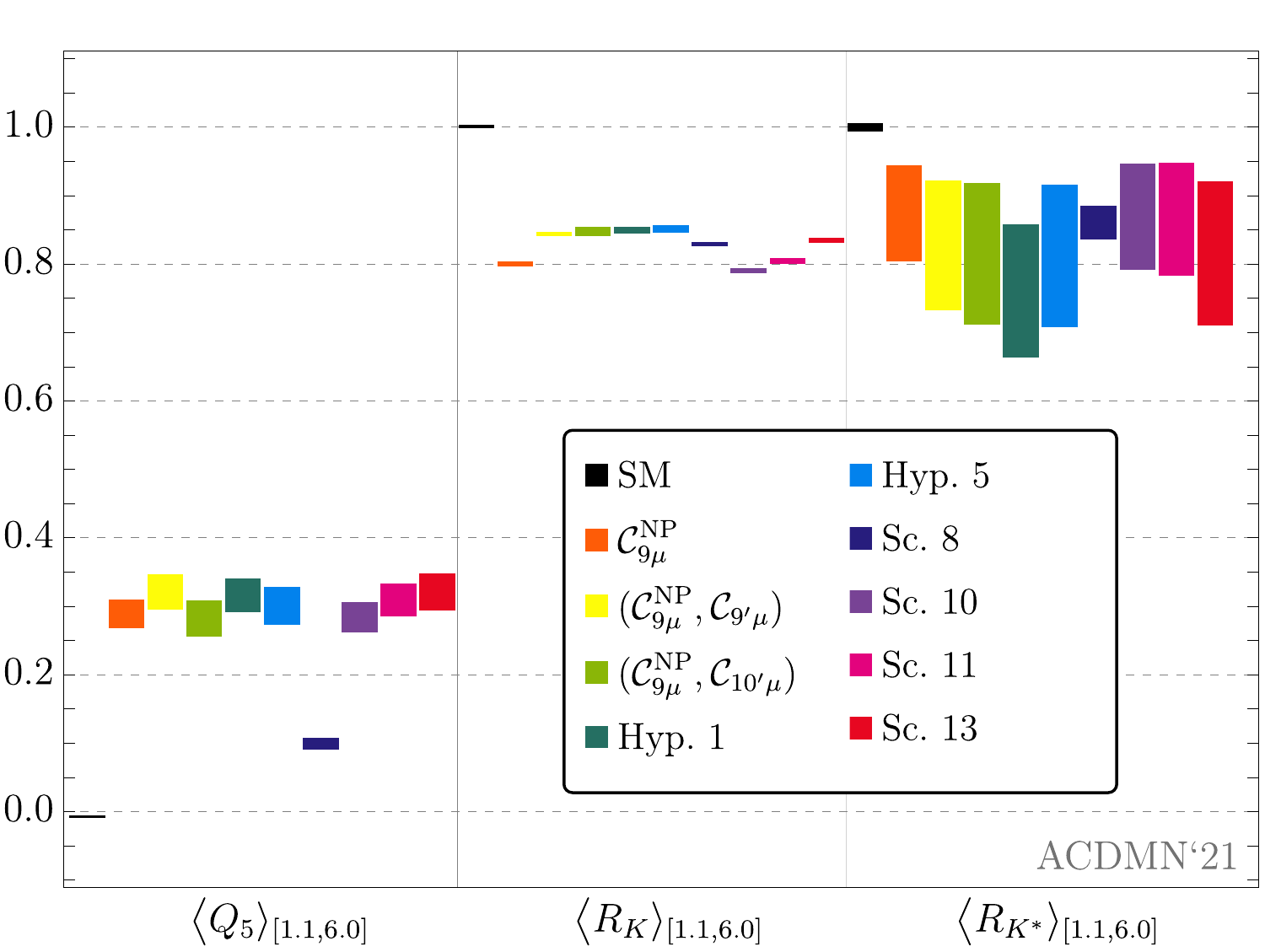}
    \caption{Values of  $\langle Q_5 \rangle_{[1.1,6]}$, $\langle R_K \rangle_{[1.1,6]}$, $\langle R_{K^\ast} \rangle_{[1.1,6]}$ in the SM and nine different scenarios: 
    %${\rm SM}$ (black), $\mathcal{C}_{9\mu }^{{\rm NP}}$ (orange), $(\mathcal{C}_{9\mu }^{{\rm NP}},\mathcal{C}_{9'\mu })$ (yellow), $(\mathcal{C}_{9\mu }^{{\rm NP}},\mathcal{C}_{10'\mu })$ (light green), ${\rm Hyp.\ 1}$ (dark green), ${\rm Hyp.\ 5}$ (light blue), ${\rm Sc.\ 8}$ (dark blue), ${\rm Sc.\ 10}$ (purple), ${\rm Sc.\ 11}$ (pink), ${\rm Sc.\ 13}$ (red). 
    ${\rm SM}$ (black), $\mathcal{C}_{9\mu }^{{\rm NP}}$ (orange), $(\mathcal{C}_{9\mu }^{{\rm NP}},\mathcal{C}_{9'\mu })$ (yellow), $(\mathcal{C}_{9\mu }^{{\rm NP}},\mathcal{C}_{10'\mu })$ (light green), $({\mathcal{C}_{9\mu}^{\rm NP}=-\mathcal{C}_{9^\prime\mu},\mathcal{C}_{10\mu}^{\rm NP}=\mathcal{C}_{10^\prime\mu}})$ (dark green), $(\mathcal{C}_{9\mu}^{\rm NP} , \mathcal{C}_{9^\prime\mu}=-\mathcal{C}_{10^\prime\mu})$ (light blue), $(\mathcal{C}_{9\mu}^{\rm V}=-\mathcal{C}_{10\mu}^{\rm V},\mathcal{C}_{9}^{\rm U})$ (dark blue), $(\mathcal{C}_{9\mu}^{\rm V},\mathcal{C}_{10}^{\rm U})$ (purple), $(\mathcal{C}_{9\mu}^{\rm V},\mathcal{C}_{10'}^{\rm U})$ (pink), $(\mathcal{C}_{9\mu}^{\rm V},\mathcal{C}_{9'\mu}^{\rm V},\mathcal{C}_{10}^{\rm U},\mathcal{C}_{10'}^{\rm U})$ (red).
    The boxes correspond to the predictions of the 1~$\sigma$ regions at the b.f.p. value of the Wilson coefficients in each of the scenarios for the fit to the ``All'' data set.}
    \label{fig:BoxPlotQ5RKRKs}
\end{figure}

\begin{figure}[!ht]
     \begin{subfigure}[b]{0.45\textwidth}
         \centering
         \includegraphics[width=\textwidth]{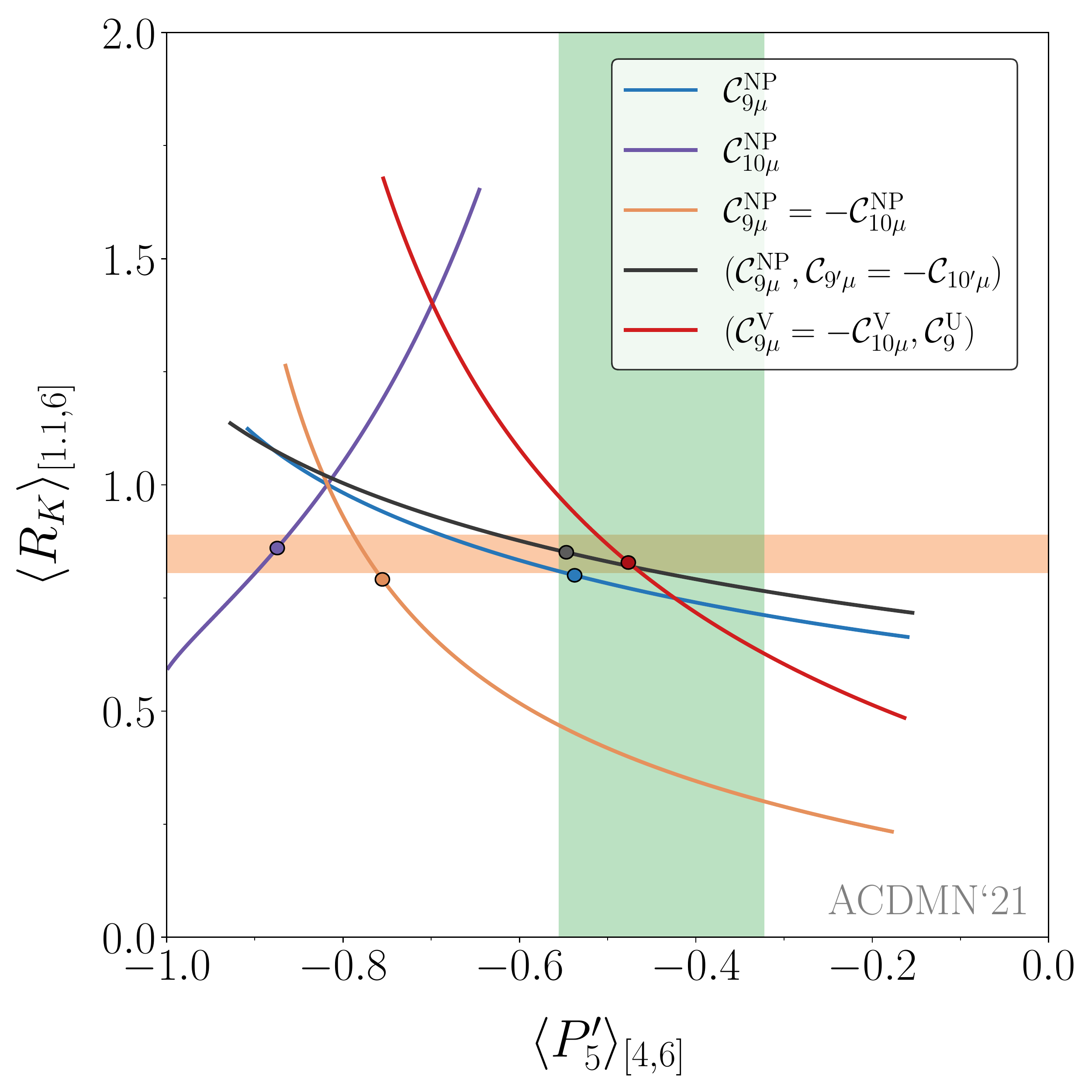}
         \caption{}
         \label{fig:RKP5pPlot}
     \end{subfigure}
     \hfill
     \begin{subfigure}[b]{0.45\textwidth}
         \centering
         \includegraphics[width=\textwidth]{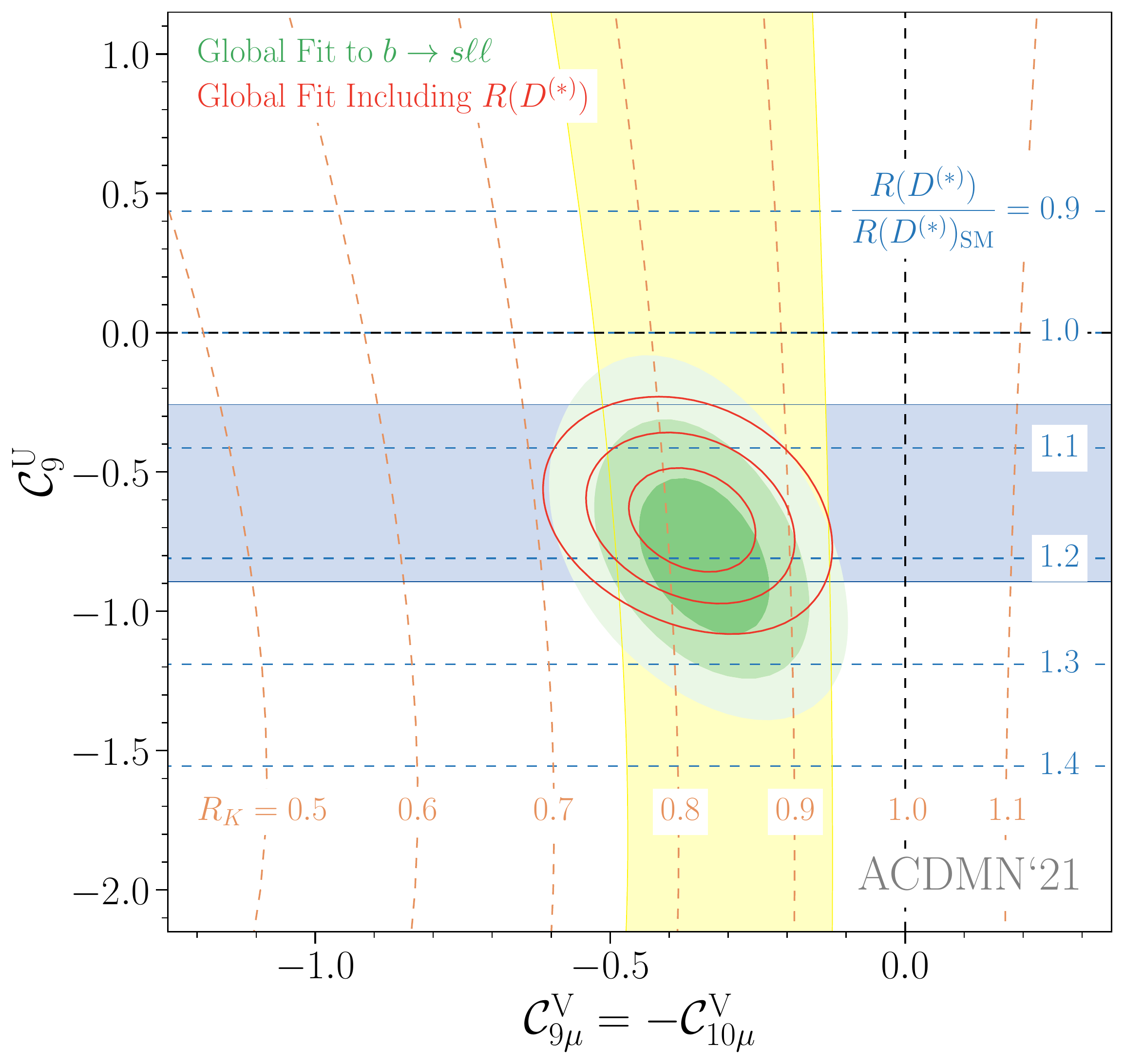}
         \caption{}
         \label{fig:RDRD*}
     \end{subfigure}
\caption{ Left: $\langle R_K \rangle_{[1.1,6]}$ versus $\langle P_5' \rangle_{[4,6]}$ in five different scenarios: ${\cal C}_{9\mu}^{\rm NP}$ (blue), ${\cal C}_{9\mu}^{\rm NP} = - {\cal C}_{10\mu}^{\rm NP}$ (orange), and $({\cal C}^{\rm V}_{9\mu}= - {\cal C}^{\rm V}_{10\mu}, {\cal C}^{\rm U}_{9})$ (red), $({\cal C}_{9\mu}^{\rm NP}, {\cal C}_{9'\mu}= - {\cal C}_{10'\mu})$ (black), and  ${\cal C}_{10\mu}^{\rm NP}$ (purple). The curves correspond only to the predictions for central values. In the 2D scenarios (red and black) the Wilson coefficient not shown is set to its b.f.p. value. The current experimental values from the LHCb collaboration are also indicated (orange horizontal and green vertical bands respectively). The dots correspond to the b.f.p. values of the corresponding scenario for the fit to the ``All'' data set.
Right: Preferred regions at the 1, 2 and 3$\,\sigma$ level (green) in the $(\Cc{9\mu}^{\rm V}=-\Cc{10\mu}^{\rm V},\,\Cc{9}^{\rm U})$ plane from $b\to s\ell\ell$ data. The red contour lines show the corresponding regions once $R_{D^{(*)}}$ is included in the fit (for $\Lambda=2$~TeV). The horizontal blue (vertical yellow) band is consistent with $R_{D^{(*)}}$ ($R_{K}$) at the $2\,\sigma$ level and the contour lines show the predicted values for these ratios.}
\end{figure}

Several scenarios exhibit a significant improvement in the description of the data compared to the SM.
Fig.~\ref{fig:BoxPlotQ5RKRKs} shows the predictions for the observables $Q_5$, $R_K$ and $R_{K^*}$ in several of these scenarios. The large uncertainties for $R_{K^*}$ in most NP scenarios come from the presence of three different helicity amplitudes involving different combinations of form factors: if the $SU(2)_L$ symmetry of the SM is respected, one amplitude dominates leading to reduced uncertainties for the prediction of $R_{K^*}$, but in other cases, the presence of several helicity amplitudes leads to larger uncertainties. One can also notice that $Q_5$ is able to separate three cases of interest: the SM, scenario 8 $(\Cc{9\mu}^{\rm V}=-\Cc{10\mu}^{\rm V},\Cc{9}^{\rm U})$, and the scenarios with right-handed couplings and a large negative contribution to $\Cc{9\mu}$ (Fig.~\ref{fig:RKP5pPlot} illustrates the importance of $R_K$ and $P_5'$ in highlighting these scenarios compared to others considered in the previous section).

As discussed in Ref.~\cite{Alguero:2019ptt}, scenario 8 allows for a model-independent connection between the anomalies in $b\to s\ell\ell$ decays and  those in $b\to c\tau\nu$ transitions~\cite{HFLAV:2019otj}. This connection arises in the SMEFT scenario where $\Cc{}^{(1)}=\Cc{}^{(3)}$ expressed in terms of gauge-invariant dimension-6 operators~\cite{Grzadkowski:2010es,Capdevila:2017iqn}. The operator involving third-generation leptons explains $R_{D^{(*)}}$ and the one involving the second generation gives a LFUV effect in $b\to s\mu\mu$ processes. The constraint from $b\to c\tau\nu$ and $SU(2)_L$ invariance leads to large contributions enhancing $b\to s\tau^+\tau^-$ processes~\cite{Capdevila:2017iqn}, whereas the mixing into ${\cal O}_9$ generates $\Cc{9}^{\rm U}$ at $\mu=m_b$~\cite{Crivellin:2018yvo}. 
Therefore, the SMEFT scenario described above reproduces scenario 8 with an additional correlation between $\Cc{9}^{\rm U}$ and $R_{D^{(*)}}$~\cite{Capdevila:2017iqn,Crivellin:2018yvo}:
\begin{equation}
\Cc{9}^{\rm U}\! \approx \! 7.5\left(1-\sqrt{\frac{R_{D^{(*)}}}{R_{D^{(*)}{\rm SM}}}}\right)\!\! \left(1+\frac{\log(\Lambda^2/(1{\rm TeV}^2))}{10.5}\right)\, , \\  \\
\end{equation}
where $\Lambda$ is the typical scale of NP involved.
We show the global fit of scenario 8 without and with the additional input on $R_{D^{(*)}}$ from Ref.~\cite{HFLAV:2019otj} in Fig.~\ref{fig:RDRD*}, taking the scale $\Lambda=2$ TeV. 
The best-fit point for
$(\Cc{9\mu}^{\rm V}=-\Cc{10\mu}^{\rm V},\Cc{9}^{\rm U})$ is $(-0.36,-0.68)$, with 1~$\sigma$ intervals
$[-0.43,-0.29]$ and $[-0.80,-0.55]$ respectively.
The agreement among all data is very good, shown by the fact that scenario 8 supplemented with $R_{D^{(*)}}$ exhibits a pull with respect to the SM of 8.0~$\sigma$ and a $p$-value of $33.1\%$. Interestingly, the agreement between scenario 8 and the allowed region for $R_{D^{(*)}}$ has increased with the addition of $R_{K_S}$, $R_{K^{*+}}$ and $B_s\to \phi\mu^+\mu^-$ into the global analysis, with a fit favouring less negative values for $\Cc{9}^{\rm U}$. An even better agreement could be reached if $R_{D^{(*)}}$ is slightly further away from the SM expectations, or if the scale of New Physics is increased.

%\newpage

\section{Discussion}\label{section6}

We have presented in this paper our most complete and updated results of the global fit to $b\to s\ell\ell$ data including 254 observables.
We see that the recent measurements  of LFUV observables  $R_K$, $R_{K_S}$, $R_{K^{*+}}$ by the LHCb collaboration together with the $B_s \to \phi\mu^+\mu^-$ update confirms the main conclusions of the previous update of $R_K$
and $B_s\to\mu^+\mu^-$ with only marginal changes. Indeed, the slight reduction of significances in most scenarios is mostly driven by the inclusion of more SM-like observables  coming from the update of $B_s \to \phi\mu^+\mu^-$ (new bins) with little sensitivity to ${\cal C}_{9\mu}$ and higher experimental precision. On the other side, even if the scenario ${\cal C}_{9\mu}=-{\cal C}_{9^\prime\mu}$ can explain neither $R_K$ nor $R_{K_S}$, it yields an acceptable solution for $R_{K^*}$ and $R_{K^{*+}}$ leading to a marginal increase of its significance in the LFUV fit. 
  
The overall hierarchy of preferences for specific scenarios remains unchanged. In our previous update~\cite{Alguero:2019ptt}  we observed an increase in the consistency among the data analysed in the framework of the favoured scenarios. More specifically, we saw that the most favoured 1D scenario remains
the case of a vector coupling to muons encoded in $\Cc{9\mu}$.
 The LHCb update of the $B_s\to \mu^+\mu^-$ branching ratio, in better agreement with the SM expectation, reduced marginally the room available for NP in $\Cc{10\mu}$  for the scenarios considered here, which do not feature NP contributions from (pseudo)scalar operators. 
%, which is compatible with zero at 1 $\sigma$ in the 2D hypothesis $(\Cc{9\mu},\Cc{10\mu})$. 
%All those conclusions are confirmed.
\newpage
Finally, the two classes of favoured scenarios of Ref.~\cite{Alguero:2019ptt}
find their status strengthened, namely
\begin{itemize}
    \item The purely muonic hypotheses with right handed currents $(\Cc{9\mu}^{\rm NP},\Cc{10'\mu})$ and $(\Cc{9\mu}^{\rm NP},\Cc{9'\mu}=-\Cc{10'\mu})$. The latter scenario (called Hypothesis 5 in Table~\ref{tab:results2D}) features a right-handed contribution which becomes compatible with zero once the $2\sigma$ confidence region is considered.
Such right-handed currents tend to counterbalance the impact on $R_K$ of a large negative $\Cc{9\mu}$ which is preferred by many observables considered in the global fit. 

    \item Scenario 8 $(\Cc{9\mu}^{\rm V}=-\Cc{10\mu}^{\rm V},\Cc{9}^{\rm U})$ with a universal component $\Cc{9}^{\rm U}$ together with a muonic component obeying $SU(2)_L$ invariance. As illustrated in Fig.~\ref{fig:RDRD*}, this scenario reaches 8.0~$\sigma$ once combined with $R_{D}$ and $R_{D^*}$ in an EFT framework explaining $b\to c\ell\nu$ and $b\to s \ell^+\ell^-$ through correlated singlet and triplet dimension-6 operators combining quark and lepton bilinears.
\end{itemize}

As an outlook for the future, besides the importance of updating the LFU ratios $R_{K^{(\ast)}}$ and the angular distributions of $B\to K^*\ell^+\ell^-$ and $B_s\to\phi\ell^+\ell^-$ modes, two experimental inputs can help guiding future analyses. First, the observation of enhanced $b \to s \tau^+\tau^-$ transitions would favour naturally a scenario with a LFU contribution in $\Cc{9}^{\rm U}$. Second, the measurement of a large $Q_5$ would favour a scenario with a large negative vector coupling $\Cc{9\mu}$, possibly with additional right-handed currents. Indeed, as illustrated by Fig.~\ref{fig:BoxPlotQ5RKRKs}, the observable $Q_5$~\cite{Capdevila:2016ivx} can distinguish between the purely muonic hypotheses with right handed currents (e.g. Hypothesis 5) 
and scenario 8 with a universal component in $\Cc{9}^{\rm U}$, with a higher value in the former case and a slightly lower value in the latter~\cite{Alguero:2019pjc}.

Further progress may also be achieved through a better understanding of the theoretical uncertainties involved~\cite{Gubernari:2018wyi,Gubernari:2020eft,
Descotes-Genon:2019bud}, more data on other modes and with other experimental setups (in particular Belle II~\cite{Abe:2010gxa}), but also the determination of additional observables~\cite{Descotes-Genon:2019dbw,Descotes-Genon:2020tnz, Alguero:2021yus}.
This supplementary information should help us to corner the actual NP pattern hinted at by the $b\to s\ell\ell$ anomalies currently observed and confirmed as an evidence in $R_K$ by the LHCb collaboration. 

Such identification at the EFT level is the first and mandatory step to build viable phenomenological models for New Physics, to be probed and confirmed through decays involving other families of quarks and leptons, as well as direct production experiments.

\section*{Acknowledgements}

We would like to thank Diego Martinez Santos for providing us with an average of the results on the $B_s\to \mu^+\mu^-$ branching ratios.
This project has received support from the European Union’s Horizon 2020 research and innovation programme under the Marie Sklodowska-Curie grant agreement No 860881-HIDDeN [S.D.G. and M.N.B.]. This work received financial support from the Spanish Ministry of Science, Innovation and Universities (FPA2017-86989-P) and the Research Grant Agency of the Government of Catalonia (SGR 1069) [M.A., J.M.]. J.M. also acknowledges the financial support by ICREA under the ICREA Academia programme. B.C. is supported by the Italian Ministry of Research (MIUR) under the Grant No. PRIN 20172LNEEZ.

\newpage

\appendix
\section{Inputs to our global fits}\label{app:predictions}

We provide here the observables included in our Fits ``All'' (254 observables) and ``LFUV'' (24 observables, replacing $P'_{4e,\mu}, P'_{5e,\mu}$ by $Q_4, Q_5$, measured by Belle). 
In the following table, we provide %the $b\to s\ell\ell$
all the observables considered in both types of fits with the corresponding legend:
no mark for observables for the fit ``All'' only, $\ddag$ for the fit ``LFUV'' only, and $\dag$ for both fits ``LFUV'' and ``All''. 
The theoretical predictions of the observables in the SM as well as the individual tension with respect to the experimental value are also provided. 

Our angle convention and definition of the angular observables for the $B\to K^*\ell^+\ell^-$ decay differs from the usual LHCb convention \cite{LHCb:2013zuf,LHCb:2015svh}. We follow the conventions given in Ref. \cite{Descotes-Genon:2015uva} where a dictionary relating both conventions can be found in Eq.~(16).

\begin{longtable}{@{}cccr@{}}
\toprule[1.6pt] 
\multicolumn{4}{c}{Standard Model Predictions}\\ 
\midrule 
$ 10^7 \times BR(B^+\to K^+\mu^+\mu^-)\text{[LHCb]} $ & Standard Model & Experiment \cite{LHCb:2014cxe} & Pull \\ 
 \midrule 
 $ [0.1,0.98] $ & $ 0.32 \pm 0.10 $ & $ 0.29 \pm 0.02 $ & $ +0.3 $ \\ 
 $ [1.1,2] $ & $ 0.33 \pm 0.10 $ & $ 0.21 \pm 0.02 $ & $ +1.2 $ \\ 
 $ [2,3] $ & $ 0.37 \pm 0.11 $ & $ 0.28 \pm 0.02 $ & $ +0.7 $ \\ 
 $ [3,4] $ & $ 0.36 \pm 0.12 $ & $ 0.25 \pm 0.02 $ & $ +0.9 $ \\ 
 $ [4,5] $ & $ 0.36 \pm 0.12 $ & $ 0.22 \pm 0.02 $ & $ +1.2 $ \\ 
 $ [5,6] $ & $ 0.36 \pm 0.12 $ & $ 0.23 \pm 0.02 $ & $ +1.0 $ \\ 
 $ [6,7] $ & $ 0.36 \pm 0.13 $ & $ 0.25 \pm 0.02 $ & $ +0.9 $ \\ 
 $ [7,8] $ & $ 0.36 \pm 0.13 $ & $ 0.23 \pm 0.02 $ & $ +0.9 $ \\ 
 $ [15,22] $ & $ 1.02 \pm 0.14 $ & $ 0.85 \pm 0.05 $ & $ +1.2 $ \\ 

\midrule[1.6pt] 
$ 10^7 \times BR(B^0\to K^0\mu^+\mu^-)\text{[LHCb]} $ & Standard Model & Experiment \cite{LHCb:2014cxe}  & Pull \\ 
 \midrule 
 $ [0.1,2] $ & $ 0.65 \pm 0.20 $ & $ 0.23 \pm 0.11 $ & $ +1.9 $ \\ 
 $ [2,4] $ & $ 0.68 \pm 0.21 $ & $ 0.37 \pm 0.11 $ & $ +1.3 $ \\ 
 $ [4,6] $ & $ 0.67 \pm 0.22 $ & $ 0.35 \pm 0.10 $ & $ +1.3 $ \\ 
 $ [6,8] $ & $ 0.66 \pm 0.24 $ & $ 0.54 \pm 0.12 $ & $ +0.5 $ \\ 
 $ [15,22] $ & $ 0.94 \pm 0.13 $ & $ 0.67 \pm 0.12 $ & $ +1.6 $ \\ 
\midrule[1.6pt] 
$ 10^7 \times BR(B^0\to K^{*0}\mu^+\mu^-)\text{[LHCb]} $ & Standard Model & Experiment \cite{LHCb:2016ykl}  & Pull \\ 
 \midrule 
 $ [0.1,0.98] $ & $ 0.92 \pm 0.80 $ & $ 0.89 \pm 0.09 $ & $ +0.0 $ \\ 
 $ [1.1,2.5] $ & $ 0.56 \pm 0.35 $ & $ 0.46 \pm 0.06 $ & $ +0.3 $ \\ 
 $ [2.5,4] $ & $ 0.58 \pm 0.40 $ & $ 0.50 \pm 0.06 $ & $ +0.2 $ \\ 
 $ [4,6] $ & $ 0.91 \pm 0.66 $ & $ 0.71 \pm 0.07 $ & $ +0.3 $ \\ 
 $ [6,8] $ & $ 1.12 \pm 0.89 $ & $ 0.86 \pm 0.08 $ & $ +0.3 $ \\ 
 $ [15,19] $ & $ 2.50 \pm 0.21 $ & $ 1.74 \pm 0.14 $ & $ +3.0 $ \\ 
\midrule[1.6pt] 
$ 10^7 \times BR(B^+\to K^{*+}\mu^+\mu^-)\text{[LHCb]} $ & Standard Model & Experiment \cite{LHCb:2014cxe} & Pull \\ 
 \midrule 
 $ [0.1,2] $ & $ 1.40 \pm 1.08 $ & $ 1.12 \pm 0.27 $ & $ +0.3 $ \\ 
 $ [2,4] $ & $ 0.84 \pm 0.56 $ & $ 1.12 \pm 0.32 $ & $ -0.4 $ \\ 
 $ [4,6] $ & $ 0.99 \pm 0.72 $ & $ 0.50 \pm 0.20 $ & $ +0.7 $ \\ 
 $ [6,8] $ & $ 1.22 \pm 0.96 $ & $ 0.66 \pm 0.22 $ & $ +0.6 $ \\ 
 $ [15,19] $ & $ 2.69 \pm 0.23 $ & $ 1.60 \pm 0.32 $ & $ +2.8 $ \\ 
\midrule[1.6pt] 
$ 10^7 \times BR(B_s\to \phi\mu^+\mu^-)\text{[LHCb]} $ & Standard Model & Experiment \cite{LHCb:2021zwz} & Pull \\ 
 \midrule 
 $ [0.1,0.98] $ & $ 1.06 \pm 0.23 $ & $ 0.68 \pm 0.06 $ & $ +1.6 $ \\ 
 $ [1.1,2.5] $ & $ 0.71 \pm 0.15 $ & $ 0.44 \pm 0.05 $ & $ +1.7 $ \\ 
 $ [2.5,4] $ & $ 0.71 \pm 0.15 $ & $ 0.35 \pm 0.04 $ & $ +2.3 $ \\ 
 $ [4,6] $ & $ 1.04 \pm 0.21 $ & $ 0.62 \pm 0.06 $ & $ +1.9 $ \\ 
 $ [6,8] $ & $ 1.21 \pm 0.25 $ & $ 0.63 \pm 0.06 $ & $ +2.2 $ \\ 
 $ [15,19] $ & $ 2.29 \pm 0.15 $ & $ 1.85 \pm 0.13 $ & $ +1.9 $ \\ 
\midrule[1.6pt] 
$ F_L (B^{0}\to K^{*0}\mu^+\mu^-)\text{[LHCb]} $ & Standard Model & Experiment \cite{LHCb:2020lmf} & Pull \\ 
 \midrule 
 $ [0.1,0.98] $ & $ 0.23 \pm 0.24 $ & $ 0.26 \pm 0.03 $ & $ -0.1 $ \\ 
 $ [1.1,2.5] $ & $ 0.68 \pm 0.26 $ & $ 0.66 \pm 0.05 $ & $ +0.1 $ \\ 
 $ [2.5,4] $ & $ 0.77 \pm 0.23 $ & $ 0.76 \pm 0.05 $ & $ +0.0 $ \\ 
 $ [4,6] $ & $ 0.71 \pm 0.28 $ & $ 0.68 \pm 0.04 $ & $ +0.1 $ \\ 
 $ [6,8] $ & $ 0.63 \pm 0.32 $ & $ 0.65 \pm 0.03 $ & $ -0.0 $ \\ 
 $ [15,19] $ & $ 0.34 \pm 0.03 $ & $ 0.35 \pm 0.02 $ & $ -0.1 $ \\ 
\midrule[1.6pt] 
$ P_1 (B^{0}\to K^{*0}\mu^+\mu^-)\text{[LHCb]} $ & Standard Model & Experiment \cite{LHCb:2020lmf} & Pull \\ 
 \midrule 
 $ [0.1,0.98] $ & $ 0.03 \pm 0.08 $ & $ 0.09 \pm 0.12 $ & $ -0.4 $ \\ 
 $ [1.1,2.5] $ & $ -0.00 \pm 0.05 $ & $ -0.62 \pm 0.30 $ & $ +2.0 $ \\ 
 $ [2.5,4] $ & $ 0.00 \pm 0.06 $ & $ 0.17 \pm 0.37 $ & $ -0.4 $ \\ 
 $ [4,6] $ & $ 0.02 \pm 0.12 $ & $ 0.09 \pm 0.24 $ & $ -0.2 $ \\ 
 $ [6,8] $ & $ 0.02 \pm 0.13 $ & $ -0.07 \pm 0.21 $ & $ +0.4 $ \\ 
 $ [15,19] $ & $ -0.64 \pm 0.06 $ & $ -0.58 \pm 0.10 $ & $ -0.6 $ \\ 
\midrule[1.6pt] 
$ P_2 (B^{0}\to K^{*0}\mu^+\mu^-)\text{[LHCb]} $ & Standard Model & Experiment \cite{LHCb:2020lmf} & Pull \\ 
 \midrule 
 $ [0.1,0.98] $ & $ 0.12 \pm 0.02 $ & $ 0.00 \pm 0.04 $ & $ +2.8 $ \\ 
 $ [1.1,2.5] $ & $ 0.44 \pm 0.03 $ & $ 0.44 \pm 0.10 $ & $ -0.0 $ \\ 
 $ [2.5,4] $ & $ 0.23 \pm 0.13 $ & $ 0.19 \pm 0.12 $ & $ +0.2 $ \\ 
 $ [4,6] $ & $ -0.19 \pm 0.11 $ & $ -0.11 \pm 0.07 $ & $ -0.6 $ \\ 
 $ [6,8] $ & $ -0.38 \pm 0.07 $ & $ -0.21 \pm 0.05 $ & $ -2.1 $ \\ 
 $ [15,19] $ & $ -0.36 \pm 0.02 $ & $ -0.36 \pm 0.02 $ & $ -0.1 $ \\ 
\midrule[1.6pt] 
$ P_3 (B^{0}\to K^{*0}\mu^+\mu^-)\text{[LHCb]} $ & Standard Model & Experiment \cite{LHCb:2020lmf}& Pull \\ 
 \midrule 
 $ [0.1,0.98] $ & $ -0.00 \pm 0.00 $ & $ -0.07 \pm 0.06 $ & $ +1.3 $ \\ 
 $ [1.1,2.5] $ & $ 0.00 \pm 0.00 $ & $ -0.32 \pm 0.15 $ & $ +2.2 $ \\ 
 $ [2.5,4] $ & $ 0.00 \pm 0.01 $ & $ -0.05 \pm 0.20 $ & $ +0.3 $ \\ 
 $ [4,6] $ & $ 0.00 \pm 0.01 $ & $ 0.09 \pm 0.14 $ & $ -0.6 $ \\ 
 $ [6,8] $ & $ 0.00 \pm 0.00 $ & $ 0.07 \pm 0.10 $ & $ -0.6 $ \\ 
 $ [15,19] $ & $ 0.00 \pm 0.02 $ & $ -0.05 \pm 0.05 $ & $ +1.0 $ \\ 
\midrule[1.6pt] 
$ P'_4 (B^{0}\to K^{*0}\mu^+\mu^-)\text{[LHCb]} $ & Standard Model & Experiment \cite{LHCb:2020lmf}& Pull \\ 
 \midrule 
 $ [0.1,0.98] $ & $ -0.50 \pm 0.16 $ & $ -0.27 \pm 0.24 $ & $ -0.8 $ \\ 
 $ [1.1,2.5] $ & $ -0.07 \pm 0.16 $ & $ 0.16 \pm 0.29 $ & $ -0.7 $ \\ 
 $ [2.5,4] $ & $ 0.53 \pm 0.21 $ & $ 0.87 \pm 0.35 $ & $ -0.9 $ \\ 
 $ [4,6] $ & $ 0.82 \pm 0.15 $ & $ 0.62 \pm 0.23 $ & $ +0.7 $ \\ 
 $ [6,8] $ & $ 0.93 \pm 0.11 $ & $ 1.15 \pm 0.19 $ & $ -1.0 $ \\ 
 $ [15,19] $ & $ 1.28 \pm 0.02 $ & $ 1.28 \pm 0.12 $ & $ +0.0 $ \\ 
\midrule[1.6pt] 
$ P'_5 (B^{0}\to K^{*0}\mu^+\mu^-)\text{[LHCb]} $ & Standard Model & Experiment \cite{LHCb:2020lmf} & Pull \\ 
 \midrule 
 $ [0.1,0.98] $ & $ 0.67 \pm 0.13 $ & $ 0.52 \pm 0.10 $ & $ +0.9 $ \\ 
 $ [1.1,2.5] $ & $ 0.19 \pm 0.11 $ & $ 0.37 \pm 0.12 $ & $ -1.0 $ \\ 
 $ [2.5,4] $ & $ -0.47 \pm 0.12 $ & $ -0.15 \pm 0.15 $ & $ -1.7 $ \\ 
 $ [4,6] $ & $ -0.82 \pm 0.08 $ & $ -0.44 \pm 0.12 $ & $ -2.7 $ \\ 
 $ [6,8] $ & $ -0.94 \pm 0.08 $ & $ -0.58 \pm 0.09 $ & $ -2.9 $ \\ 
 $ [15,19] $ & $ -0.57 \pm 0.05 $ & $ -0.67 \pm 0.06 $ & $ +1.2 $ \\ 
\midrule[1.6pt] 
$ P'_6 (B^{0}\to K^{*0}\mu^+\mu^-)\text{[LHCb]} $ & Standard Model & Experiment \cite{LHCb:2020lmf} & Pull \\ 
 \midrule 
 $ [0.1,0.98] $ & $ -0.06 \pm 0.02 $ & $ 0.02 \pm 0.09 $ & $ -0.7 $ \\ 
 $ [1.1,2.5] $ & $ -0.07 \pm 0.03 $ & $ -0.23 \pm 0.13 $ & $ +1.2 $ \\ 
 $ [2.5,4] $ & $ -0.06 \pm 0.03 $ & $ -0.16 \pm 0.15 $ & $ +0.6 $ \\ 
 $ [4,6] $ & $ -0.04 \pm 0.02 $ & $ -0.29 \pm 0.12 $ & $ +2.2 $ \\ 
 $ [6,8] $ & $ -0.02 \pm 0.01 $ & $ -0.16 \pm 0.10 $ & $ +1.4 $ \\ 
 $ [15,19] $ & $ -0.00 \pm 0.07 $ & $ 0.07 \pm 0.07 $ & $ -0.8 $ \\ 
\midrule[1.6pt] 
$ P'_8 (B^{0}\to K^{*0}\mu^+\mu^-)\text{[LHCb]} $ & Standard Model & Experiment \cite{LHCb:2020lmf} & Pull \\ 
 \midrule 
 $ [0.1,0.98] $ & $ 0.02 \pm 0.02 $ & $ 0.01 \pm 0.24 $ & $ +0.0 $ \\ 
 $ [1.1,2.5] $ & $ 0.04 \pm 0.03 $ & $ 0.73 \pm 0.32 $ & $ -2.2 $ \\ 
 $ [2.5,4] $ & $ 0.05 \pm 0.03 $ & $ -0.07 \pm 0.34 $ & $ +0.4 $ \\ 
 $ [4,6] $ & $ 0.03 \pm 0.02 $ & $ -0.33 \pm 0.25 $ & $ +1.4 $ \\ 
 $ [6,8] $ & $ 0.02 \pm 0.01 $ & $ 0.26 \pm 0.20 $ & $ -1.2 $ \\ 
 $ [15,19] $ & $ -0.00 \pm 0.03 $ & $ -0.02 \pm 0.14 $ & $ +0.2 $ \\ 
\midrule[1.6pt] 
$ F_L (B^{+}\to K^{*+}\mu^+\mu^-)\text{[LHCb]} $ & Standard Model & Experiment \cite{LHCb:2020gog} & Pull \\ 
 \midrule 
 $ [0.1,0.98] $ & $ 0.23 \pm 0.24 $ & $ 0.34 \pm 0.12 $ & $ -0.4 $ \\ 
 $ [1.1,2.5] $ & $ 0.68 \pm 0.26 $ & $ 0.54 \pm 0.19 $ & $ +0.5 $ \\ 
 $ [2.5,4] $ & $ 0.77 \pm 0.23 $ & $ 0.17 \pm 0.24 $ & $ +1.8 $ \\ 
 $ [4,6] $ & $ 0.71 \pm 0.28 $ & $ 0.67 \pm 0.14 $ & $ +0.1 $ \\ 
 $ [6,8] $ & $ 0.63 \pm 0.32 $ & $ 0.39 \pm 0.21 $ & $ +0.6 $ \\ 
 $ [15,19] $ & $ 0.34 \pm 0.03 $ & $ 0.40 \pm 0.13 $ & $ -0.4 $ \\ 
\midrule[1.6pt] 
$ P_1 (B^{+}\to K^{*+}\mu^+\mu^-)\text{[LHCb]} $ & Standard Model & Experiment \cite{LHCb:2020gog} & Pull \\ 
 \midrule 
 $ [0.1,0.98] $ & $ 0.03 \pm 0.08 $ & $ 0.44 \pm 0.41 $ & $ -1.0 $ \\ 
 $ [1.1,2.5] $ & $ -0.00 \pm 0.05 $ & $ 1.60 \pm 4.93 $ & $ -0.3 $ \\ 
 $ [2.5,4] $ & $ 0.00 \pm 0.06 $ & $ -0.29 \pm 1.45 $ & $ +0.2 $ \\ 
 $ [4,6] $ & $ 0.02 \pm 0.12 $ & $ -1.24 \pm 1.21 $ & $ +1.0 $ \\ 
 $ [6,8] $ & $ 0.02 \pm 0.13 $ & $ -0.78 \pm 0.70 $ & $ +1.1 $ \\ 
 $ [15,19] $ & $ -0.64 \pm 0.06 $ & $ -0.70 \pm 0.44 $ & $ +0.1 $ \\ 

\midrule[1.6pt] 
$ P_2 (B^{+}\to K^{*+}\mu^+\mu^-)\text{[LHCb]} $ & Standard Model & Experiment \cite{LHCb:2020gog}& Pull \\ 
 \midrule 
 $ [0.1,0.98] $ & $ 0.12 \pm 0.02 $ & $ 0.05 \pm 0.12 $ & $ +0.6 $ \\ 
 $ [1.1,2.5] $ & $ 0.44 \pm 0.03 $ & $ 0.28 \pm 0.45 $ & $ +0.4 $ \\ 
 $ [2.5,4] $ & $ 0.23 \pm 0.13 $ & $ -0.03 \pm 0.28 $ & $ +0.8 $ \\ 
 $ [4,6] $ & $ -0.19 \pm 0.11 $ & $ 0.15 \pm 0.21 $ & $ -1.5 $ \\ 
 $ [6,8] $ & $ -0.38 \pm 0.07 $ & $ 0.06 \pm 0.14 $ & $ -2.9 $ \\ 
 $ [15,19] $ & $ -0.36 \pm 0.02 $ & $ -0.34 \pm 0.10 $ & $ -0.2 $ \\ 
\midrule[1.6pt] 
$ P_3 (B^{+}\to K^{*+}\mu^+\mu^-)\text{[LHCb]} $ & Standard Model & Experiment \cite{LHCb:2020gog}& Pull \\ 
 \midrule 
 $ [0.1,0.98] $ & $ -0.00 \pm 0.00 $ & $ 0.42 \pm 0.22 $ & $ -2.0 $ \\ 
 $ [1.1,2.5] $ & $ 0.00 \pm 0.00 $ & $ 0.09 \pm 1.01 $ & $ -0.1 $ \\ 
 $ [2.5,4] $ & $ 0.00 \pm 0.01 $ & $ 0.45 \pm 0.65 $ & $ -0.7 $ \\ 
 $ [4,6] $ & $ 0.00 \pm 0.01 $ & $ 0.52 \pm 0.83 $ & $ -0.6 $ \\ 
 $ [6,8] $ & $ 0.00 \pm 0.00 $ & $ -0.17 \pm 0.34 $ & $ +0.5 $ \\ 
 $ [15,19] $ & $ 0.00 \pm 0.02 $ & $ 0.07 \pm 0.13 $ & $ -0.5 $ \\ 
\midrule[1.6pt] 
$ P'_4 (B^{+}\to K^{*+}\mu^+\mu^-)\text{[LHCb]} $ & Standard Model & Experiment \cite{LHCb:2020gog}& Pull \\ 
 \midrule 
 $ [0.1,0.98] $ & $ -0.50 \pm 0.16 $ & $ 0.18 \pm 0.76 $ & $ -0.8 $ \\ 
 $ [1.1,2.5] $ & $ -0.07 \pm 0.16 $ & $ -1.16 \pm 1.26 $ & $ +0.9 $ \\ 
 $ [2.5,4] $ & $ 0.53 \pm 0.21 $ & $ 1.62 \pm 2.20 $ & $ -0.5 $ \\ 
 $ [4,6] $ & $ 0.82 \pm 0.15 $ & $ 1.58 \pm 0.96 $ & $ -0.8 $ \\ 
 $ [6,8] $ & $ 0.93 \pm 0.11 $ & $ 0.86 \pm 0.91 $ & $ +0.1 $ \\ 
 $ [15,19] $ & $ 1.28 \pm 0.02 $ & $ 0.78 \pm 0.47 $ & $ +1.1 $ \\ 
\midrule[1.6pt] 
$ P'_5 (B^{+}\to K^{*+}\mu^+\mu^-)\text{[LHCb]} $ & Standard Model & Experiment \cite{LHCb:2020gog}& Pull \\ 
 \midrule 
 $ [0.1,0.98] $ & $ 0.67 \pm 0.13 $ & $ 0.51 \pm 0.32 $ & $ +0.5 $ \\ 
 $ [1.1,2.5] $ & $ 0.19 \pm 0.11 $ & $ 0.88 \pm 0.72 $ & $ -1.0 $ \\ 
 $ [2.5,4] $ & $ -0.47 \pm 0.12 $ & $ -0.87 \pm 1.68 $ & $ +0.2 $ \\ 
 $ [4,6] $ & $ -0.82 \pm 0.08 $ & $ -0.25 \pm 0.41 $ & $ -1.4 $ \\ 
 $ [6,8] $ & $ -0.94 \pm 0.08 $ & $ -0.15 \pm 0.41 $ & $ -1.9 $ \\ 
 $ [15,19] $ & $ -0.57 \pm 0.05 $ & $ -0.24 \pm 0.17 $ & $ -1.9 $ \\ 
\midrule[1.6pt] 
$ P'_6 (B^{+}\to K^{*+}\mu^+\mu^-)\text{[LHCb]} $ & Standard Model & Experiment \cite{LHCb:2020gog}& Pull \\ 
 \midrule 
 $ [0.1,0.98] $ & $ -0.06 \pm 0.02 $ & $ -0.02 \pm 0.40 $ & $ -0.1 $ \\ 
 $ [1.1,2.5] $ & $ -0.07 \pm 0.03 $ & $ 0.25 \pm 1.32 $ & $ -0.2 $ \\ 
 $ [2.5,4] $ & $ -0.06 \pm 0.03 $ & $ -0.37 \pm 3.91 $ & $ +0.1 $ \\ 
 $ [4,6] $ & $ -0.04 \pm 0.02 $ & $ -0.09 \pm 0.41 $ & $ +0.1 $ \\ 
 $ [6,8] $ & $ -0.02 \pm 0.01 $ & $ -0.74 \pm 0.40 $ & $ +1.8 $ \\ 
 $ [15,19] $ & $ -0.00 \pm 0.07 $ & $ -0.28 \pm 0.19 $ & $ +1.4 $ \\ 
\midrule[1.6pt] 
$ P'_8 (B^{+}\to K^{*+}\mu^+\mu^-)\text{[LHCb]} $ & Standard Model & Experiment \cite{LHCb:2020gog}& Pull \\ 
 \midrule 
 $ [0.1,0.98] $ & $ 0.02 \pm 0.02 $ & $ -0.90 \pm 1.02 $ & $ +1.0 $ \\ 
 $ [1.1,2.5] $ & $ 0.04 \pm 0.03 $ & $ -0.24 \pm 1.52 $ & $ +0.2 $ \\ 
 $ [2.5,4] $ & $ 0.05 \pm 0.03 $ & $ -0.24 \pm 15.80 $ & $ +0.0 $ \\ 
 $ [4,6] $ & $ 0.03 \pm 0.02 $ & $ 0.30 \pm 0.97 $ & $ -0.3 $ \\ 
 $ [6,8] $ & $ 0.02 \pm 0.01 $ & $ 0.78 \pm 0.78 $ & $ -1.0 $ \\ 
 $ [15,19] $ & $ -0.00 \pm 0.03 $ & $ 0.22 \pm 0.38 $ & $ -0.6 $ \\ 
\midrule[1.6pt] 
$ P_1 (B_s\to \phi\mu^+\mu^-)\text{[LHCb]} $ & Standard Model & Experiment \cite{LHCb:2021xxq}& Pull \\ 
 \midrule 
 $ [0.1,0.98] $ & $ 0.11 \pm 0.08 $ & $ -0.01 \pm 0.19 $ & $ +0.6 $ \\ 
 $ [1.1,4] $ & $ 0.01 \pm 0.06 $ & $ -0.22 \pm 0.42 $ & $ +0.5 $ \\ 
 $ [4,6] $ & $ -0.17 \pm 0.11 $ & $ -1.09 \pm 0.47 $ & $ +1.9 $ \\ 
 $ [6,8] $ & $ -0.21 \pm 0.11 $ & $ 0.07 \pm 0.43 $ & $ -0.6 $ \\ 
 $ [15,18.9] $ & $ -0.69 \pm 0.03 $ & $ -0.77 \pm 0.14 $ & $ +0.6 $ \\ 
\midrule[1.6pt] 
$ P'_4 (B_s\to \phi\mu^+\mu^-)\text{[LHCb]} $ & Standard Model & Experiment \cite{LHCb:2021xxq}& Pull \\ 
 \midrule 
 $ [0.1,0.98] $ & $ -0.45 \pm 0.15 $ & $ -0.98 \pm 0.38 $ & $ +1.3 $ \\ 
 $ [1.1,4] $ & $ 0.44 \pm 0.15 $ & $ 0.49 \pm 0.35 $ & $ -0.1 $ \\ 
 $ [4,6] $ & $ 1.01 \pm 0.08 $ & $ 0.97 \pm 0.41 $ & $ +0.1 $ \\ 
 $ [6,8] $ & $ 1.08 \pm 0.06 $ & $ 0.73 \pm 0.32 $ & $ +1.1 $ \\ 
 $ [15,18.9] $ & $ 1.30 \pm 0.01 $ & $ 0.87 \pm 0.20 $ & $ +2.2 $ \\ 
\midrule[1.6pt] 
$ P'_6 (B_s\to \phi\mu^+\mu^-)\text{[LHCb]} $ & Standard Model & Experiment \cite{LHCb:2021xxq}& Pull \\ 
 \midrule 
 $ [0.1,0.98] $ & $ -0.07 \pm 0.02 $ & $ -0.41 \pm 0.16 $ & $ +2.1 $ \\ 
 $ [1.1,4] $ & $ -0.07 \pm 0.02 $ & $ -0.23 \pm 0.17 $ & $ +0.9 $ \\ 
 $ [4,6] $ & $ -0.03 \pm 0.01 $ & $ 0.38 \pm 0.20 $ & $ -2.1 $ \\ 
 $ [6,8] $ & $ -0.02 \pm 0.01 $ & $ 0.07 \pm 0.17 $ & $ -0.5 $ \\ 
 $ [15,18.9] $ & $ -0.00 \pm 0.07 $ & $ 0.01 \pm 0.10 $ & $ -0.1 $ \\ 
\midrule[1.6pt] 
$ F_L (B_s\to \phi\mu^+\mu^-)\text{[LHCb]} $ & Standard Model & Experiment \cite{LHCb:2021xxq}& Pull \\ 
 \midrule 
 $ [0.1,0.98] $ & $ 0.28 \pm 0.09 $ & $ 0.25 \pm 0.05 $ & $ +0.3 $ \\ 
 $ [1.1,4] $ & $ 0.77 \pm 0.05 $ & $ 0.72 \pm 0.06 $ & $ +0.6 $ \\ 
 $ [4,6] $ & $ 0.71 \pm 0.05 $ & $ 0.70 \pm 0.05 $ & $ +0.1 $ \\ 
 $ [6,8] $ & $ 0.60 \pm 0.06 $ & $ 0.62 \pm 0.05 $ & $ -0.3 $ \\ 
 $ [15,18.9] $ & $ 0.36 \pm 0.02 $ & $ 0.36 \pm 0.04 $ & $ -0.1 $ \\ 
\midrule[1.6pt] 
{$ B^0\to K^{*0}e^+e^-\text{[LHCb]} $} & Standard Model & Experiment \cite{LHCb:2020dof} & Pull \\ 
 \midrule 
 $ F_L[0.008,0.257] $ & $ 0.03 \pm 0.06 $ & $ 0.04 \pm 0.03 $ & $ -0.2 $ \\ 
 $ P_1[0.008,0.257] $ & $ 0.03 \pm 0.08 $ & $ 0.11 \pm 0.10 $ & $ -0.6 $ \\ 
 $ P_2[0.008,0.257] $ & $ 0.01 \pm 0.00 $ & $ 0.03 \pm 0.04 $ & $ -0.5 $ \\ 
 % 
%  $ P^{CP}_3[0.008,0.257] $ & $ 0.00 \pm 0.00 $ & $ 0.01 \pm 0.05 $ & $ -0.2 $ \\ 
 % 
\midrule[1.6pt] 
$ R_{K^+}\text{[LHCb]} ^\dag$ & Standard Model & Experiment \cite{LHCb:2021trn} & Pull \\ 
 \midrule 
 $ [1.1,6.0] $ & $ 1.00 \pm 0.01 $ & $ 0.85 \pm 0.04 $ & $ +3.4 $ \\  
\midrule[1.6pt] 
$ R_{K^0}\text{[LHCb]} ^\dag $ & Standard Model & Experiment \cite{LHCb:2021lvy}& Pull \\ 
 \midrule 
 $ [1.1,6.0] $ & $ 1.00 \pm 0.01 $ & $ 0.66 \pm 0.20 $ & $ +1.7 $ \\ 
\midrule[1.6pt] 
$ R_{K}\text{[Belle]} ^\dag $ & Standard Model & Experiment \cite{BELLE:2019xld} & Pull \\ 
 \midrule 
 $ [1.0,6.0] $ & $ 1.00 \pm 0.01 $ & $ 1.03 \pm 0.28 $ & $ -0.1 $ \\ 
 $ [14.18,22.90] $ & $ 1.00 \pm 0.01 $ & $ 1.16 \pm 0.30 $ & $ -0.6 $ \\ 
\midrule[1.6pt] 
$ R_{K^{*0}}\text{[LHCb]} ^\dag $ & Standard Model & Experiment  \cite{LHCb:2017avl} & Pull \\ 
 \midrule 
 $ [0.045,1.1] $ & $ 0.91 \pm 0.02 $ & $ 0.66 \pm 0.11 $ & $ +2.2 $ \\ 
 $ [1.1,6.0] $ & $ 1.00 \pm 0.01 $ & $ 0.69 \pm 0.12 $ & $ +2.6 $ \\ 
\midrule[1.6pt] 
$ R_{K^{*+}}\text{[LHCb]}^\dag $ & Standard Model & Experiment  \cite{LHCb:2021lvy}& Pull \\ 
 \midrule 
 $ [0.045,6.0] $ & $ 0.93 \pm 0.05 $ & $ 0.70 \pm 0.18 $ & $ +1.2 $ \\ 
\midrule[1.6pt] 
$ R_{K^*}\text{[Belle]}^\dag $ & Standard Model & Experiment \cite{Belle:2019oag} & Pull \\ 
 \midrule 
 $ [0.045,1.1] $ & $ 0.92 \pm 0.02 $ & $ 0.52 \pm 0.36 $ & $ +1.1 $ \\ 
 $ [1.1,6.0] $ & $ 1.00 \pm 0.01 $ & $ 0.96 \pm 0.46 $ & $ +0.1 $ \\ 
 $ [15,19] $ & $ 1.00 \pm 0.00 $ & $ 1.18 \pm 0.53 $ & $ -0.5 $ \\ 
\midrule[1.6pt] 
$ P'_4 (B^{0}\to K^{*0}e^+e^-)\text{[Belle]} $ & Standard Model & Experiment \cite{Belle:2016fev} & Pull \\ 
 \midrule 
 $ [0.1,4] $ & $ -0.09 \pm 0.15 $ & $ -0.68 \pm 0.93 $ & $ +0.6 $ \\ 
 $ [4,8] $ & $ 0.88 \pm 0.13 $ & $ 1.04 \pm 0.48 $ & $ -0.3 $ \\ 
 $ [14.18,19] $ & $ 1.26 \pm 0.03 $ & $ 0.30 \pm 0.82 $ & $ +1.2 $ \\ 
\midrule[1.6pt] 
$ P'_4 (B^{0}\to K^{*0}\mu^+\mu^-)\text{[Belle]} $ & Standard Model & Experiment \cite{Belle:2016fev} & Pull \\ 
 \midrule 
 $ [0.1,4] $ & $ -0.06 \pm 0.16 $ & $ 0.76 \pm 1.03 $ & $ -0.8 $ \\ 
 $ [4,8] $ & $ 0.88 \pm 0.13 $ & $ 0.14 \pm 0.66 $ & $ +1.1 $ \\ 
 $ [14.18,19] $ & $ 1.26 \pm 0.03 $ & $ 0.20 \pm 0.79 $ & $ +1.3 $ \\ 
\midrule[1.6pt] 
$ P'_5 (B^{0}\to K^{*0}e^+e^-)\text{[Belle]} $ & Standard Model & Experiment \cite{Belle:2016fev} & Pull \\ 
 \midrule 
 $ [0.1,4] $ & $ 0.18 \pm 0.09 $ & $ 0.51 \pm 0.47 $ & $ -0.7 $ \\ 
 $ [4,8] $ & $ -0.88 \pm 0.07 $ & $ -0.52 \pm 0.28 $ & $ -1.3 $ \\ 
 $ [14.18,19] $ & $ -0.60 \pm 0.05 $ & $ -0.91 \pm 0.36 $ & $ +0.9 $ \\ 
\midrule[1.6pt] 
$ P'_5 (B^{0}\to K^{*0}\mu^+\mu^-)\text{[Belle]} $ & Standard Model & Experiment \cite{Belle:2016fev} & Pull \\ 
 \midrule 
 $ [0.1,4] $ & $ 0.17 \pm 0.10 $ & $ 0.42 \pm 0.41 $ & $ -0.6 $ \\ 
 $ [4,8] $ & $ -0.89 \pm 0.07 $ & $ -0.03 \pm 0.32 $ & $ -2.7 $ \\ 
 $ [14.18,19] $ & $ -0.60 \pm 0.05 $ & $ -0.13 \pm 0.39 $ & $ -1.3 $ \\ 
\midrule[1.6pt] 
$ Q_4 (B\to K^{*}\mu^+\mu^-)\text{[Belle]} ^\ddag$ & Standard Model & Experiment \cite{Belle:2016fev} & Pull \\ 
 \midrule 
 $ [0.1,4] $ & $ 0.03 \pm 0.01 $ & $ 1.45 \pm 1.39 $ & $ -1.0 $ \\ 
 $ [4,8] $ & $ 0.00 \pm 0.01 $ & $ -0.90 \pm 0.80 $ & $ +1.1 $ \\ 
 $ [14.18,19] $ & $ 0.00 \pm 0.01 $ & $ -0.08 \pm 1.14 $ & $ +0.1 $ \\ 

\midrule[1.6pt] 
$ Q_5 (B\to K^{*}\mu^+\mu^-)\text{[Belle]} ^\ddag$ & Standard Model & Experiment \cite{Belle:2016fev} & Pull \\ 
 \midrule 
 $ [0.1,4] $ & $ -0.02 \pm 0.01 $ & $ -0.10 \pm 0.62 $ & $ +0.1 $ \\ 
 $ [4,8] $ & $ -0.00 \pm 0.01 $ & $ 0.50 \pm 0.42 $ & $ -1.2 $ \\ 
 $ [14.18,19] $ & $ -0.00 \pm 0.01 $ & $ 0.78 \pm 0.51 $ & $ -1.5 $ \\ 
\midrule[1.6pt] 
$ 10^7 \times BR(B^+\to K^{+}\mu^+\mu^-)\text{[Belle]} $ & Standard Model & Experiment \cite{BELLE:2019xld} & Pull \\ 
 \midrule 
 $ [1,6] $ & $ 1.82 \pm 0.58 $ & $ 2.30 \pm 0.40 $ & $ -0.7 $ \\ 
 $ [14.18,22.9] $ & $ 1.23 \pm 0.17 $ & $ 1.34 \pm 0.23 $ & $ -0.4 $ \\ 
\midrule[1.6pt] 
$ 10^7 \times BR(B^0\to K^{0}\mu^+\mu^-)\text{[Belle]} $ & Standard Model & Experiment \cite{BELLE:2019xld} & Pull \\ 
 \midrule 
 $ [1,6] $ & $ 1.69 \pm 0.54 $ & $ 0.62 \pm 0.38 $ & $ +1.6 $ \\ 
 $ [14.18,22.9] $ & $ 1.14 \pm 0.15 $ & $ 0.98 \pm 0.40 $ & $ +0.4 $ \\ 
\midrule[1.6pt] 
{$ F_L (B^{0}\to K^{*0}\mu^+\mu^-)\text{[ATLAS]} $} & Standard Model & Experiment  \cite{ATLAS:2018gqc} & Pull \\ 
 \midrule 
 $ [0.04,2] $ & $ 0.36 \pm 0.30 $ & $ 0.44 \pm 0.11 $ & $ -0.3 $ \\ 
 $ [2,4] $ & $ 0.76 \pm 0.23 $ & $ 0.64 \pm 0.12 $ & $ +0.5 $ \\ 
 $ [4,6] $ & $ 0.71 \pm 0.28 $ & $ 0.42 \pm 0.18 $ & $ +0.9 $ \\ 
\midrule[1.6pt] 
{$ P_1 (B^{0}\to K^{*0}\mu^+\mu^-)\text{[ATLAS]} $} & Standard Model & Experiment  \cite{ATLAS:2018gqc} & Pull \\ 
 \midrule 
 $ [0.04,2] $ & $ 0.02 \pm 0.07 $ & $ -0.05 \pm 0.31 $ & $ +0.2 $ \\ 
 $ [2,4] $ & $ -0.00 \pm 0.05 $ & $ -0.78 \pm 0.61 $ & $ +1.3 $ \\ 
 $ [4,6] $ & $ 0.02 \pm 0.12 $ & $ 0.14 \pm 0.50 $ & $ -0.2 $ \\ 
\midrule[1.6pt] 
{$ P'_4 (B^{0}\to K^{*0}\mu^+\mu^-)\text{[ATLAS]} $} & Standard Model & Experiment  \cite{ATLAS:2018gqc} & Pull \\ 
 \midrule 
 $ [0.04,2] $ & $ -0.35 \pm 0.14 $ & $ -0.62 \pm 0.89 $ & $ +0.3 $ \\ 
 $ [2,4] $ & $ 0.43 \pm 0.21 $ & $ 1.52 \pm 0.75 $ & $ -1.4 $ \\ 
 $ [4,6] $ & $ 0.82 \pm 0.15 $ & $ -1.28 \pm 0.75 $ & $ +2.7 $ \\ 
\midrule[1.6pt] 
{$ P'_5 (B^{0}\to K^{*0}\mu^+\mu^-)\text{[ATLAS]} $} & Standard Model & Experiment  \cite{ATLAS:2018gqc} & Pull \\ 
 \midrule 
 $ [0.04,2] $ & $ 0.50 \pm 0.10 $ & $ 0.67 \pm 0.31 $ & $ -0.5 $ \\ 
 $ [2,4] $ & $ -0.36 \pm 0.12 $ & $ -0.33 \pm 0.34 $ & $ -0.1 $ \\ 
 $ [4,6] $ & $ -0.82 \pm 0.08 $ & $ 0.26 \pm 0.39 $ & $ -2.7 $ \\ 
\midrule[1.6pt] 
{$ P'_6 (B^{0}\to K^{*0}\mu^+\mu^-)\text{[ATLAS]} $} & Standard Model & Experiment  \cite{ATLAS:2018gqc} & Pull \\ 
 \midrule 
 $ [0.04,2] $ & $ -0.06 \pm 0.02 $ & $ -0.18 \pm 0.21 $ & $ +0.6 $ \\ 
 $ [2,4] $ & $ -0.06 \pm 0.03 $ & $ 0.31 \pm 0.34 $ & $ -1.1 $ \\ 
 $ [4,6] $ & $ -0.04 \pm 0.02 $ & $ 0.06 \pm 0.30 $ & $ -0.3 $ \\ 
\midrule[1.6pt] 
{$ P'_8 (B^{0}\to K^{*0}\mu^+\mu^-)\text{[ATLAS]} $} & Standard Model & Experiment  \cite{ATLAS:2018gqc} & Pull \\ 
 \midrule 
 $ [0.04,2] $ & $ 0.03 \pm 0.02 $ & $ 0.58 \pm 1.03 $ & $ -0.5 $ \\ 
 $ [2,4] $ & $ 0.05 \pm 0.03 $ & $ -2.14 \pm 1.13 $ & $ +1.9 $ \\ 
 $ [4,6] $ & $ 0.03 \pm 0.02 $ & $ 0.48 \pm 0.86 $ & $ -0.5 $ \\ 
\midrule[1.6pt] 
{$ P_1 (B^{0}\to K^{*0}\mu^+\mu^-)\text{[CMS 8 TeV]} $} & Standard Model & Experiment  \cite{CMS:2017rzx} & Pull \\ 
 \midrule 
 $ [1,2] $ & $ 0.00 \pm 0.06 $ & $ 0.12 \pm 0.48 $ & $ -0.2 $ \\ 
 $ [2,4.3] $ & $ 0.00 \pm 0.05 $ & $ -0.69 \pm 0.62 $ & $ +1.1 $ \\ 
 $ [4.3,6] $ & $ 0.03 \pm 0.12 $ & $ 0.53 \pm 0.38 $ & $ -1.3 $ \\ 
 $ [6,8.68] $ & $ 0.02 \pm 0.14 $ & $ -0.47 \pm 0.31 $ & $ +1.4 $ \\ 
 $ [16,19] $ & $ -0.70 \pm 0.05 $ & $ -0.53 \pm 0.25 $ & $ -0.7 $ \\ 
\midrule[1.6pt] 
{$ P'_5 (B^{0}\to K^{*0}\mu^+\mu^-)\text{[CMS 8 TeV]} $} & Standard Model & Experiment  \cite{CMS:2017rzx} & Pull \\ 
 \midrule 
 $ [1,2] $ & $ 0.33 \pm 0.11 $ & $ 0.10 \pm 0.33 $ & $ +0.7 $ \\ 
 $ [2,4.3] $ & $ -0.41 \pm 0.12 $ & $ -0.57 \pm 0.38 $ & $ +0.4 $ \\ 
 $ [4.3,6] $ & $ -0.84 \pm 0.08 $ & $ -0.96 \pm 0.33 $ & $ +0.4 $ \\ 
 $ [6,8.68] $ & $ -0.95 \pm 0.08 $ & $ -0.64 \pm 0.23 $ & $ -1.3 $ \\ 
 $ [16,19] $ & $ -0.53 \pm 0.04 $ & $ -0.56 \pm 0.14 $ & $ +0.2 $ \\ 
\midrule[1.6pt] 
$ F_L (B^{0}\to K^{*0}\mu^+\mu^-)\text{[CMS 8 TeV]} $ & Standard Model & Experiment \cite{CMS:2015bcy} & Pull \\ 
 \midrule 
 $ [1,2] $ & $ 0.63 \pm 0.28 $ & $ 0.64 \pm 0.12 $ & $ -0.0 $ \\ 
 $ [2,4.3] $ & $ 0.76 \pm 0.23 $ & $ 0.80 \pm 0.10 $ & $ -0.2 $ \\ 
 $ [4.3,6] $ & $ 0.71 \pm 0.28 $ & $ 0.62 \pm 0.12 $ & $ +0.3 $ \\ 
 $ [6,8.68] $ & $ 0.62 \pm 0.32 $ & $ 0.50 \pm 0.08 $ & $ +0.3 $ \\ 
 $ [16,19] $ & $ 0.34 \pm 0.03 $ & $ 0.38 \pm 0.07 $ & $ -0.6 $ \\ 
\midrule[1.6pt] 
$ A_{FB} (B^{0}\to K^{*0}\mu^+\mu^-)\text{[CMS 8 TeV]} $ & Standard Model & Experiment \cite{CMS:2015bcy} & Pull \\ 
 \midrule 
 $ [1,2] $ & $ -0.20 \pm 0.18 $ & $ -0.27 \pm 0.41 $ & $ +0.3 $ \\ 
 $ [2,4.3] $ & $ -0.08 \pm 0.08 $ & $ -0.12 \pm 0.18 $ & $ +0.2 $ \\ 
 $ [4.3,6] $ & $ 0.09 \pm 0.11 $ & $ 0.01 \pm 0.15 $ & $ +0.4 $ \\ 
 $ [6,8.68] $ & $ 0.22 \pm 0.21 $ & $ 0.03 \pm 0.10 $ & $ +0.8 $ \\ 
 $ [16,19] $ & $ 0.34 \pm 0.03 $ & $ 0.35 \pm 0.07 $ & $ -0.2 $ \\ 
\midrule[1.6pt] 
$ 10^7 \times BR(B^{0}\to K^{*0}\mu^+\mu^-)\text{[CMS 8 TeV]} $ & Standard Model & Experiment  \cite{CMS:2015bcy} & Pull \\ 
 \midrule 
 $ [1,2] $ & $ 0.42 \pm 0.26 $ & $ 0.46 \pm 0.08 $ & $ -0.1 $ \\ 
 $ [2,4.3] $ & $ 0.89 \pm 0.61 $ & $ 0.76 \pm 0.12 $ & $ +0.2 $ \\ 
 $ [4.3,6] $ & $ 0.78 \pm 0.58 $ & $ 0.58 \pm 0.10 $ & $ +0.4 $ \\ 
 $ [6,8.68] $ & $ 1.57 \pm 1.25 $ & $ 1.26 \pm 0.13 $ & $ +0.2 $ \\ 
 $ [16,19] $ & $ 1.73 \pm 0.14 $ & $ 1.26 \pm 0.13 $ & $ +2.5 $ \\ 
\midrule[1.6pt] 
$ F_H (B^{+}\to K^{+}\mu^+\mu^-)\text{[CMS 8 TeV]} $ & Standard Model & Experiment \cite{CMS:2018qih} & Pull \\ 
 \midrule 
 $ [1,2] $ & $ 0.05 \pm 0.00 $ & $ 0.21 \pm 0.49 $ & $ -0.4 $ \\ 
 $ [2,4.3] $ & $ 0.02 \pm 0.00 $ & $ 0.85 \pm 0.37 $ & $ -2.4 $ \\ 
 $ [4.3,8.68] $ & $ 0.01 \pm 0.00 $ & $ 0.01 \pm 0.04 $ & $ +0.0 $ \\ 
 $ [16,18] $ & $ 0.01 \pm 0.00 $ & $ 0.07 \pm 0.10 $ & $ -0.6 $ \\ 
 $ [18,22] $ & $ 0.01 \pm 0.00 $ & $ 0.10 \pm 0.13 $ & $ -0.7 $ \\ 
\midrule[1.6pt] 
$ A_{FB} (B^{+}\to K^{+}\mu^+\mu^-)\text{[CMS 8 TeV]} $ & Standard Model & Experiment \cite{CMS:2018qih} & Pull \\ 
 \midrule 
 $ [1,2] $ & $ 0 \pm 0.00 $ & $ 0.08 \pm 0.23 $ & $ -0.4 $ \\ 
 $ [2,4.3] $ & $ 0 \pm 0.00 $ & $ -0.04 \pm 0.14 $ & $ +0.3 $ \\ 
 $ [4.3,8.68] $ & $ 0 \pm 0.00 $ & $ 0.00 \pm 0.04 $ & $ +0.0 $ \\ 
 $ [16,18] $ & $ 0 \pm 0.00 $ & $ 0.04 \pm 0.06 $ & $ -0.8 $ \\ 
 $ [18,22] $ & $ 0 \pm 0.00 $ & $ 0.05 \pm 0.05 $ & $ -1.1 $ \\ 
\midrule[1.6pt] 
$ F_L (B^{+}\to K^{*+}\mu^+\mu^-)\text{[CMS 8 TeV]} $ & Standard Model & Experiment  \cite{CMS:2020oqb} & Pull \\ 
 \midrule 
 $ [1,8.68] $ & $ 0.67 \pm 0.29 $ & $ 0.60 \pm 0.34 $ & $ +0.2 $ \\ 
 $ [14.18,19] $ & $ 0.35 \pm 0.04 $ & $ 0.55 \pm 0.14 $ & $ -1.7 $ \\ 
\midrule[1.6pt] 
$ A_{FB} (B^{+}\to K^{*+}\mu^+\mu^-)\text{[CMS 8 TeV]} $ & Standard Model & Experiment  \cite{CMS:2020oqb} & Pull \\ 
 \midrule 
 $ [1,8.68] $ & $ 0.08 \pm 0.09 $ & $ -0.14 \pm 0.39 $ & $ +0.6 $ \\ 
 $ [14.18,19] $ & $ 0.37 \pm 0.03 $ & $ 0.33 \pm 0.12 $ & $ +0.3 $ \\ 
\midrule[1.6pt] 
$ F_L (B^{0}\to K^{*0}\mu^+\mu^-)\text{[CMS 7 TeV]} $ & Standard Model & Experiment \cite{CMS:2013mkz} & Pull \\ 
 \midrule 
 $ [1,2] $ & $ 0.63 \pm 0.28 $ & $ 0.60 \pm 0.34 $ & $ +0.1 $ \\ 
 $ [2,4.3] $ & $ 0.76 \pm 0.23 $ & $ 0.65 \pm 0.17 $ & $ +0.4 $ \\ 
 $ [4.3,8.68] $ & $ 0.65 \pm 0.31 $ & $ 0.81 \pm 0.14 $ & $ -0.5 $ \\ 
 $ [16,19] $ & $ 0.34 \pm 0.03 $ & $ 0.44 \pm 0.08 $ & $ -1.3 $ \\ 
\midrule[1.6pt] 
$ A_{FB} (B^{0}\to K^{*0}\mu^+\mu^-)\text{[CMS 7 TeV]} $ & Standard Model & Experiment \cite{CMS:2013mkz} & Pull \\ 
 \midrule 
 $ [1,2] $ & $ -0.20 \pm 0.18 $ & $ -0.29 \pm 0.41 $ & $ +0.2 $ \\ 
 $ [2,4.3] $ & $ -0.08 \pm 0.08 $ & $ -0.07 \pm 0.20 $ & $ -0.0 $ \\ 
 $ [4.3,8.68] $ & $ 0.18 \pm 0.18 $ & $ -0.01 \pm 0.11 $ & $ +0.9 $ \\ 
 $ [16,19] $ & $ 0.34 \pm 0.03 $ & $ 0.41 \pm 0.06 $ & $ -1.1 $ \\ 
\midrule[1.6pt] 
$ 10^7 \times BR(B^{0}\to K^{*0}\mu^+\mu^-)\text{[CMS 7 TeV]} $ & Standard Model & Experiment \cite{CMS:2013mkz} & Pull \\ 
 \midrule 
 $ [1,2] $ & $ 0.42 \pm 0.26 $ & $ 0.48 \pm 0.15 $ & $ -0.2 $ \\ 
 $ [2,4.3] $ & $ 0.89 \pm 0.61 $ & $ 0.87 \pm 0.18 $ & $ +0.0 $ \\ 
 $ [4.3,8.68] $ & $ 2.35 \pm 1.82 $ & $ 1.62 \pm 0.35 $ & $ +0.4 $ \\ 
 $ [16,19] $ & $ 1.73 \pm 0.14 $ & $ 1.56 \pm 0.23 $ & $ +0.6 $ \\ 

\midrule[1.6pt] 
{$ 10^5 \times BR(B^0\to K^{*0}\gamma)\text{[PDG]}^\dag $} & Standard Model & Experiment \cite{ParticleDataGroup:2020ssz} & Pull \\ 
 \midrule 
 & $ 4.57 \pm 5.27 $ & $ 4.18 \pm 0.25 $ & $ { +0.1} $ \\ 
\midrule[1.6pt] 
{$ 10^5 \times BR(B^+\to K^{*+}\gamma)\text{[PDG]}^\dag $} & Standard Model & Experiment \cite{ParticleDataGroup:2020ssz} & Pull \\ 
 \midrule 
 & $ 4.61 \pm 5.49 $ & $ 3.92 \pm 0.22 $ & $ { +0.1} $ \\ 
\midrule[1.6pt] 
$ 10^5 \times BR(B_s\to \phi\gamma)\text{[PDG]}^\dag $ & Standard Model & Experiment \cite{ParticleDataGroup:2020ssz} & Pull \\ 
 \midrule 
 & $ 4.86 \pm 1.35 $ & $ 3.40 \pm 0.40 $ & $ { +1.0} $ \\ 
 \midrule[1.6pt] 
$ 10^4 \times BR(B\to X_s\gamma)\text{[HFLAV]}^\dag $ & Standard Model \cite{Misiak:2020vlo} & Experiment \cite{HFLAV:2019otj} & Pull \\ 
 \midrule 
 & $ 3.32 \pm 0.15 $ & $ 3.40 \pm 0.17 $ & $ -0.4 $ \\ 
 \midrule[1.6pt] 
 
$  S(B \to K^* \gamma)\text{[BaBar+Belle]}^\dag $ & Standard Model \cite{Descotes-Genon:2011nqe} & Experiment \cite{HFLAV:2019otj} & Pull \\ 
 \midrule 
 & $-0.03\pm 0.01$ & $-0.16 \pm 0.22$ & ${+0.6}$  \\ 
 \midrule[1.6pt] 
$  AI(B \to K^* \gamma)\text{[BaBar+Belle]}^\dag $ & Standard Model \cite{Descotes-Genon:2011nqe} & Experiment \cite{HFLAV:2019otj} & Pull \\ 
 \midrule 
 & $0.041\pm0.025$  & $0.063 \pm 0.017$  & ${-0.7}$ \\ 

 \midrule[1.6pt] 
 
 %1311.0903 
$ 10^9\times BR(B_s \to \mu^+\mu^-) \text{[LHCb+CMS+ATLAS]}^\dag $ & Standard Model \cite{MisiakOrsay} & Experiment \cite{Hurth:2021nsi} & Pull \\ 
 \midrule 
 & $3.64\pm0.14$ & $2.85\pm0.34$  & $+2.2$  \\ 

 \midrule[1.6pt] 
$ 10^{6} \times BR(B\to X_s\mu^+\mu^-)\text{[BaBar]}^\dag $ & Standard Model \cite{Huber:2020vup} & Experiment \cite{BaBar:2013qry} & Pull \\ 
 \midrule 
  $[1,6]$ & $1.73 \pm 0.13$  & $0.66\pm 0.88$ & $+1.2$  \\ 
 \midrule[1.6pt] 
$ 10^{6} \times BR(B\to X_s e^+e^-)\text{[BaBar]}^\dag $ & Standard Model \cite{Huber:2020vup} & Experiment \cite{BaBar:2013qry} & Pull \\ 
 \midrule 
  $[1,6]$ & $1.78 \pm 0.13$  & $1.93\pm 0.55$ & $-0.3$  \\ 
\bottomrule[1.6pt]
\label{table_predictions_SM}
\end{longtable}

\bibliography{bibliography}

\begin{thebibliography}{10}
\expandafter\ifx\csname url\endcsname\relax
  \def\url#1{\texttt{#1}}\fi
\expandafter\ifx\csname urlprefix\endcsname\relax\def\urlprefix{URL }\fi
\expandafter\ifx\csname href\endcsname\relax
  \def\href#1#2{#2} \def\path#1{#1}\fi

\bibitem{Bifani:2018zmi}
S.~Bifani, S.~Descotes-Genon, A.~Romero~Vidal, M.-H. Schune, {Review of Lepton
  Universality tests in $B$ decays}, J. Phys. G 46~(2) (2019) 023001.
\newblock \href {http://arxiv.org/abs/1809.06229} {\path{arXiv:1809.06229}},
  \href {https://doi.org/10.1088/1361-6471/aaf5de}
  {\path{doi:10.1088/1361-6471/aaf5de}}.

\bibitem{Albrecht:2021tul}
J.~Albrecht, D.~van Dyk, C.~Langenbruch, {Flavour anomalies in heavy quark
  decays}, Prog. Part. Nucl. Phys. 120 (2021) 103885.
\newblock \href {http://arxiv.org/abs/2107.04822} {\path{arXiv:2107.04822}},
  \href {https://doi.org/10.1016/j.ppnp.2021.103885}
  {\path{doi:10.1016/j.ppnp.2021.103885}}.

\bibitem{London:2021lfn}
D.~London, J.~Matias, {$B$ Flavour Anomalies: 2021 Theoretical Status Report}
  (10 2021).
\newblock \href {http://arxiv.org/abs/2110.13270} {\path{arXiv:2110.13270}},
  \href {https://doi.org/10.1146/annurev-nucl-102020-090209}
  {\path{doi:10.1146/annurev-nucl-102020-090209}}.

\bibitem{Aaij:2021vac}
R.~Aaij, et~al., {Test of lepton universality in beauty-quark decays} (3 2021).
\newblock \href {http://arxiv.org/abs/2103.11769} {\path{arXiv:2103.11769}}.

\bibitem{LHCb:2021lvy}
R.~Aaij, et~al., {Tests of lepton universality using $B^0\to K^0_S \ell^+
  \ell^-$ and $B^+\to K^{*+} \ell^+ \ell^-$ decays} (10 2021).
\newblock \href {http://arxiv.org/abs/2110.09501} {\path{arXiv:2110.09501}}.

\bibitem{Descotes-Genon:2015uva}
S.~Descotes-Genon, L.~Hofer, J.~Matias, J.~Virto, {Global analysis of $b\to
  s\ell\ell$ anomalies}, JHEP 06 (2016) 092.
\newblock \href {http://arxiv.org/abs/1510.04239} {\path{arXiv:1510.04239}},
  \href {https://doi.org/10.1007/JHEP06(2016)092}
  {\path{doi:10.1007/JHEP06(2016)092}}.

\bibitem{Capdevila:2017bsm}
B.~Capdevila, A.~Crivellin, S.~Descotes-Genon, J.~Matias, J.~Virto, {Patterns
  of New Physics in $b\to s\ell^+\ell^-$ transitions in the light of recent
  data}, JHEP 01 (2018) 093.
\newblock \href {http://arxiv.org/abs/1704.05340} {\path{arXiv:1704.05340}},
  \href {https://doi.org/10.1007/JHEP01(2018)093}
  {\path{doi:10.1007/JHEP01(2018)093}}.

\bibitem{Alguero:2018nvb}
M.~Alguer\'o, B.~Capdevila, S.~Descotes-Genon, P.~Masjuan, J.~Matias, {Are we
  overlooking lepton flavour universal new physics in $b\to s\ell\ell$ ?},
  Phys. Rev. D 99~(7) (2019) 075017.
\newblock \href {http://arxiv.org/abs/1809.08447} {\path{arXiv:1809.08447}},
  \href {https://doi.org/10.1103/PhysRevD.99.075017}
  {\path{doi:10.1103/PhysRevD.99.075017}}.

\bibitem{Alguero:2019ptt}
M.~Alguer\'o, B.~Capdevila, A.~Crivellin, S.~Descotes-Genon, P.~Masjuan,
  J.~Matias, M.~Novoa~Brunet, J.~Virto, {Emerging patterns of New Physics with
  and without Lepton Flavour Universal contributions}, Eur. Phys. J. C 79~(8)
  (2019) 714, [Addendum: Eur.Phys.J.C 80, 511 (2020)].
\newblock \href {http://arxiv.org/abs/1903.09578} {\path{arXiv:1903.09578}},
  \href {https://doi.org/10.1140/epjc/s10052-019-7216-3}
  {\path{doi:10.1140/epjc/s10052-019-7216-3}}.

\bibitem{Geng:2021nhg}
L.-S. Geng, B.~Grinstein, S.~J\"ager, S.-Y. Li, J.~Martin~Camalich, R.-X. Shi,
  {Implications of new evidence for lepton-universality violation in $b\to
  s\ell\ell$ decays}, Phys. Rev. D 104~(3) (2021) 035029.
\newblock \href {http://arxiv.org/abs/2103.12738} {\path{arXiv:2103.12738}},
  \href {https://doi.org/10.1103/PhysRevD.104.035029}
  {\path{doi:10.1103/PhysRevD.104.035029}}.

\bibitem{Altmannshofer:2021qrr}
W.~Altmannshofer, P.~Stangl, {New physics in rare B decays after Moriond 2021},
  Eur. Phys. J. C 81~(10) (2021) 952.
\newblock \href {http://arxiv.org/abs/2103.13370} {\path{arXiv:2103.13370}},
  \href {https://doi.org/10.1140/epjc/s10052-021-09725-1}
  {\path{doi:10.1140/epjc/s10052-021-09725-1}}.

\bibitem{Hurth:2020ehu}
T.~Hurth, F.~Mahmoudi, S.~Neshatpour, {Model independent analysis of the
  angular observables in $B^{0} \to K^{*0} \mu^+ \mu^-$ and $B^{+} \to K^{*+}
  \mu^+ \mu^-$}, Phys. Rev. D 103 (2021) 095020.
\newblock \href {http://arxiv.org/abs/2012.12207} {\path{arXiv:2012.12207}},
  \href {https://doi.org/10.1103/PhysRevD.103.095020}
  {\path{doi:10.1103/PhysRevD.103.095020}}.

\bibitem{Alok:2019ufo}
A.~K. Alok, A.~Dighe, S.~Gangal, D.~Kumar, {Continuing search for new physics
  in $b \to s \mu \mu$ decays: two operators at a time}, JHEP 06 (2019) 089.
\newblock \href {http://arxiv.org/abs/1903.09617} {\path{arXiv:1903.09617}},
  \href {https://doi.org/10.1007/JHEP06(2019)089}
  {\path{doi:10.1007/JHEP06(2019)089}}.

\bibitem{Ciuchini:2020gvn}
M.~Ciuchini, M.~Fedele, E.~Franco, A.~Paul, L.~Silvestrini, M.~Valli, {Lessons
  from the $B^{0,+}\to K^{*0,+}\mu^+\mu^-$ angular analyses}, Phys. Rev. D
  103~(1) (2021) 015030.
\newblock \href {http://arxiv.org/abs/2011.01212} {\path{arXiv:2011.01212}},
  \href {https://doi.org/10.1103/PhysRevD.103.015030}
  {\path{doi:10.1103/PhysRevD.103.015030}}.

\bibitem{Datta:2019zca}
A.~Datta, J.~Kumar, D.~London, {The $B$ anomalies and new physics in $b \to s
  e^+ e^-$}, Phys. Lett. B 797 (2019) 134858.
\newblock \href {http://arxiv.org/abs/1903.10086} {\path{arXiv:1903.10086}},
  \href {https://doi.org/10.1016/j.physletb.2019.134858}
  {\path{doi:10.1016/j.physletb.2019.134858}}.

\bibitem{Hurth:2020rzx}
T.~Hurth, F.~Mahmoudi, S.~Neshatpour, {Implications of the new LHCb angular
  analysis of $B \to K^* \mu^+ \mu^-$ : Hadronic effects or new physics?},
  Phys. Rev. D 102~(5) (2020) 055001.
\newblock \href {http://arxiv.org/abs/2006.04213} {\path{arXiv:2006.04213}},
  \href {https://doi.org/10.1103/PhysRevD.102.055001}
  {\path{doi:10.1103/PhysRevD.102.055001}}.

\bibitem{LHCb:2021trn}
R.~Aaij, et~al., {Test of lepton universality in beauty-quark decays} (3 2021).
\newblock \href {http://arxiv.org/abs/2103.11769} {\path{arXiv:2103.11769}}.

\bibitem{Abdesselam:2019lab}
S.~Choudhury, et~al., {Test of lepton flavor universality and search for lepton
  flavor violation in $B \rightarrow K\ell \ell$ decays}, JHEP 03 (2021) 105.
\newblock \href {http://arxiv.org/abs/1908.01848} {\path{arXiv:1908.01848}},
  \href {https://doi.org/10.1007/JHEP03(2021)105}
  {\path{doi:10.1007/JHEP03(2021)105}}.

\bibitem{LHCb:2021vsc}
R.~Aaij, et~al., {Analysis of neutral $B$-meson decays into two muons} (8
  2021).
\newblock \href {http://arxiv.org/abs/2108.09284} {\path{arXiv:2108.09284}}.

\bibitem{Sirunyan:2019xdu}
A.~M. Sirunyan, et~al., {Measurement of properties of
  B$^0_\mathrm{s}\to\mu^+\mu^-$ decays and search for B$^0\to\mu^+\mu^-$ with
  the CMS experiment}, JHEP 04 (2020) 188.
\newblock \href {http://arxiv.org/abs/1910.12127} {\path{arXiv:1910.12127}},
  \href {https://doi.org/10.1007/JHEP04(2020)188}
  {\path{doi:10.1007/JHEP04(2020)188}}.

\bibitem{Aaboud:2018mst}
M.~Aaboud, et~al., {Study of the rare decays of $B^0_s$ and $B^0$ mesons into
  muon pairs using data collected during 2015 and 2016 with the ATLAS
  detector}, JHEP 04 (2019) 098.
\newblock \href {http://arxiv.org/abs/1812.03017} {\path{arXiv:1812.03017}},
  \href {https://doi.org/10.1007/JHEP04(2019)098}
  {\path{doi:10.1007/JHEP04(2019)098}}.

\bibitem{Hurth:2021nsi}
T.~Hurth, F.~Mahmoudi, D.~M. Santos, S.~Neshatpour, {More indications for
  lepton nonuniversality in $b \to s\ell^+\ell^-$}, Phys. Lett. B 824 (2022)
  136838.
\newblock \href {http://arxiv.org/abs/2104.10058} {\path{arXiv:2104.10058}},
  \href {https://doi.org/10.1016/j.physletb.2021.136838}
  {\path{doi:10.1016/j.physletb.2021.136838}}.

\bibitem{MisiakOrsay}
M.~Misiak, Talk given at the workshop “New physics at the low-energy
  precision frontier”, Orsay (https://indico.cern.ch/event/815529/) and
  private communication (2019).

\bibitem{LHCb:2020gog}
R.~Aaij, et~al., {Angular Analysis of the $B^{+}\rightarrow
  K^{\ast+}\mu^{+}\mu^{-}$ Decay}, Phys. Rev. Lett. 126~(16) (2021) 161802.
\newblock \href {http://arxiv.org/abs/2012.13241} {\path{arXiv:2012.13241}},
  \href {https://doi.org/10.1103/PhysRevLett.126.161802}
  {\path{doi:10.1103/PhysRevLett.126.161802}}.

\bibitem{Descotes-Genon:2013vna}
S.~Descotes-Genon, T.~Hurth, J.~Matias, J.~Virto, {Optimizing the basis of
  $B\to K^*ll$ observables in the full kinematic range}, JHEP 05 (2013) 137.
\newblock \href {http://arxiv.org/abs/1303.5794} {\path{arXiv:1303.5794}},
  \href {https://doi.org/10.1007/JHEP05(2013)137}
  {\path{doi:10.1007/JHEP05(2013)137}}.

\bibitem{CMS:2020oqb}
A.~M. Sirunyan, et~al., {Angular analysis of the decay B$^+$ $\to$
  K$^*$(892)$^+\mu^+\mu^-$ in proton-proton collisions at $\sqrt{s} =$ 8 TeV},
  JHEP 04 (2021) 124.
\newblock \href {http://arxiv.org/abs/2010.13968} {\path{arXiv:2010.13968}},
  \href {https://doi.org/10.1007/JHEP04(2021)124}
  {\path{doi:10.1007/JHEP04(2021)124}}.

\bibitem{Beneke:2004dp}
M.~Beneke, T.~Feldmann, D.~Seidel, {Exclusive radiative and electroweak $b \to
  d$ and $b \to s$ penguin decays at NLO}, Eur. Phys. J. C 41 (2005) 173--188.
\newblock \href {http://arxiv.org/abs/hep-ph/0412400}
  {\path{arXiv:hep-ph/0412400}}, \href
  {https://doi.org/10.1140/epjc/s2005-02181-5}
  {\path{doi:10.1140/epjc/s2005-02181-5}}.

\bibitem{Sirunyan:2018jll}
A.~M. Sirunyan, et~al., {Angular analysis of the decay B$^+\to$ K$^+\mu^+\mu^-$
  in proton-proton collisions at $\sqrt{s} =$ 8 TeV}, Phys. Rev. D 98~(11)
  (2018) 112011.
\newblock \href {http://arxiv.org/abs/1806.00636} {\path{arXiv:1806.00636}},
  \href {https://doi.org/10.1103/PhysRevD.98.112011}
  {\path{doi:10.1103/PhysRevD.98.112011}}.

\bibitem{Aaij:2020umj}
R.~Aaij, et~al., {Strong constraints on the $b \to s\gamma$ photon polarisation
  from $B^0 \to K^{*0} e^+ e^-$ decays}, JHEP 12 (2020) 081.
\newblock \href {http://arxiv.org/abs/2010.06011} {\path{arXiv:2010.06011}},
  \href {https://doi.org/10.1007/JHEP12(2020)081}
  {\path{doi:10.1007/JHEP12(2020)081}}.

\bibitem{Aaij:2015dea}
R.~Aaij, et~al., {Angular analysis of the $B^{0} \to K^{*0} e^{+} e^{-}$ decay
  in the low-q$^{2}$ region}, JHEP 04 (2015) 064.
\newblock \href {http://arxiv.org/abs/1501.03038} {\path{arXiv:1501.03038}},
  \href {https://doi.org/10.1007/JHEP04(2015)064}
  {\path{doi:10.1007/JHEP04(2015)064}}.

\bibitem{LHCb:2021xxq}
R.~Aaij, et~al., {Angular analysis of the rare decay $ {B}_s^0
  \to\phi\mu^+\mu^-$}, JHEP 11 (2021) 043.
\newblock \href {http://arxiv.org/abs/2107.13428} {\path{arXiv:2107.13428}},
  \href {https://doi.org/10.1007/JHEP11(2021)043}
  {\path{doi:10.1007/JHEP11(2021)043}}.

\bibitem{LHCb:2021zwz}
R.~Aaij, et~al., {Branching Fraction Measurements of the Rare
  $B^0_s\rightarrow\phi\mu^+\mu^-$ and $B^0_s\rightarrow
  f_2^\prime(1525)\mu^+\mu^-$- Decays}, Phys. Rev. Lett. 127~(15) (2021)
  151801.
\newblock \href {http://arxiv.org/abs/2105.14007} {\path{arXiv:2105.14007}},
  \href {https://doi.org/10.1103/PhysRevLett.127.151801}
  {\path{doi:10.1103/PhysRevLett.127.151801}}.

\bibitem{Aaij:2015esa}
R.~Aaij, et~al., {Angular analysis and differential branching fraction of the
  decay $B^0_s\to\phi\mu^+\mu^-$}, JHEP 09 (2015) 179.
\newblock \href {http://arxiv.org/abs/1506.08777} {\path{arXiv:1506.08777}},
  \href {https://doi.org/10.1007/JHEP09(2015)179}
  {\path{doi:10.1007/JHEP09(2015)179}}.

\bibitem{LHCb:2018jna}
R.~Aaij, et~al., {Angular moments of the decay $\Lambda_b^0 \rightarrow \Lambda
  \mu^{+} \mu^{-}$ at low hadronic recoil}, JHEP 09 (2018) 146.
\newblock \href {http://arxiv.org/abs/1808.00264} {\path{arXiv:1808.00264}},
  \href {https://doi.org/10.1007/JHEP09(2018)146}
  {\path{doi:10.1007/JHEP09(2018)146}}.

\bibitem{Blake:2019guk}
T.~Blake, S.~Meinel, D.~van Dyk, {Bayesian Analysis of $b\to s\mu^+\mu^-$
  Wilson Coefficients using the Full Angular Distribution of $\Lambda_b\to
  \Lambda(\to p\, \pi^-)\mu^+\mu^-$ Decays}, Phys. Rev. D 101~(3) (2020)
  035023.
\newblock \href {http://arxiv.org/abs/1912.05811} {\path{arXiv:1912.05811}},
  \href {https://doi.org/10.1103/PhysRevD.101.035023}
  {\path{doi:10.1103/PhysRevD.101.035023}}.

\bibitem{Grinstein:1987vj}
B.~Grinstein, R.~P. Springer, M.~B. Wise, {Effective Hamiltonian for Weak
  Radiative B Meson Decay}, Phys. Lett. B 202 (1988) 138--144.
\newblock \href {https://doi.org/10.1016/0370-2693(88)90868-4}
  {\path{doi:10.1016/0370-2693(88)90868-4}}.

\bibitem{Buchalla:1995vs}
G.~Buchalla, A.~J. Buras, M.~E. Lautenbacher, {Weak decays beyond leading
  logarithms}, Rev. Mod. Phys. 68 (1996) 1125--1144.
\newblock \href {http://arxiv.org/abs/hep-ph/9512380}
  {\path{arXiv:hep-ph/9512380}}, \href
  {https://doi.org/10.1103/RevModPhys.68.1125}
  {\path{doi:10.1103/RevModPhys.68.1125}}.

\bibitem{Jager:2012uw}
S.~J\"ager, J.~Martin~Camalich, {On $B \to V \ell \ell$ at small dilepton
  invariant mass, power corrections, and new physics}, JHEP 05 (2013) 043.
\newblock \href {http://arxiv.org/abs/1212.2263} {\path{arXiv:1212.2263}},
  \href {https://doi.org/10.1007/JHEP05(2013)043}
  {\path{doi:10.1007/JHEP05(2013)043}}.

\bibitem{Descotes-Genon:2014uoa}
S.~Descotes-Genon, L.~Hofer, J.~Matias, J.~Virto, {On the impact of power
  corrections in the prediction of $B \to K^*\mu^+\mu^-$ observables}, JHEP 12
  (2014) 125.
\newblock \href {http://arxiv.org/abs/1407.8526} {\path{arXiv:1407.8526}},
  \href {https://doi.org/10.1007/JHEP12(2014)125}
  {\path{doi:10.1007/JHEP12(2014)125}}.

\bibitem{Jager:2014rwa}
S.~J\"ager, J.~Martin~Camalich, {Reassessing the discovery potential of the $B
  \to K^{*} \ell^+\ell^-$ decays in the large-recoil region: SM challenges and
  BSM opportunities}, Phys. Rev. D 93~(1) (2016) 014028.
\newblock \href {http://arxiv.org/abs/1412.3183} {\path{arXiv:1412.3183}},
  \href {https://doi.org/10.1103/PhysRevD.93.014028}
  {\path{doi:10.1103/PhysRevD.93.014028}}.

\bibitem{Ciuchini:2015qxb}
M.~Ciuchini, M.~Fedele, E.~Franco, S.~Mishima, A.~Paul, L.~Silvestrini,
  M.~Valli, {$B\to K^* \ell^+ \ell^-$ decays at large recoil in the Standard
  Model: a theoretical reappraisal}, JHEP 06 (2016) 116.
\newblock \href {http://arxiv.org/abs/1512.07157} {\path{arXiv:1512.07157}},
  \href {https://doi.org/10.1007/JHEP06(2016)116}
  {\path{doi:10.1007/JHEP06(2016)116}}.

\bibitem{Capdevila:2017ert}
B.~Capdevila, S.~Descotes-Genon, L.~Hofer, J.~Matias, {Hadronic uncertainties
  in $B \to K^* \mu^+ \mu^-$: a state-of-the-art analysis}, JHEP 04 (2017) 016.
\newblock \href {http://arxiv.org/abs/1701.08672} {\path{arXiv:1701.08672}},
  \href {https://doi.org/10.1007/JHEP04(2017)016}
  {\path{doi:10.1007/JHEP04(2017)016}}.

\bibitem{Ciuchini:2017mik}
M.~Ciuchini, A.~M. Coutinho, M.~Fedele, E.~Franco, A.~Paul, L.~Silvestrini,
  M.~Valli, {On Flavourful Easter eggs for New Physics hunger and Lepton
  Flavour Universality violation}, Eur. Phys. J. C 77~(10) (2017) 688.
\newblock \href {http://arxiv.org/abs/1704.05447} {\path{arXiv:1704.05447}},
  \href {https://doi.org/10.1140/epjc/s10052-017-5270-2}
  {\path{doi:10.1140/epjc/s10052-017-5270-2}}.

\bibitem{Arbey:2018ics}
A.~Arbey, T.~Hurth, F.~Mahmoudi, S.~Neshatpour, {Hadronic and New Physics
  Contributions to $b \to s$ Transitions}, Phys. Rev. D 98~(9) (2018) 095027.
\newblock \href {http://arxiv.org/abs/1806.02791} {\path{arXiv:1806.02791}},
  \href {https://doi.org/10.1103/PhysRevD.98.095027}
  {\path{doi:10.1103/PhysRevD.98.095027}}.

\bibitem{Ball:2004rg}
P.~Ball, R.~Zwicky, {$B_{d,s} \to \rho, \omega, K^*, \phi$ decay form-factors
  from light-cone sum rules revisited}, Phys. Rev. D 71 (2005) 014029.
\newblock \href {http://arxiv.org/abs/hep-ph/0412079}
  {\path{arXiv:hep-ph/0412079}}, \href
  {https://doi.org/10.1103/PhysRevD.71.014029}
  {\path{doi:10.1103/PhysRevD.71.014029}}.

\bibitem{Bharucha:2015bzk}
A.~Bharucha, D.~M. Straub, R.~Zwicky, {$B\to V\ell^+\ell^-$ in the Standard
  Model from light-cone sum rules}, JHEP 08 (2016) 098.
\newblock \href {http://arxiv.org/abs/1503.05534} {\path{arXiv:1503.05534}},
  \href {https://doi.org/10.1007/JHEP08(2016)098}
  {\path{doi:10.1007/JHEP08(2016)098}}.

\bibitem{Khodjamirian:2010vf}
A.~Khodjamirian, T.~Mannel, A.~A. Pivovarov, Y.~M. Wang, {Charm-loop effect in
  $B \to K^{(*)} \ell^{+} \ell^{-}$ and $B\to K^*\gamma$}, JHEP 09 (2010) 089.
\newblock \href {http://arxiv.org/abs/1006.4945} {\path{arXiv:1006.4945}},
  \href {https://doi.org/10.1007/JHEP09(2010)089}
  {\path{doi:10.1007/JHEP09(2010)089}}.

\bibitem{Khodjamirian:2012rm}
A.~Khodjamirian, T.~Mannel, Y.~M. Wang, {$B \to K \ell^{+}\ell^{-}$ decay at
  large hadronic recoil}, JHEP 02 (2013) 010.
\newblock \href {http://arxiv.org/abs/1211.0234} {\path{arXiv:1211.0234}},
  \href {https://doi.org/10.1007/JHEP02(2013)010}
  {\path{doi:10.1007/JHEP02(2013)010}}.

\bibitem{Gubernari:2018wyi}
N.~Gubernari, A.~Kokulu, D.~van Dyk, {$B\to P$ and $B\to V$ Form Factors from
  $B$-Meson Light-Cone Sum Rules beyond Leading Twist}, JHEP 01 (2019) 150.
\newblock \href {http://arxiv.org/abs/1811.00983} {\path{arXiv:1811.00983}},
  \href {https://doi.org/10.1007/JHEP01(2019)150}
  {\path{doi:10.1007/JHEP01(2019)150}}.

\bibitem{Gubernari:2020eft}
N.~Gubernari, D.~Van~Dyk, J.~Virto, {Non-local matrix elements in $B_{(s)}\to
  \{K^{(*)},\phi\}\ell^+\ell^-$}, JHEP 02 (2021) 088.
\newblock \href {http://arxiv.org/abs/2011.09813} {\path{arXiv:2011.09813}},
  \href {https://doi.org/10.1007/JHEP02(2021)088}
  {\path{doi:10.1007/JHEP02(2021)088}}.

\bibitem{Bouchard:2013eph}
C.~Bouchard, G.~P. Lepage, C.~Monahan, H.~Na, J.~Shigemitsu, {Rare decay $B \to
  K \ell^+ \ell^-$ form factors from lattice QCD}, Phys. Rev. D 88~(5) (2013)
  054509, [Erratum: Phys.Rev.D 88, 079901 (2013)].
\newblock \href {http://arxiv.org/abs/1306.2384} {\path{arXiv:1306.2384}},
  \href {https://doi.org/10.1103/PhysRevD.88.054509}
  {\path{doi:10.1103/PhysRevD.88.054509}}.

\bibitem{Horgan:2013hoa}
R.~R. Horgan, Z.~Liu, S.~Meinel, M.~Wingate, {Lattice QCD calculation of form
  factors describing the rare decays $B \to K^* \ell^+ \ell^-$ and $B_s \to
  \phi \ell^+ \ell^-$}, Phys. Rev. D 89~(9) (2014) 094501.
\newblock \href {http://arxiv.org/abs/1310.3722} {\path{arXiv:1310.3722}},
  \href {https://doi.org/10.1103/PhysRevD.89.094501}
  {\path{doi:10.1103/PhysRevD.89.094501}}.

\bibitem{Grinstein:2002rm}
B.~Grinstein, D.~Pirjol, {Subleading corrections to the |V(ub)| determination
  from exclusive B decays}, Phys. Lett. B 549 (2002) 314--320.
\newblock \href {http://arxiv.org/abs/hep-ph/0209211}
  {\path{arXiv:hep-ph/0209211}}, \href
  {https://doi.org/10.1016/S0370-2693(02)02929-5}
  {\path{doi:10.1016/S0370-2693(02)02929-5}}.

\bibitem{Bobeth:2010wg}
C.~Bobeth, G.~Hiller, D.~van Dyk, {The Benefits of $\bar{B} -> \bar{K}^* l^+
  l^-$ Decays at Low Recoil}, JHEP 07 (2010) 098.
\newblock \href {http://arxiv.org/abs/1006.5013} {\path{arXiv:1006.5013}},
  \href {https://doi.org/10.1007/JHEP07(2010)098}
  {\path{doi:10.1007/JHEP07(2010)098}}.

\bibitem{Beylich:2011aq}
M.~Beylich, G.~Buchalla, T.~Feldmann, {Theory of $B \to K^{(*)}\ell^+ \ell^-$
  decays at high $q^2$: OPE and quark-hadron duality}, Eur. Phys. J. C 71
  (2011) 1635.
\newblock \href {http://arxiv.org/abs/1101.5118} {\path{arXiv:1101.5118}},
  \href {https://doi.org/10.1140/epjc/s10052-011-1635-0}
  {\path{doi:10.1140/epjc/s10052-011-1635-0}}.

\bibitem{Bobeth:2012vn}
C.~Bobeth, G.~Hiller, D.~van Dyk, {General analysis of $\bar{B} \to
  \bar{K}^{(*)}\ell^+ \ell^-$ decays at low recoil}, Phys. Rev. D 87~(3) (2013)
  034016.
\newblock \href {http://arxiv.org/abs/1212.2321} {\path{arXiv:1212.2321}},
  \href {https://doi.org/10.1103/PhysRevD.87.034016}
  {\path{doi:10.1103/PhysRevD.87.034016}}.

\bibitem{Bobeth:2017vxj}
C.~Bobeth, M.~Chrzaszcz, D.~van Dyk, J.~Virto, {Long-distance effects in
  $B\rightarrow K^*\ell \ell $ from analyticity}, Eur. Phys. J. C 78~(6) (2018)
  451.
\newblock \href {http://arxiv.org/abs/1707.07305} {\path{arXiv:1707.07305}},
  \href {https://doi.org/10.1140/epjc/s10052-018-5918-6}
  {\path{doi:10.1140/epjc/s10052-018-5918-6}}.

\bibitem{Ciuchini:2021smi}
M.~Ciuchini, M.~Fedele, E.~Franco, A.~Paul, L.~Silvestrini, M.~Valli, {New
  Physics without bias: Charming Penguins and Lepton Universality Violation in
  $b \to s \ell^+ \ell^-$ decays} (10 2021).
\newblock \href {http://arxiv.org/abs/2110.10126} {\path{arXiv:2110.10126}}.

\bibitem{iminuit}
H.~Dembinski, P.~O. et~al.,
  \href{https://doi.org/10.5281/zenodo.3949207}{scikit-hep/iminuit} (Dec 2020).
\newblock \href {https://doi.org/10.5281/zenodo.3949207}
  {\path{doi:10.5281/zenodo.3949207}}.
\newline\urlprefix\url{https://doi.org/10.5281/zenodo.3949207}

\bibitem{Bhom:2020lmk}
J.~Bhom, M.~Chrzaszcz, F.~Mahmoudi, M.~T. Prim, P.~Scott, M.~White, {A
  model-independent analysis of $b{\rightarrow }s\mu ^{+}\mu ^{-}$transitions
  with GAMBIT\textquoteright{}s FlavBit}, Eur. Phys. J. C 81~(12) (2021) 1076.
\newblock \href {http://arxiv.org/abs/2006.03489} {\path{arXiv:2006.03489}},
  \href {https://doi.org/10.1140/epjc/s10052-021-09840-z}
  {\path{doi:10.1140/epjc/s10052-021-09840-z}}.

\bibitem{Wilks:1938dza}
S.~S. Wilks, {The Large-Sample Distribution of the Likelihood Ratio for Testing
  Composite Hypotheses}, Annals Math. Statist. 9~(1) (1938) 60--62.
\newblock \href {https://doi.org/10.1214/aoms/1177732360}
  {\path{doi:10.1214/aoms/1177732360}}.

\bibitem{Isidori:2021vtc}
G.~Isidori, D.~Lancierini, P.~Owen, N.~Serra, {On the significance of new
  physics in $b\to s\ell^+\ell^-$ decays}, Phys. Lett. B 822 (2021) 136644.
\newblock \href {http://arxiv.org/abs/2104.05631} {\path{arXiv:2104.05631}},
  \href {https://doi.org/10.1016/j.physletb.2021.136644}
  {\path{doi:10.1016/j.physletb.2021.136644}}.

\bibitem{Isidori:2021tzd}
G.~Isidori, D.~Lancierini, A.~Mathad, P.~Owen, N.~Serra, R.~S. Coutinho, {A
  general effective field theory description of $b \to s l^+ l^-$ lepton
  universality ratios} (10 2021).
\newblock \href {http://arxiv.org/abs/2110.09882} {\path{arXiv:2110.09882}}.

\bibitem{Hurth:2018kcq}
T.~Hurth, A.~Arbey, F.~Mahmoudi, S.~Neshatpour, {New global fits to $b \to s$
  data with all relevant parameters}, Nucl. Part. Phys. Proc. 303-305 (2018)
  2--7.
\newblock \href {http://arxiv.org/abs/1812.07602} {\path{arXiv:1812.07602}},
  \href {https://doi.org/10.1016/j.nuclphysbps.2019.03.002}
  {\path{doi:10.1016/j.nuclphysbps.2019.03.002}}.

\bibitem{Alguero:2019pjc}
M.~Alguer\'o, B.~Capdevila, S.~Descotes-Genon, P.~Masjuan, J.~Matias, {What
  $R_K$ and $Q_5$ can tell us about New Physics in $b\to s\ell\ell$
  transitions?}, JHEP 07 (2019) 096.
\newblock \href {http://arxiv.org/abs/1902.04900} {\path{arXiv:1902.04900}},
  \href {https://doi.org/10.1007/JHEP07(2019)096}
  {\path{doi:10.1007/JHEP07(2019)096}}.

\bibitem{Crivellin:2019dun}
A.~Crivellin, D.~M\"uller, C.~Wiegand, {$b\to s\ell^+\ell^-$ transitions in
  two-Higgs-doublet models}, JHEP 06 (2019) 119.
\newblock \href {http://arxiv.org/abs/1903.10440} {\path{arXiv:1903.10440}},
  \href {https://doi.org/10.1007/JHEP06(2019)119}
  {\path{doi:10.1007/JHEP06(2019)119}}.

\bibitem{Bobeth:2016llm}
C.~Bobeth, A.~J. Buras, A.~Celis, M.~Jung, {Patterns of Flavour Violation in
  Models with Vector-Like Quarks}, JHEP 04 (2017) 079.
\newblock \href {http://arxiv.org/abs/1609.04783} {\path{arXiv:1609.04783}},
  \href {https://doi.org/10.1007/JHEP04(2017)079}
  {\path{doi:10.1007/JHEP04(2017)079}}.

\bibitem{HFLAV:2019otj}
Y.~S. Amhis, et~al., {Averages of b-hadron, c-hadron, and $\tau $-lepton
  properties as of 2018}, Eur. Phys. J. C 81~(3) (2021) 226.
\newblock \href {http://arxiv.org/abs/1909.12524} {\path{arXiv:1909.12524}},
  \href {https://doi.org/10.1140/epjc/s10052-020-8156-7}
  {\path{doi:10.1140/epjc/s10052-020-8156-7}}.

\bibitem{Grzadkowski:2010es}
B.~Grzadkowski, M.~Iskrzynski, M.~Misiak, J.~Rosiek, {Dimension-Six Terms in
  the Standard Model Lagrangian}, JHEP 10 (2010) 085.
\newblock \href {http://arxiv.org/abs/1008.4884} {\path{arXiv:1008.4884}},
  \href {https://doi.org/10.1007/JHEP10(2010)085}
  {\path{doi:10.1007/JHEP10(2010)085}}.

\bibitem{Capdevila:2017iqn}
B.~Capdevila, A.~Crivellin, S.~Descotes-Genon, L.~Hofer, J.~Matias, {Searching
  for New Physics with $b\to s\tau^+\tau^-$ processes}, Phys. Rev. Lett.
  120~(18) (2018) 181802.
\newblock \href {http://arxiv.org/abs/1712.01919} {\path{arXiv:1712.01919}},
  \href {https://doi.org/10.1103/PhysRevLett.120.181802}
  {\path{doi:10.1103/PhysRevLett.120.181802}}.

\bibitem{Crivellin:2018yvo}
A.~Crivellin, C.~Greub, D.~M\"uller, F.~Saturnino, {Importance of Loop Effects
  in Explaining the Accumulated Evidence for New Physics in B Decays with a
  Vector Leptoquark}, Phys. Rev. Lett. 122~(1) (2019) 011805.
\newblock \href {http://arxiv.org/abs/1807.02068} {\path{arXiv:1807.02068}},
  \href {https://doi.org/10.1103/PhysRevLett.122.011805}
  {\path{doi:10.1103/PhysRevLett.122.011805}}.

\bibitem{Capdevila:2016ivx}
B.~Capdevila, S.~Descotes-Genon, J.~Matias, J.~Virto, {Assessing lepton-flavour
  non-universality from $B\to K^*\ell\ell$ angular analyses}, JHEP 10 (2016)
  075.
\newblock \href {http://arxiv.org/abs/1605.03156} {\path{arXiv:1605.03156}},
  \href {https://doi.org/10.1007/JHEP10(2016)075}
  {\path{doi:10.1007/JHEP10(2016)075}}.

\bibitem{Descotes-Genon:2019bud}
S.~Descotes-Genon, A.~Khodjamirian, J.~Virto, {Light-cone sum rules for $B\to
  K\pi$ form factors and applications to rare decays}, JHEP 12 (2019) 083.
\newblock \href {http://arxiv.org/abs/1908.02267} {\path{arXiv:1908.02267}},
  \href {https://doi.org/10.1007/JHEP12(2019)083}
  {\path{doi:10.1007/JHEP12(2019)083}}.

\bibitem{Abe:2010gxa}
T.~Abe, et~al., {Belle II Technical Design Report} (11 2010).
\newblock \href {http://arxiv.org/abs/1011.0352} {\path{arXiv:1011.0352}}.

\bibitem{Descotes-Genon:2019dbw}
S.~Descotes-Genon, M.~Novoa-Brunet, {Angular analysis of the rare decay
  $\Lambda_b\to \Lambda(1520)(\to NK)\ell^+\ell^-$}, JHEP 06 (2019) 136,
  [Erratum: JHEP 06, 102 (2020)].
\newblock \href {http://arxiv.org/abs/1903.00448} {\path{arXiv:1903.00448}},
  \href {https://doi.org/10.1007/JHEP06(2019)136}
  {\path{doi:10.1007/JHEP06(2019)136}}.

\bibitem{Descotes-Genon:2020tnz}
S.~Descotes-Genon, M.~Novoa-Brunet, K.~K. Vos, {The time-dependent angular
  analysis of $B_d\to K_S\ell\ell$, a new benchmark for new physics}, JHEP 02
  (2021) 129.
\newblock \href {http://arxiv.org/abs/2008.08000} {\path{arXiv:2008.08000}},
  \href {https://doi.org/10.1007/JHEP02(2021)129}
  {\path{doi:10.1007/JHEP02(2021)129}}.

\bibitem{Alguero:2021yus}
M.~Alguer\'o, P.~A. Cartelle, A.~M. Marshall, P.~Masjuan, J.~Matias, M.~A.
  McCann, M.~Patel, K.~A. Petridis, M.~Smith, {A complete description of P- and
  S-wave contributions to the $B^{0}\to K^+\pi^-\ell^+\ell^-$ decay}, JHEP 12
  (2021) 085.
\newblock \href {http://arxiv.org/abs/2107.05301} {\path{arXiv:2107.05301}},
  \href {https://doi.org/10.1007/JHEP12(2021)085}
  {\path{doi:10.1007/JHEP12(2021)085}}.

\bibitem{LHCb:2013zuf}
R.~Aaij, et~al., {Differential branching fraction and angular analysis of the
  decay $B^{0} \to K^{*0} \mu^{+}\mu^{-}$}, JHEP 08 (2013) 131.
\newblock \href {http://arxiv.org/abs/1304.6325} {\path{arXiv:1304.6325}},
  \href {https://doi.org/10.1007/JHEP08(2013)131}
  {\path{doi:10.1007/JHEP08(2013)131}}.

\bibitem{LHCb:2015svh}
R.~Aaij, et~al., {Angular analysis of the $B^{0} \to K^{*0} \mu^{+} \mu^{-}$
  decay using 3 fb$^{-1}$ of integrated luminosity}, JHEP 02 (2016) 104.
\newblock \href {http://arxiv.org/abs/1512.04442} {\path{arXiv:1512.04442}},
  \href {https://doi.org/10.1007/JHEP02(2016)104}
  {\path{doi:10.1007/JHEP02(2016)104}}.

\bibitem{LHCb:2014cxe}
R.~Aaij, et~al., {Differential branching fractions and isospin asymmetries of
  $B \to K^{(*)} \mu^+ \mu^-$ decays}, JHEP 06 (2014) 133.
\newblock \href {http://arxiv.org/abs/1403.8044} {\path{arXiv:1403.8044}},
  \href {https://doi.org/10.1007/JHEP06(2014)133}
  {\path{doi:10.1007/JHEP06(2014)133}}.

\bibitem{LHCb:2016ykl}
R.~Aaij, et~al., {Measurements of the S-wave fraction in $B^{0}\rightarrow
  K^{+}\pi^{-}\mu^{+}\mu^{-}$ decays and the $B^{0}\rightarrow
  K^{\ast}(892)^{0}\mu^{+}\mu^{-}$ differential branching fraction}, JHEP 11
  (2016) 047, [Erratum: JHEP 04, 142 (2017)].
\newblock \href {http://arxiv.org/abs/1606.04731} {\path{arXiv:1606.04731}},
  \href {https://doi.org/10.1007/JHEP11(2016)047}
  {\path{doi:10.1007/JHEP11(2016)047}}.

\bibitem{LHCb:2020lmf}
R.~Aaij, et~al., {Measurement of $CP$-Averaged Observables in the
  $B^{0}\rightarrow K^{*0}\mu^{+}\mu^{-}$ Decay}, Phys. Rev. Lett. 125~(1)
  (2020) 011802.
\newblock \href {http://arxiv.org/abs/2003.04831} {\path{arXiv:2003.04831}},
  \href {https://doi.org/10.1103/PhysRevLett.125.011802}
  {\path{doi:10.1103/PhysRevLett.125.011802}}.

\bibitem{LHCb:2020dof}
R.~Aaij, et~al., {Strong constraints on the $b \to s\gamma$ photon polarisation
  from $B^0 \to K^{*0} e^+ e^-$ decays}, JHEP 12 (2020) 081.
\newblock \href {http://arxiv.org/abs/2010.06011} {\path{arXiv:2010.06011}},
  \href {https://doi.org/10.1007/JHEP12(2020)081}
  {\path{doi:10.1007/JHEP12(2020)081}}.

\bibitem{BELLE:2019xld}
S.~Choudhury, et~al., {Test of lepton flavor universality and search for lepton
  flavor violation in $B \rightarrow K\ell \ell$ decays}, JHEP 03 (2021) 105.
\newblock \href {http://arxiv.org/abs/1908.01848} {\path{arXiv:1908.01848}},
  \href {https://doi.org/10.1007/JHEP03(2021)105}
  {\path{doi:10.1007/JHEP03(2021)105}}.

\bibitem{LHCb:2017avl}
R.~Aaij, et~al., {Test of lepton universality with $B^{0} \rightarrow
  K^{*0}\ell^{+}\ell^{-}$ decays}, JHEP 08 (2017) 055.
\newblock \href {http://arxiv.org/abs/1705.05802} {\path{arXiv:1705.05802}},
  \href {https://doi.org/10.1007/JHEP08(2017)055}
  {\path{doi:10.1007/JHEP08(2017)055}}.

\bibitem{Belle:2019oag}
A.~Abdesselam, et~al., {Test of Lepton-Flavor Universality in ${B\to
  K^\ast\ell^+\ell^-}$ Decays at Belle}, Phys. Rev. Lett. 126~(16) (2021)
  161801.
\newblock \href {http://arxiv.org/abs/1904.02440} {\path{arXiv:1904.02440}},
  \href {https://doi.org/10.1103/PhysRevLett.126.161801}
  {\path{doi:10.1103/PhysRevLett.126.161801}}.

\bibitem{Belle:2016fev}
S.~Wehle, et~al., {Lepton-Flavor-Dependent Angular Analysis of $B\to K^\ast
  \ell^+\ell^-$}, Phys. Rev. Lett. 118~(11) (2017) 111801.
\newblock \href {http://arxiv.org/abs/1612.05014} {\path{arXiv:1612.05014}},
  \href {https://doi.org/10.1103/PhysRevLett.118.111801}
  {\path{doi:10.1103/PhysRevLett.118.111801}}.

\bibitem{ATLAS:2018gqc}
M.~Aaboud, et~al., {Angular analysis of $B^0_d \rightarrow K^{*}\mu^+\mu^-$
  decays in $pp$ collisions at $\sqrt{s}= 8$ TeV with the ATLAS detector}, JHEP
  10 (2018) 047.
\newblock \href {http://arxiv.org/abs/1805.04000} {\path{arXiv:1805.04000}},
  \href {https://doi.org/10.1007/JHEP10(2018)047}
  {\path{doi:10.1007/JHEP10(2018)047}}.

\bibitem{CMS:2017rzx}
A.~M. Sirunyan, et~al., {Measurement of angular parameters from the decay
  $\mathrm{B}^0 \to \mathrm{K}^{*0} \mu^+ \mu^-$ in proton-proton collisions at
  $\sqrt{s} = $ 8 TeV}, Phys. Lett. B 781 (2018) 517--541.
\newblock \href {http://arxiv.org/abs/1710.02846} {\path{arXiv:1710.02846}},
  \href {https://doi.org/10.1016/j.physletb.2018.04.030}
  {\path{doi:10.1016/j.physletb.2018.04.030}}.

\bibitem{CMS:2015bcy}
V.~Khachatryan, et~al., {Angular analysis of the decay $B^0 \to K^{*0} \mu^+
  \mu^-$ from pp collisions at $\sqrt s = 8$ TeV}, Phys. Lett. B 753 (2016)
  424--448.
\newblock \href {http://arxiv.org/abs/1507.08126} {\path{arXiv:1507.08126}},
  \href {https://doi.org/10.1016/j.physletb.2015.12.020}
  {\path{doi:10.1016/j.physletb.2015.12.020}}.

\bibitem{CMS:2018qih}
A.~M. Sirunyan, et~al., {Angular analysis of the decay B$^+\to$ K$^+\mu^+\mu^-$
  in proton-proton collisions at $\sqrt{s} =$ 8 TeV}, Phys. Rev. D 98~(11)
  (2018) 112011.
\newblock \href {http://arxiv.org/abs/1806.00636} {\path{arXiv:1806.00636}},
  \href {https://doi.org/10.1103/PhysRevD.98.112011}
  {\path{doi:10.1103/PhysRevD.98.112011}}.

\bibitem{CMS:2013mkz}
S.~Chatrchyan, et~al., {Angular Analysis and Branching Fraction Measurement of
  the Decay $B^0 \to K^{*0} \mu^+\mu^-$}, Phys. Lett. B 727 (2013) 77--100.
\newblock \href {http://arxiv.org/abs/1308.3409} {\path{arXiv:1308.3409}},
  \href {https://doi.org/10.1016/j.physletb.2013.10.017}
  {\path{doi:10.1016/j.physletb.2013.10.017}}.

\bibitem{ParticleDataGroup:2020ssz}
P.~A. Zyla, et~al., {Review of Particle Physics}, PTEP 2020~(8) (2020) 083C01.
\newblock \href {https://doi.org/10.1093/ptep/ptaa104}
  {\path{doi:10.1093/ptep/ptaa104}}.

\bibitem{Misiak:2020vlo}
M.~Misiak, A.~Rehman, M.~Steinhauser, {Towards $ \overline{B}\to {X}_s\gamma $
  at the NNLO in QCD without interpolation in m$_{c}$}, JHEP 06 (2020) 175.
\newblock \href {http://arxiv.org/abs/2002.01548} {\path{arXiv:2002.01548}},
  \href {https://doi.org/10.1007/JHEP06(2020)175}
  {\path{doi:10.1007/JHEP06(2020)175}}.

\bibitem{Descotes-Genon:2011nqe}
S.~Descotes-Genon, D.~Ghosh, J.~Matias, M.~Ramon, {Exploring New Physics in the
  C7-C7' plane}, JHEP 06 (2011) 099.
\newblock \href {http://arxiv.org/abs/1104.3342} {\path{arXiv:1104.3342}},
  \href {https://doi.org/10.1007/JHEP06(2011)099}
  {\path{doi:10.1007/JHEP06(2011)099}}.

\bibitem{Huber:2020vup}
T.~Huber, T.~Hurth, J.~Jenkins, E.~Lunghi, Q.~Qin, K.~K. Vos, {Phenomenology of
  inclusive $ \overline{B}\to {X}_s{\mathrm{\ell}}^{+}{\mathrm{\ell}}^{-} $ for
  the Belle II era}, JHEP 10 (2020) 088.
\newblock \href {http://arxiv.org/abs/2007.04191} {\path{arXiv:2007.04191}},
  \href {https://doi.org/10.1007/JHEP10(2020)088}
  {\path{doi:10.1007/JHEP10(2020)088}}.

\bibitem{BaBar:2013qry}
J.~P. Lees, et~al., {Measurement of the $B \to X_s l^+l^-$ branching fraction
  and search for direct CP violation from a sum of exclusive final states},
  Phys. Rev. Lett. 112 (2014) 211802.
\newblock \href {http://arxiv.org/abs/1312.5364} {\path{arXiv:1312.5364}},
  \href {https://doi.org/10.1103/PhysRevLett.112.211802}
  {\path{doi:10.1103/PhysRevLett.112.211802}}.

\end{thebibliography}

\end{document}